\documentclass[10pt,twocolumn]{article}
\usepackage{cite}
\usepackage{hyperref}
\usepackage{graphicx} % Required for the inclusion of images
\usepackage{amsmath} % Required for some math elements 
\usepackage{microtype}
\usepackage{xcolor}
\usepackage{times}
\usepackage{eurosym}
\usepackage{amssymb}
\usepackage{amsmath}
\usepackage{latexsym}
\usepackage{mathrsfs}
\usepackage{amsfonts}
\usepackage{bm}
\usepackage{geometry}
\begin{document}
\thispagestyle{empty}

\title{\textbf{Finite Size Scaling in Time Evolution During the Colorless-QCD Confining Phase Transition}}
\author{Salah Cherif$^{1,2,3}$, Madjid Lakhdar Hamou Ladrem$^{2,3,4}$\footnote{Corespanding autor: mladrem@yahoo.fr}, Zainab Zaki Mohammed Alfull$^4$ \\
and Mohammed Ahmed Abdulmalek Ahmed$^{2,5,6}$}

\maketitle
\begin{center}
$^1$Physics Department, Ghardaia University, Ghardaia, Algeria,\\
$^2$\textbf{L}aboratoire de \textbf{P}hysique et de \textbf{M}ath\'ematiques et \textbf{A}pplications (LPMA), ENS-Vieux Kouba (Bachir El-Ibrahimi), Algiers, Algeria.\\
$^3$Physics Department, ENS Elbachir El-Ibrahimi, Kouba, Algiers, Algeria. \\
$^4$Physics Department, Faculty of Science, Taibah University, Al-Madinah Al-Munawwarah,\\
Kingdom of Saudi Arabia,\\
$^5$Laboratory of Computational Sciences and Mathematical Physics, Institute for Mathematical Research, Universiti Putra Malaysia, 43400 Serdang, Selangor, Malaysia, \\
$^6$Physics Department,Taiz University in Turba,Taiz,Yemen.\\
\end{center} 

\abstract{
The time evolution of the expanding Colorless Partonic Matter, created in Ultra-Relativistic Heavy Ion Collisions and undergoing the confining phase transition towards a Hadronic Gas, is discussed in the context of a unified model combining our Colorless QCD-MIT Bag Model with the boost invariant Bjorken expansion. The Bjorken Equation in the case of a longitudinal expansion scenario of a non-ideal relativistic medium in finite volume is solved using certain initial conditions $(\tau_i,T(\tau_i))$ and their effect is studied in detail. The evolution of the temperature as a function of the proper time $T(\tau,V)$ is then obtained at different volumes. Different times characterising different scales of the whole time evolution, like the time of the finite volume transition point $\tau_0(V)$, the hadronic time $\tau_H(V)$ at which the hadronization is completed, the lifetime of the Colorless Partonic Plasma $\Delta \tau_{CPP}(V)$ and the lifetime of the confining phase transition $\Delta \tau_{PT}(V)$ are calculated and their finite size scaling properties are studied in detail. New finite size scaling laws are derived. Also, the time evolution of some Thermal Response Functions as the order parameter $\mathcal{H}(\tau,V)$, energy density $\epsilon(\tau,V)$, pressure $\mathscr{P}(\tau,V)$ and the sound velocity $\mathscr{C}_{s}(\tau,V)$ are investigated and studied in detail. We find that the time evolution of our system is really affected by the colorlessness requirement and the initial conditions of the partonic matter: the closer the volume is to the thermodynamic limit, the longer are the times and the lifetimes of the system. A detailed analysis of the temporal decreasing, in negative power, of the energy density $\epsilon(\tau,V) \propto \tau^{-\theta}$ in each of the three stages of the Bjorken expansion is carried out. It has been noticed that pressure anisotropy, which appears during the pre-equilibrium stage of the system, really affects the subsequent temporal evolution of the system.

\section{Introduction} \label{intro}

The Quantum ChromoDynamics (QCD) Equation of State (EoS) is of crucial importance for a better comprehension of the strongly interacting matter created in the Ultra-Relativistic Heavy-Ion Collisions (URHIC)\cite{QGP2018}.
Due to the non-perturbative nature of QCD at low temperature and small chemical potential where the system is strongly coupled, lattice QCD approach is the most successful method for determining the EoS in this part of the phase diagram. However, since QCD at high temperature is asymptotically free, an agreement between the results from lattice QCD calculations and predictions from perturbative theory is always realized.

It was noted even before the advent of QCD \cite{QCD1998,QCD2013}as the underlying theory of strongly interacting matter, that Hadronic Matter (HM) cannot exist as hadrons at an arbitrarily high temperature or density.
The existence of a limiting temperature was formulated in the context of the Hagedorn Resonance Gas Model\cite{HAG2016}. Indeed, in the context of QCD, the existence of a hadronic limiting temperature is synonymous with a
phase transition separating ordinary HM from a new phase of elementary strongly interacting Partonic Matter (PM).\\
The properties of the PM are currently under active experimental investigation using the URHIC at the Relativistic Heavy Ion Collider (RHIC) and the Large Hadron Collider (LHC). To make a connection between hot QCD predictions and experimental data, it is essential to formulate a general framework in order to describe the space-time evolution of strongly hot/dense matter produced in URHIC. The Relativistic Hydrodynamics (RH) using appropriate initial conditions is known to be one such framework. The hypothesis of considering dissipative effects beyond the approximation of an ideal PM fluid has been recognized recently both theoretically and experimentally. Among the most important results obtained from the RHIC and LHC experiments\cite{JetQuenchingExp,EllipticFlowExp}, is the fact that the PM was a nearly perfect relativistic fluid, very different from what was supposed to be, a partonic gas, due to the asymptotic freedom of hot QCD. In the case of peripheral collisions, the PM typically has an elliptic shape in the transverse plane. The appearance of spatial azimuthal anisotropy in the momentum distribution is intimately related to the parton-parton interaction. The stronger the parton-parton interaction, the more apparent the asymmetry. Thus, the asymmetry property is nothing but an indication that the system is not gas-like.

It is obvious and with great importance, that to understand the physical properties at finite temperature of strongly interacting matter, to study its EoS. It highlights the intimate relationship between the degrees of freedom and the different phases of the system. Over the years, we have acquired a solid knowledge concerning this EoS, in the high temperature regime from perturbation theory and in the low temperature regime from hadron gas phenomenology.

Hydrodynamics is an approach which can describe the motion of continuous medium based on their local properties and using the conservation laws of energy, momentum, and other conserved quantities. The most important role of hydrodynamics is to reduce the high number of degrees of freedom in the microscopic level to macroscopic variables describing the local properties of the fluid.

Precisely, in the present work we want to perform a detailed study of some interesting observables, outshining the physics of their time evolution using the EoS of the system undergoing a colorless QCD confining phase transition. At the transition point, the system exhibits, by definition, a singular behavior in some Thermal Response Function (TRF), which appears only in the thermodynamic limit. Any TRF, with the duality temperature-time relation, is translated to time Response Function (tRF). When using the time as a variable, this tRF manifest different behaviors from that of TRF and in some cases drastic changes appear like the disappearance of the finite discontinuity. Since Bjorken equation will be solved using the finite volume EoS of our system, certainly Finite Size Effects (FSE) will be observed in the extracted times and lifetimes. We believe that this is the first time that this kind of calculation dealing with the Finite Size Scaling (FSS) properties in times and lifetimes is done successfully and the corresponding FSE are quantitatively studied. Therefore, bulk times and lifetimes are calculated. Due to the colorlessness requirement, the model is not simple enough to obtain analytical results. Therefore, numerical methods using Mathematica software are employed to perform some calculations.

\section{Conversion of Partonic Matter into Hadronic Matter} \label{sec:01}

Due to the color charge confinement property, only the colorless part of the quark configurations would manifest themselves as physically observed particles. All hadrons created in the final stage of an URHIC are colorless.  Therefore the whole partonic plasma fireball needs to be in a colorless state. In URHIC processes ranging from low to high energies, there is a production of a multitude of particles (multiparticle production). The stronger growth of the multiplicity of these particles with the center of mass energy $\sqrt{s_{NN}}$ is generally described with a power law: $\propto s^{\alpha}$\cite{ParticleProduction}. Final-state interactions between the produced particles determine the dynamical evolution of the system. Most of these particles (hadrons) are created not only just in the first stage of the collision but also in the final stage after a subsequent evolution of the parton states.  We foresee the multiparticle processes proceeding through the production of unobservable partonic states in a first stage followed by the final stage consisting of colorless hadronic states. The evolution from the first to the final stage is described via the hadronization phenomena, where the parton fragmentation refers to the process of converting high-energy, colored partons into colorless hadronic jets. Fragmentation phenomena incorporate the non-perturbative effects at long distances in QCD of the hadronization process, which cannot be calculated analytically. At present, these can only be measured experimentally. The creation of the finite volume hot PM is strongly indicated because some important signatures are observed. One of these signatures is precisely the jet quenching phenomenon, which inform us what happens when a very energetic parton plows through the hot PM. These energetic partons are produced within the same collision that produces the hot PM itself. The physical interest is focused on how rapidly the energetic parton loses its energy when moving in the hot PM. The energy reduction of the jets is what we call "jet quenching". Also, and due to thermal effects of the hot medium the cross section of the hadronization and the fragmentation process decrease. This what it was calculated, in the context of TFD\cite{Ladrem2011,Ladrem2013}, showing that at the lowest-order that, as the temperature is increased, the hadron production rate decreases. It would be interesting to measure the role of temperature in converting partonic degrees of freedom to hadronic ones.

\section{MIT-Bag Model with Colorlessness Condition}
\subsection{Non-ideality from  Colorlessness Condition}
The confinement phenomenon concerns any partonic many-body system, and then  the Colorlessness Condition (CC) can be considered as an effect of the color interaction between partons rendering the system to be in a colorless state. One shall account for this parton-parton interaction by requiring that all hadronic states to be colorless with respect to the $SU(3)$ color gauge group. Not only at low temperatures the different hadrons are colorless, even at fairly high temperatures there was multiple evidence provided by lattice QCD calculations, that colorless many-parton clusters can propagate in the PM\cite{DeTar1985,DeTar1987}.
Another important point in favor of CC comes from the study of the $SU(n)$ configurations of $qq$, $q\bar{q}$ and $qqq$ systems using the explicit projection operators method, in which it has proved that color confinement phenomenon occurs only in colorless state\cite{Milton1983}.

This CC manifests itself as a non-ideal character in the EoS. The ideal gas approximation is quite relevant at high temperatures due to the property of asymptotic freedom in this regime. However, during the phase transition, when the hadronic system is prepared for the colorless deconfinement or just after, things are not so simple, in which the non-abelian character of the CPP manifests itself in an important manner. Consequently, the
consideration of perturbative and non-perturbative corrections of higher order is more than necessary.
The CC can be then considered as an interaction effect represented by different terms added to the ideal plasma EoS, improving the approximation of the ideal gas and generating an EoS of a non-ideal gas\cite{Ladrem2019}. This non-perturbative effect is visible in particular in the transition region. In the vicinity of the transition point, this strong non-perturbative effect dominates the deconfined state. Several models have been proposed to describe this phenomenon by assuming the appearance of massive quasi-particles, namely massive partons. Such a quasi-particle model has also been invoked in solid state physics and other fields of physics to study phase transitions in which a large part of the interaction between the real particles and the medium can be viewed as an effective masses of quasi-particles moving freely.

\subsection{Total Partition Function with Colorlessness Condition  }

The most important effect of the partonic interaction is that only the colorless states exist. Thus, imposing this condition on the total partition function of the partonic system should retain a large part of the non-perturbative aspects of the interaction. The color projection has to be carried out simultaneously for the whole partonic system, leading to an overall color correlation.

Logically, any theoretical approach which is intended to describe the QCD deconfinement phase transition rigorously must contain the CC in its formalism. In our previous work, a new method was developed which has allowed us to accurately calculate physical quantities which describe the deconfinement phase transition efficiently within the QCD-MIT bag model, including the CC and using a mixed phase system evolving in a finite total volume $V$: this is what we call the Colorless QCD MIT-Bag Model\cite{Ladrem2005,Ladrem2015}. The fraction of volume (defined by the parameter $\bm{h}$) occupied by the HM phase is given by: $V_{HM}=\bm{h}\,V$, and the remaining volume:
$V_{PM}=(1-\bm{h})\,V$ contains then the colorless PM phase. We neglect the different parton-parton, hadron-hadron and parton-hadron interactions, even if they must interact to achieve thermal equilibrium and in this case the total colorless partition function factorized in the final form reads:

\begin{eqnarray}
Ln\mathscr{Z}(h,T,V,\mu)= Ln\mathscr{Z}_{CPP}(h,T,V,\mu) \\ \notag
+ Ln\mathscr{Z}_{Vac}(h,T,V,\mu)+ Ln\mathscr{Z}_{HG}(h,T,V,\mu) 
\end{eqnarray}

where,
\begin{equation}
Ln\mathscr{Z}_{Vac}(h,T,V)=-(1-h)\mathfrak{B}V/T,
\end{equation}%
accounts for the difference between the real vacuum and the perturbative vacuum due to the color confinement meaning that the constant $\left( \mathfrak{B}\right) $ of the bag model represents the pressure on the surface of the bag in order to balance the outward pressure exerted by the partons moving inside the bag.
A non-variable Bag constant $\mathfrak{B}$ is not sufficient because the non-abelian nature of QCD leads to a complicated non-perturbative structure of QCD vacuum. For the hadronic phase, we simply assumed only pionic degrees of freedom and the corresponding partition function is just given by,
\begin{equation}
Ln\mathscr{Z}_{HG}(\mathnormal{h},T,V)=\mathbf{a}_{HG}\mathnormal{h}VT^{3}.
\end{equation}

Due to the internal symmetry of the color charge confinement, which still remains valid, one must reduce the number of the states contributing really to the total partition function of the partonic system by imposing specific constraint for a desired conﬁguration. Therefore, the grand canonical partition function of QCD many-parton system subject to colorlessness condition, by means of the color projection operator $\hat{\mathcal{P}}$ which selects those configurations that are allowed by the colorlessness constraint, can be written as,

\begin{eqnarray}
\mathscr{Z}_{CPP}(T,V_{PP},\mu )= \mathbf{Tr}(\hat{\mathcal{P}} e^{-\beta (\hat{\mathcal{H}} -\mu \hat{\mathcal{N}})}).
\end{eqnarray}

When considering the color symmetry group $\mathcal{SU}(N_c)$ with unitary representation $\hat{\mathcal{U}}(g)$ in a Hilbert space having the structure of a tensor product of all the Fock spaces as $\mathscr{H}=\mathscr{H}_{q} \otimes \mathscr{H}_{\bar{q}} \otimes \mathscr{H}_{g}$, the projection operator $\hat{\mathcal{P}}$ for the colorless conﬁguration is given as
\begin{equation}
\hat{\mathcal{P}}= d_{j} \int_{\mathcal{SU}(N_c)} d\mu(g) \chi_{j}^{*}(g) \hat{\mathcal{U}}(g).
\end{equation}
where $d_{j}=1$ and $\chi_{j}=1$ are the dimension and the character of the irreducible representation $j$ of $\mathcal{SU}(N_c)$ and $d\mu(g)$ is the invariant Haar measure, respectively. The colorless partition function for the partonic system contained in a volume $V_{PP},$ at a temperature $T$ and with quark chemical potential $\mu$, after a considerable amount of algebraic calculations using the group-theoretical projection method \cite{Redlich1980,Turko1981,ElzeCC,Islam2014,Zakout2,Abir2009} and without any formal details, becomes in its final form,
\begin{eqnarray}
\mathscr{Z}_{CPP}(T,V_{PP},\mu )&=& \frac{4}{9\pi ^{2}}\int_{-\pi }^{+\pi }\int_{-\pi}^{+\pi }d\varphi d\psi\hspace*{1cm}  \\ \notag
& & M(\varphi ,\psi ) e^{\mathcal{G}\left(\varphi ,\psi ,\frac{\mu }{T}\right)V_{PP}T^{3} },
\end{eqnarray}

where $M(\varphi ,\psi )$ is the weight function (Haar measure) given by:
\begin{equation}
M(\varphi ,\psi )=\left[ \sin \left( \frac{1}{2}(\psi +\frac{\varphi }{2})\right) \sin (\frac{\varphi }{2})\sin \left( \frac{1}{2}(\psi -\frac{%
\varphi }{2})\right) \right] ^{2},
\end{equation}
with,
\begin{equation}
\mathcal{G}(\varphi ,\psi ,\frac{\mu }{T})=\mathcal{G}(0 ,0 ,\frac{\mu }{T})
+\mathcal{G}_{QG}(\varphi ,\psi ,\frac{\mu }{T}).
\label{Gfunc}
\end{equation}
The two functions are given in terms of ($T,V,\mu$) variables as follows:
\begin{equation}
\mathcal{G}(0,0,\frac{\mu }{T})=\mathbf{a}_{QG}+\frac{N_{f}N_{c}}{6\pi ^{2}}
( \frac{\mu^{4}}{2T^{4}}+\frac{\mu^{2}\pi ^{2}}{T^{2}} )
\label{G00}
\end{equation}%
and
\begin{eqnarray}
\mathcal{G}_{QG}(\varphi ,\psi ,\frac{\mu }{T})= \hspace*{4cm} \notag \\
\frac{\pi ^{2}N_{c}N_{f}}{36}\sum_{q=r,b,g}\left\{ -1+\left( \frac{\left( \alpha _{q}-i(\frac{\mu }{T})\right) ^{2}}{\pi ^{2}}-1\right) ^{2}\right\} \notag \\
-\frac{\pi ^{2}N_{g}}{96}\sum_{g=1}^{4}\left( \frac{\left( \alpha_{g}-\pi \right) ^{2}}{\pi ^{2}}-1\right) ^{2} -\frac{N_{f}N_{c}}{6\pi ^{2}}( \frac{\mu^{4}}{2T^{4}}+\frac{\mu^{2}\pi ^{2}}{T^{2}} ).  
\label{GQGfunc}
\end{eqnarray}
The two factors $\mathbf{a}_{HG}$ and $\mathbf{a}_{QG}$ which are related to the degeneracy number of particles in the system are given by,
\begin{equation}
\left\{
\begin{array}{l}
\mathbf{a}_{QG}=\frac{\pi ^{2}}{180}(7N_{c}N_{f}+4N_{g}) \vspace*{0.25cm} \\
\mathbf{a}_{HG}=\frac{\pi ^{2}}{90}N_{\pi}
\end{array}
\label{AHGQGP}
\right.
\end{equation}
when $N_{f}$ , $N_{c}$, $N_{g}$  and $N_{\pi}$ being the number of quark flavours, of color charges, of gluons and of pions, respectively. $\ \alpha _{q}$ $\left( q=r,\,b,\,g\right) $ are the angles
determined by the eigenvalues of the color charge operators in eq. (\ref{Gfunc}):
\begin{equation}
\alpha _{r}=\frac{\varphi }{2}+\frac{\psi }{3},\;\alpha _{g}=-\frac{%
\varphi }{2}+\frac{\psi }{3},\;\alpha _{b}=-\frac{2\psi }{3},
\end{equation}%
and $\alpha _{g}$ $\left( g=1,...,4\right) $\ being: $\alpha _{1}=\alpha
_{r}-\alpha _{g},\;\alpha _{2}=\alpha _{g}-\alpha _{b},\;\alpha _{3}=\alpha
_{b}-\alpha _{r},\;\alpha _{4}=0.$
Thus, the partition function of the CPP is then given by,
\begin{equation}
Ln\mathscr{Z}_{CPP}\left( \mathnormal{h}\right) =Ln\mathscr{Z}_{PP}\left( \mathnormal{h}\right)+ Ln\mathscr{Z}_{CC}\left( \mathnormal{h}\right),
\label{ZCQGP}
\end{equation}%
where
\begin{eqnarray}
\mathscr{Z}_{CC}\left( \mathnormal{h},T,V,\mu\right) =\frac{4}{9\pi ^{2}}  \vspace*{0.25cm} \int_{-\pi }^{+\pi }\int_{-\pi }^{+\pi }d\varphi d\psi M(\varphi,\psi ) \notag \\
e^{\mathnormal{(1-h)\mathcal{G}_{QG}(\varphi ,\psi ,\frac{\mu }{T})VT^{3}} },
\label{CCQGP}
\end{eqnarray}
is the colorless part and,
\begin{equation}
Ln\mathscr{Z}_{PP}\left( \mathnormal{h},T,V,\mu\right) =\mathnormal{(1-\mathnormal{h})VT^{3}\mathcal{G}(0,0,\frac{\mu }{T})} .
\label{ZQGP}
\end{equation}%
is the PP part without the colorlessness condition.
Finally, the exact total colorless partition function is given by,%
\begin{eqnarray}
Ln\mathscr{Z}\left( \bm{h},T,V,\mu\right)=\tilde{\mathfrak{Z}}\left( \bm{h},T,V,\mu\right) \notag \\
=Ln\mathscr{Z}_{0}\left( \bm{h},T,V,\mu\right)+Ln\mathscr{Z}_{CC}\left( \bm{h},T,V,\mu\right)
\label{ToTCQGP}
\end{eqnarray}
with $\mathscr{Z}_{0}\left( \bm{h},T,V,\mu\right)$ is only the total grand partition function of the system without the CC, which can be rewritten in its most familiar form as obtained in earliest papers(see for example Ref.\citen{CGS1986}):
\begin{eqnarray}
Ln\mathscr{Z}_{0}(\bm{h},T,V,\mu)= VT^{3}\Bigg[\hspace*{4cm}\notag  \\
 \Bigg\{ \mathbf{a}_{QG} +\frac{N_{c}N_{f}}{6\pi ^{2}} \left( \pi ^{2}\frac{\mu ^{2}}{T^{2}}+\frac{\mu ^{4}}{2T^{4}} \right) \vspace*{0.25cm}-\frac{\mathfrak{B}}{T^{4}} \Bigg\} (1-\bm{h}) +\mathbf{a}_{HG} \bm{h} \Bigg]  \label{ToTNCC}
\end{eqnarray}
where $\left( \mathfrak{B}\right) $ accounts for the confinement of quarks and gluons by the real vacuum pressure exerted on the perturbative vacuum of the bag model. A non-variable bag constant $\mathfrak{B}$ is not sufficient because the non-abelian nature of QCD leads to the complicated non-perturbative structure of the QCD vacuum. We will investigate some additional consequences of the SU(3) colorlessness condition of partonic matter beyond those studied and published previously\cite{ElzeCC,Islam2014}, especially in the context where no approximation is used.

One of the advantages of factorizing the total partition function (\ref{ToTCQGP}) to be able to easily switch on/off the CC in the calculations.
As a consequence of its definition, the colorless part of the partition function $\mathscr{Z}_{CC}\left( \bm{q},T,V,\mu\right)$ possesses a certain main property which it is desirable to state at once. This property expresses the fact that if we want to omit the CC from the calculation and in order to recover the ordinary partition function $\mathscr{Z}_{0}(\bm{h},T,V,\mu)$ (Rel.\ref{ToTNCC}), just put this part of the partition function equal to one:
\begin{equation}
\mathscr{Z}_{CC}\left( \bm{q},T,V,\mu\right)=1.
\label{ZCC}
\end{equation}
Thus the CC influence of a specified parton on its environment is to reduce the number of the possible states of the whole system.

\subsection{Unified $\mathscr{L}_{m,n}$-Method : to be Colorless and not to be Colorless}
The definition of the Hadronic Probability Density Function (hpdf) in our model is
given by,
\begin{equation}
\bm{p}(\bm{h},V,T,\mu) \int\limits_{0}^{1}\mathscr{Z}(\bm{h},V,T,\mu)d\bm{h}=\mathscr{Z}(\bm{h},V,T,\mu).  \label{hpdf}
\end{equation}
Since our hpdf is directly related to the partition function of the system, it is believed that the whole information concerning the deconfinement phase transition is self-contained in this hpdf.
Then we can perform the calculation of the mean value of any TRF $\bm{Q}(\bm{h},T,\mu ,V)$ characterizing the system in the state $\bm{h}$ by,

\begin{eqnarray}
\mathcal{Q}(T,\mu ,V) =\langle \bm{Q}\left(\bm{h},T,\mu ,V\right) \rangle\hspace{2cm} \notag \\
=\int\limits_{0}^{1}\bm{Q}\left(\bm{h},T,\mu ,V\right) \bm{p}(\bm{h},V,T,\mu) d\bm{h}.
\label{meanh}
\end{eqnarray}

For the sole purpose of simplifying the different relationships in what follows, we define the mean value of the temperature(volume) derivative of $\tilde{\mathfrak{Z}}(\bm{h},T,\mu ,V)$ as
\begin{equation}
\partial_{x}\tilde{\mathcal{Z}}\left(T,V,\mu\right)=\langle \frac{\partial \tilde{\mathfrak{Z}}(\bm{h},T,\mu ,V)}{\partial x}  \rangle , (x=T,V)
\end{equation}

In our previous work, as mentioned above, a new method was developed, which has allowed us to calculate easily physical quantities describing well the deconfinement phase transition to a PM in a finite volume $V$ \cite{Ladrem2005,Ladrem2015,Ladrem2019}, namely the $\mathscr{L}_{m,n}$ method.

This method will enable us to carry out the study of the hydrodynamical time evolution of the system using the EoS of the Colorless QCD MIT-Bag model.

The most important result consists in the fact that practically all TRF calculated in this context can be simply expressed as a function of only a certain double integral coefficient $\mathscr{L}_{m,n}\left( \bm{q},T,V\right)$, as defined in our previous articles \cite{Ladrem2005,Ladrem2015,Ladrem2019} and given by:

\begin{eqnarray}
\mathscr{L}_{m,n}\left( \mathnormal{q},T,V\right) =\int_{-\pi }^{+\pi }\int_{-\pi }^{+\pi}d\varphi d\psi M(\varphi ,\psi )(\mathcal{G}(\varphi ,\psi,0 ))^{m} \vspace*{0.25cm}\notag \\
\frac{e^{\mathnormal{q\ \mathscr{R}}\left( \varphi ,\psi ;T,V\right) }}{\left( \mathbf{\mathscr{R}}\left( \varphi ,\psi ;T,V\right) \right) ^{n}}, \hspace*{3cm} \label{Lmn}
\end{eqnarray}
where the function $\mathscr{R}\left( \varphi ,\psi ;T,V\right)$ is given by,
\begin{equation}
\mathscr{R}\left( \varphi ,\psi ;T,V\right) =\left( \mathcal{G}(\varphi ,\psi,0 )-\mathbf{a}_{HG}-\frac{\mathfrak{B}}{T^{4}}\right) VT^{3}.
\label{Rfunc0}
\end{equation}

It's obvious that these $\mathscr{L}_{m,n}(\bm{q},T,V)$ are just state functions depending on ($T$ ,$V$) and of course on state variable $ \bm{q}$, and they can be calculated numerically at each value of temperature $T$ and volume $V$.

The idea is to try to rewrite the $\mathscr{L}_{m,n}\left( \bm{q},T,V\right)$, in a way to make appear the $Z_{CC}$ term, and thus be able to play easily on the taking into account of the CC. We extrapolate in a very transparent way the colorless$\mathscr{L}_{m,n}$-method towards non-colorless case. In this sense, we build a unified treatment of both cases, such that there is no fundamental difference defining an artificial border between colorless confinement phase transition and non-colorless confinement phase transition.
In order to have the same mathematical description of our system with and without the CC, one starts to re-write these $\mathscr{L}_{m,n}\left( \bm{q},T,V\right)$ in an important form,
\begin{eqnarray}
\mathscr{L}_{m,n}\left( \bm{q},T,V\right) &=& \frac{9\pi^2}{4}\sum_{k=0}^{m}C_m^k\frac{\mathcal{K}^k}{(VT^3)^{m-k}}\Big[\int \mathcal{D}q\Big]^{n+k-m} \notag \\
& & e^{\bm{q}\mathscr{R}\left(0,0;T,V\right)}\mathscr{Z}_{CC}\left( \bm{q},T,V,0\right) 
\label{Lmnq}
\end{eqnarray}
where $C_m^k=\frac{m!}{k!(m-k)!}$, $\mathcal{K}=\mathbf{a}_{HG}+\frac{\mathfrak{B}}{T^{4}}$ and

\begin{equation}
\Big[\int \mathcal{D}q\Big]^{p}= \underbrace{\int \int \int ....\int dq dq dq....dq}_{p-times}
\label{DLmnq}
\end{equation}
The function $\mathscr{R}\left( 0 ,0 ;T,V\right)$, can be deduced from the general form given by the relation (\ref{Rfunc0}),
\begin{equation}
\mathscr{R}\left( 0 ,0 ;T,V\right) =\left( \mathbf{a}_{QG}-\mathcal{K}\right) VT^{3}.
\label{Rfunc}
\end{equation}
As a consequence of the new definition of the $\mathscr{L}_{m,n}$ given by the relation (\ref{Lmnq}) and using the property (\ref{ZCC}), we can derive the appropriate relation giving the $\mathscr{L}_{m,n}$ without the CC,
\begin{eqnarray}
\mathscr{L}_{m,n}^{NCC}\left( \bm{q},T,V\right) = \frac{9\pi^2}{4}\sum_{k=0}^{m}C_m^k\frac{\mathcal{K}^k}{(VT^3)^{m-k}}\notag \\
\Big[\int \mathcal{D}q\Big]^{n+k-m}e^{\bm{q}\mathscr{R}\left(0,0;T,V\right)}.
\label{LmnqNCC}
\end{eqnarray}
Therefore, we summarize the two definitions of the $\mathscr{L}_{m,n}$ :
\begin{equation}
\left\{
\begin{array}{c}
\mathscr{L}_{m,n}\left( \bm{q},T,V\right) \text{with CC}  \\
\mathscr{L}_{m,n}^{NCC}\left( \bm{q},T,V\right) \text{without CC}.   \\
\end{array}%
\right.  \label{UnifiedLmn}
\end{equation}

We have a potent formula allowing us a direct translation of any function of these $\mathscr{L}_{m,n}$ when the CC is taken into account into the same function using $\mathscr{L}_{m,n}^{NCC}$ when the CC is not included in the calculation and vice versa. This makes the analytical transition from colorless case to no-colorless case very simple.

\begin{equation}
\left\{
\begin{array}{c}
\mathcal{Q}(T,\mu ,V)_{CC} = f(\mathscr{L}_{m,n}).\\
\mathcal{Q}(T,\mu ,V)_{NCC} = f(\mathscr{L}_{m,n}^{NCC}).   \\
\end{array}%
\right.  \label{CCvsNCC}
\end{equation}

We will consider the case of vanishing chemical potential $\left( \mu =0\right) $ and on only two lightest quarks $u$\ and $d$ $\left( N_{f}=2\right) $ and a massless pionic matter. The common value $%
\mathfrak{B}^{1/4}=145MeV$ for the Bag constant will be used.

\section{Finite Volume Colorless QCD Thermodynamics: Results and Discussions}
\subsection{\textbf{Order Parameter $\mathcal{H}(T,V)$ and Violette Term $\mathscr{V}(T,V)$ }}
Two quantities are very important in our model which are : the mean value of the hadronic volume fraction $\mathcal{H}(\tau,V)$, which is considered as the order parameter of the phase transition under consideration and the mean value of the partonic degrees of freedom number $\mathscr{V}(T,V)$ \cite{Mhamed2014}. According to (\ref{hpdf},\ref{meanh}), these quantities are given by,
\begin{equation}
\mathcal{H}(T,V)=\int\limits_{0}^{1} \bm{h}\, \bm{p}(\bm{h},T,V) d\bm{h}.
\end{equation}
We can also define the fraction of the total volume occupied by the partonic plasma as $\bm{q}=1-\bm{h}$ and its mean value as:
\begin{equation}
\mathscr{Q}(T,V)=\int\limits_{0}^{1}\bm{q }\, \bm{p}(\bm{h},T,V) d\bm{h}.=1-\mathcal{H}(T,V).
\end{equation}
First of all, we shall define the following TRF as representing the difference of the state function $\mathscr{L}_{0,1}(q)$ in the PM state and in the HM state
\begin{equation}
\Delta \mathscr{L}_{0,1}=\mathscr{L}_{0,1}\left( 1\right) -\mathscr{L}_{0,1}\left( 0\right)
\label{DeLmn}
\end{equation}
In terms of $\mathscr{L}_{m,n}$, the three quantities thus evoked before are given by,
\begin{equation}
\mathcal{H}(T,V) \Delta \mathscr{L}_{0,1}=\mathscr{L}_{0,2}\left( 1\right)-\mathscr{L}_{0,2}\left( 0\right)
-\mathscr{L}_{0,1}\left( 0\right),
\label{MOP}
\end{equation}
\begin{equation}
\mathscr{Q}(T,V) \Delta \mathscr{L}_{0,1} =\mathscr{\mathscr{L}}_{0,2}\left( 0\right) -\mathscr{L}_{0,2}\left( 1\right)
+\mathscr{L}_{0,1}\left( 1\right),
\label{meanQTV}
\end{equation}%
\begin{equation}
\mathscr{V}(T,V) \Delta \mathscr{L}_{0,1}=
\mathscr{L}_{1,1}\left( 1\right)-\mathscr{L}_{1,2}\left( 1\right) +\mathscr{L}_{1,2}\left(0\right),
\label{meanV}
\end{equation}%
Also we can define the thermal derivatives of these functions as :
\begin{itemize}
  \item Hadronic Thermal Susceptibility
  \begin{equation}
  \chi^{H} _{T}\left( T,V\right) =\partial_{T}\mathcal{H}(T,V)=-\partial_{T}\mathscr{Q}(T,V)
  \label{ThHSusc}
  \end{equation}
  \item Partonic Thermal Susceptibility
  \begin{equation}
  \chi^{P} _{T}\left( T,V\right) =\partial_{T}\mathscr{V}(T,V)
  \label{ThQGSusc}
  \end{equation}
\end{itemize}
We have shown that any TRF investigated within our model\cite{Mhamed2014}, can be written by means of these fundamental quantities.
\subsection{\textbf{Energy Density  $\epsilon(T,V)$ and Entropy Density $\mathscr{S}(T,V)$} }
The mean value of energy density $\epsilon (T,V)$ is the most important TRF and using the definition
\begin{equation}
\epsilon (T,V) =\frac{T^{2}}{V} \partial_{T} \tilde{\mathcal{Z}}\left(T,V\right) \label{EDensity}
\end{equation}
thus, collecting terms to show up the fundamental TRF $\mathcal{H}(T,V)$, $\mathscr{Q}(T,V)$ \ and $\mathscr{V}(T,V)$\, we obtain,%
\begin{equation}
\epsilon (T,V)=3T^{4}[\mathbf{a}_{HG}\mathcal{H}(T,V)+\mathscr{V}(T,V)] +\mathfrak{B}\mathscr{Q}(T,V).  \label{EDensity2}
\end{equation}
We also can calculate the thermal derivative of the $\epsilon (T,V)$,
\begin{eqnarray}
c_{T}\left( T,V\right) =\partial_{T}\epsilon (T,V)= 3T^{4}(\mathbf{a}_{HG}\chi^{H} _{T}+\chi^{P} _{T}) \notag \\
+12T^{3}\left[\mathbf{a}_{HG}\mathcal{H}(T,V)+\mathscr{V}(T,V)\right] -\mathfrak{B}\chi^{H} _{T}.  \label{SpeHeat}
\end{eqnarray}
 representing the specific heat of the system. When, inserting (\ref{MOP},\ref{meanQTV},\ref{meanV}) in (\ref{EDensity2}), we get the final expression of $\epsilon(T,V)$ in terms of $\mathscr{L}_{m,n}\left( \bm{q},T,V\right) $. However, a problem appears when trying to calculate the Entropy Density directly from its standard definition, but we can perform the calculation easily from the energy density $\epsilon (T,V)$ or the specific heat $c_{T} (T,V)$:

\begin{eqnarray}
\mathscr{S}(T,V) =\int T^{-1} \partial_{T}\epsilon (T,V) dT\hspace*{1cm} \notag\\
= \int  T^{-1} c_{T}(T,V) dT=T^{-1} \epsilon (T,V) \notag \\
+ \int T^{-2} \epsilon (T,V) dT\hspace*{1cm}  \label{EntropyDensity}
\end{eqnarray}

\subsection{\textbf{Thermodynamic Pressure $ \mathscr{P}(T,V)$}, EoS and Sound Velocity $\mathscr{C}_{s}^{2}(T,V)$}
The mean value of the thermodynamic pressure is calculated using the standard definition:
\begin{equation}
\mathscr{P}(T,V) =T \partial_{V} \tilde{\mathcal{Z}}\left(T,V\right),  \label{Press}
\end{equation}
and after some mathematical calculation we can arrive at the final expression showing the contribution of each phase,
\begin{equation}
\mathscr{P}(T,V) =\mathbf{a}_{HG}T^{4}\mathcal{H}(T,V)-\mathfrak{B}\mathscr{Q}(T,V) +T^{4}\mathscr{V}(T,V).  \label{Pressure}
\end{equation}

A rapid growth of the pressure was clearly observed just after the transition temperature(see Ref. \citen{Ladrem2019}), especially when the volume approaches the thermodynamic limit. When expressing the pressure as a function of the energy density, the relation becomes more simple in its compact form :%
\begin{equation}
\mathscr{I}(T,V)=\epsilon(T,V)-3\mathscr{P}(T,V)=4\mathfrak{B}\mathscr{Q}(T,V).  \label{pressureandDE}
\end{equation}

\begin{figure}
\resizebox{1.0\hsize}{!}{
\includegraphics*{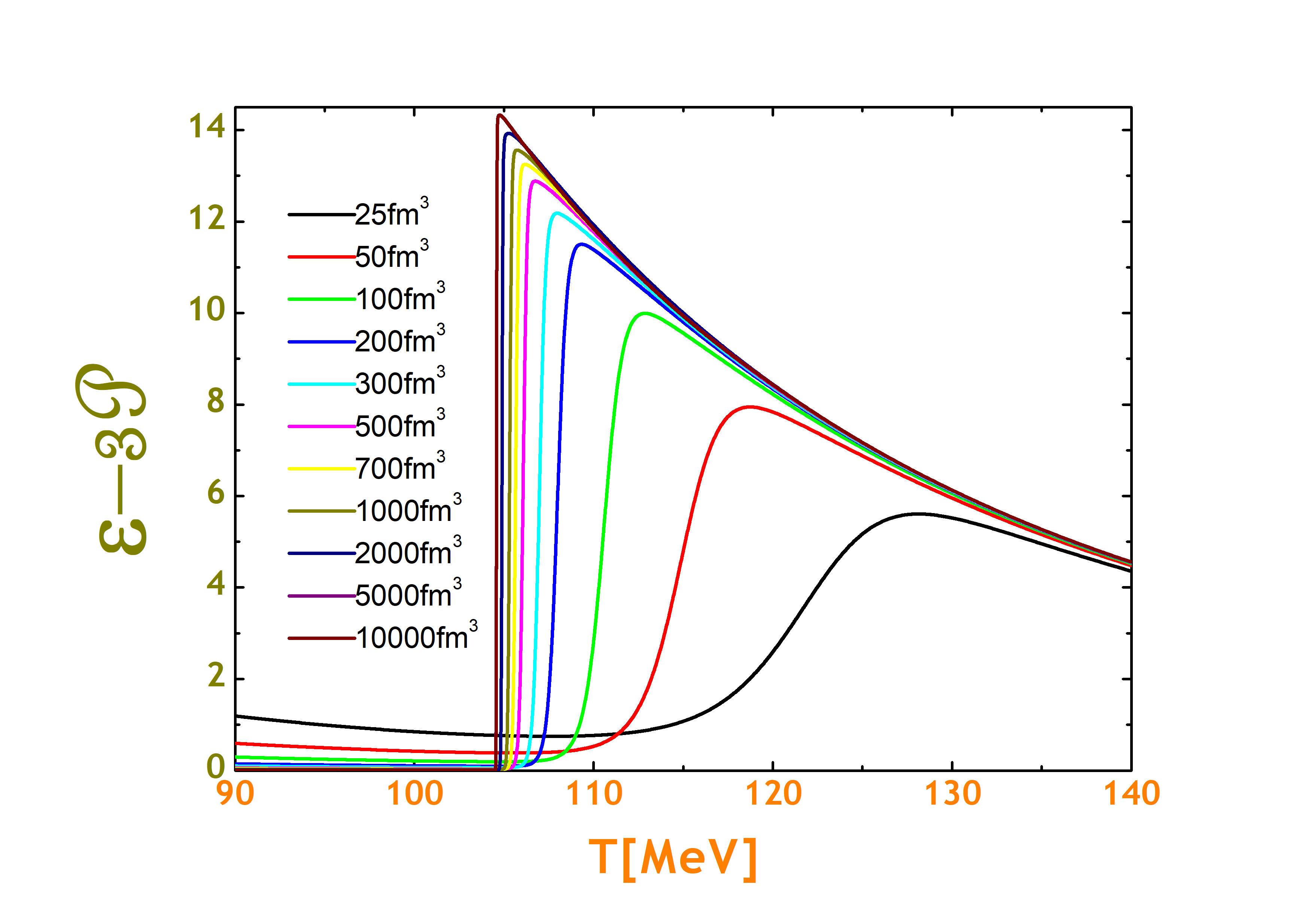}}
\caption{Trace Anomaly from our EoS.}\label{Fig:100}
\end{figure}

The difference between $\epsilon$ and $3\mathscr{P}$ is known as the interaction measure(or trace anomaly) $\mathscr{I}(T,V)$, which vanishes for a massless, non-interacting gas and is non-zero in the case of an interacting system. It has been well understood in our Colorless QCD-MIT Bag Model how the non-zero value of the interaction measure is related to the CC and the vacuum contribution. The appearance in our EoS of the term $4\mathfrak{B}\mathscr{Q}(T,V)$, in which the TRF $\mathscr{Q}(T,V)$ (Rel.\ref{meanQTV}) is written based on the complex form of the color double integral $\mathscr{L}_{m,n}\left( \bm{q},T,V\right)$ (Rel.\ref{Lmn}), makes it very complicated analytically. What we mean is that the complexity of our EoS emerges from the presence of color symmetry. This term marks well the difference with the relativistic EoS, commonly used: $3\mathscr{P}=\epsilon$. It is connected to everything that happens during the deconfinement phase transition; therefore, one recognizes its non-perturbative origin. The plots given in the figure (Fig.\ref{Fig:100}) illustrate the variation of the interaction measure $\mathscr{I}(T,V)$ as a function of temperature for different volumes\cite{Mhamed2014}. We find that our results are in good agreement with the common behavior of the trace anomaly. In particular, at low temperatures, $\mathscr{I}(T,V)$ rises rapidly, and at high temperatures, it slowly decreases, letting emerge a maximum point just beyond the finite volume transition point. The volume dependence is small in both low and high temperatures. However, this dependence is noticeable in the central region, around the maximum. Just above the finite volume transition temperature $T_0(V)$, the system is not in an ideal state. The mutual interactions between partons lead to a resulting non vanishing interaction measure, which manifests a good agreement with different models, like the hot lattice QCD calculations. The fact that even at high temperatures, the pressure, energy density, and entropy density of the partonic matter are far from their ideal gas values indicates substantial remaining interactions among the partons, highlighting the non-ideal nature of the strongly coupled PM.

Another quantity that will play an important role in the hydrodynamic expansion is the relativistic sound velocity $\mathscr{C}_{s}^{2}(T,V)$, it measures the system's tendency to expand and the speed at which linear sound waves propagate. When calculated from its definition, in the context of our model, we obtain,
\begin{equation}
3 \mathscr{C}_{s}^{2}(T,V)=3 \left(\dfrac{\partial \mathscr{P}}{\partial \epsilon }\right)= 1+4\mathfrak{B} \chi^{H} _{T} c^{-1}_{T}. \label{FirstSoundV}
\end{equation}
A relationship that shows how sound velocity $\mathscr{C}_{s}^{2}(T,V)$ is connected to both hadronic thermal susceptibility $\chi^{H} _{T}\left( T,V\right)$ and specific heat density $c_{T}\left( T,V\right)$. Finite volume TRF, such as pressure and sound velocity, show deviations from conformal ideal gas behavior even beyond the finite volume transition point $T_{0}(V)$ and approach the ideal gas limit relatively in a rapid way than in the case of the lattice QCD simulation \cite{Ladrem2019}. Here also, we can rewrite $\mathscr{P}(T,V)$ and $\mathscr{C}_{s}^{2}(T,V)$ as functions of $\mathscr{L}_{m,n}\left( \bm{q},T,V\right)$ using the relations (\ref{MOP}-\ref{ThQGSusc}). When, one contemplates our EoS given by the two relations(\ref{pressureandDE},\ref{FirstSoundV}) one notices the presence of an additional term, a complicated non-perturbative term due to the confinement phenomenon, marking a clear and important difference with the conformal EoS. The general form of our EoS is very similar to that obtained from lattice QCD calculations\cite{Ejiri2006,Cheng2008}.

\section{Hydrodynamic Evolution in the Boost Invariant Bjorken Model}\label{sec:03}
\subsection{The frame-independence or just the boost invariance symmetry}
The boost invariance property of the multiple particle production probably comes from ideas developed by Feynman \cite{Feynman1969} in its famous paper in considering the multiple hadron production in high energies as a phenomenon of field radiation. The boost invariance is just the symmetry of the physical systems with respect to the Lorentz boosts along the beam axis, from which particular constraints are imposed on the form of the related physical quantities.
The frame-independence symmetry or just the boost invariance symmetry is very frequently used in the models describing the evolution of matter created in the URHIC. It reduces the number of independent variables and facilitates theoretical calculations. Bjorken in his famous paper has succeeded to give global importance in using this symmetry \cite{Bjorken1983}, where he implemented this symmetry into hydrodynamic equations and made estimates of the initial energy density accessible in the URHIC. From the experimental point of view, one can notice that the boost-invariance may be regarded as the good approximation only for the central region of an URHIC.
In other words, there is an invariance of the system in all frames related to the center of mass frame by a boost transformation along the longitudinal direction.
In the Bjorken model, and after a certain proper time, we assume  the system acquires this boost invariance symmetry allowing a further expansion governed by the hydrodynamic equation of motion. This symmetry provides a natural mechanism which eventually leads to a flat inclusive longitudinal rapidity distribution and it also admits a sharp cutoff in the inclusive transverse momentum distribution. These features differ from those of Landau's model\cite{Wong1994,Florkowski2010}.
The hydrodynamic evolution preserves this boost-invariance, which in URHIC is expected to be realized near mid-rapidity \cite{Bjorken1983}.

\subsection{Hydrodynamical Description in Milne Space-Time}
We shall begin this section by recalling the important hydrodynamic equations of the perfect fluid by emphasizing the crucial role played by the initial conditions,  that should be very appropriate in the URHIC processes. Let's remember that the system subject to our study and formed in URHIC starts expanding soon as it is produced. In both sides of the deconfinement phase transition, it contains a very huge number of particles in small finite volume (partons in the PM and hadrons in the HM). If these particles interact strongly enough, the \ system may reach a state of local thermodynamic equilibrium. If it can be locally maintained during the subsequent
expansion, the further evolution of the PM and HM can be described conveniently by RH. Hydrodynamics is a macroscopic approach which describes the system by macroscopic variables, such as local energy density $\epsilon(T,V)$, pressure $ \mathscr{P}(T,V)$, sound velocity $ \mathscr{C}_{s}(T,V)$, the dynamical number of degrees of freedom and entropy density $\mathscr{S}(T,V)$. It requires knowledge of the EoS, which gives a relation between pressure, energy and entropy density, but not detailed knowledge of the microscopic dynamics. The simplest version is, of course the ideal RH, in which we totally neglect viscous effects and assume that local equilibrium
is always perfectly maintained during the system expansion\cite{RFD2019}. Fundamentally speaking, this is fulfilled when the scattering time is very much shorter than the macroscopic evolution time and that the mean free path is much smaller than the system size. We say that collisions between particles keep the system in approximate local thermal equilibrium. If this is not satisfied, viscous effects manifest, and one can take them into account only when the deviation from local equilibrium remains small. When the system is far away from equilibrium, one has to switch to a kinetic theory approach, such as parton or hadron cascade models. Therefore, hydrodynamics allow us to study the deconfinement phase transition to the CPP in a simple, straight forward manner. We notice that the EoS  used in our analytical calculation provide us the good behavior of the hydrodynamic evolution starting from the PM phase.
For a neutral fluid, the equations for hydrodynamics are simply the stress tensor and entropy conservation laws:

\begin{equation}
\left\{
\begin{array}{c}
\partial_{\mu}\mathscr{T}^{\nu\mu} = 0  \\
\partial_{\mu}\mathscr{S} \mathscr{U}^{\mu} = 0 \\
\end{array}%
\right.  \label{ConservationLaws}
\end{equation}

where $\mathscr{T}^{\nu\mu}$ denotes the expectation value of the quantum stress tensor operator. $\mathscr{T}^{\nu\mu}$ is in turn expressed via constitutive relations in terms of a derivative expansion of four hydrodynamic
fields which we will choose to be the temperature $T$ in the local fluid rest frame and the local fluid four-velocity $\mathscr{U}^{\nu}$, normalized according to $\mathscr{U}^{\nu}\mathscr{U}_{\nu}=-1$. Up to first order in
derivatives, $\mathscr{T}^{\nu\mu}$ can be written as
\begin{eqnarray}
\mathscr{T}^{\nu\mu} = \epsilon(T,V) \mathscr{U}^{\nu}\mathscr{U}^{\mu} +  \mathscr{P}(T,V) \Delta^{\nu\mu}\notag \\
 - \sigma(T,V) \Pi^{\nu\mu}- \Sigma(T,V) \partial_{\alpha}\mathscr{U}^{\alpha}  \Delta^{\nu\mu}  \label{HydTensor}
\end{eqnarray}
where
\begin{equation}
\left\{
\begin{array}{c}
\Delta_{\nu\mu}= \eta_{\nu\mu}+ \mathscr{U}_{\nu}\mathscr{U}_{\mu}  \\
\Pi_{\nu\mu}= (\Delta_{\nu}^{\alpha} \Delta_{\mu}^{\beta} -\frac{1}{3}\Delta_{\nu\mu}\Delta^{\alpha\beta})(\partial_{\alpha}\mathscr{U}_{\beta}+\partial_{\beta}\mathscr{U}_{\alpha}) \\
\end{array}%
\right.  \label{HydTensor2}
\end{equation}

Based on the tensor algebra rules, the indices are raised and lowered using the Minkowski metric : $ \eta_{\nu\mu}$. The coefficients $\sigma(T,V)$ and $ \Sigma(T,V)$ are the shear and bulk viscosities, respectively. We can also
continue the derivative expansion (\ref{HydTensor}) to any higher order by including all possible terms allowed by symmetries and the local second law of thermodynamics.

In URHIC at RHIC and LHC, approximate invariance under the longitudinal Lorentz boost is observed in particle rapidity distributions around mid-rapidity. The Bjorken scaling expansion of the system can best be
described by using the proper time $\tau=\sqrt{t^2 - z^2}$ and space-time rapidity $\eta =tanh(1-(z/t))$ variables, usually called Milne coordinates. The central point is the one dimension ansatz for the 4-velocity $\mathscr{U}_{\mu}=(t,0,0,z)/\tau$; indicating the proportionality between velocity in the z-direction $\mathscr{U}_{\mu}$ to $z$, which is an analogy to
three-dimensional Hubble flow of the universe. it is also called one-dimensional Hubble flow.

Note that all volume elements are expanded linearly with time and move along straight lines from the collision point. Exactly as in the big bang, for each “observer” (the volume element) the picture is just the same.
The history is also the same for all volume elements if it is expressed in its own proper time $\tau$. In the context of the Bjorken symmetry, any TRF becomes independent of $\eta$ and
depends solely on the proper time $\tau$. When composing this TRF with the duality temperature-time relation $T(\tau)$,  we get the corresponding tRF.

The boost invariance in the Bjorken model means that the central rapidity regime is initially approximately Lorentz invariant under longitudinal boosts, so that conditions at point $z$ at time $t$ are the same as those at $z=0$ at proper time $\tau$. Such invariance greatly simplifies the mathematical calculations in solving equations, since it relates the distribution function at different points $z$  in the central region and is manifested as a central rapidity plateau in the final particle distribution produced in URHIC.

The time evolution of the system can be studied by the hydrodynamical Bjorken equation\cite{Bjorken1983}. For one dimensional expansion scenario in Bjorken Model of an Ideal Relativistic fluid, from the two conservation laws (\ref{ConservationLaws}) we can deduce,

\begin{equation}
\left\{
\begin{array}{c}
\tau \partial_{\tau} \epsilon=- (\epsilon+\mathscr{P})  \\
\tau \partial_{\tau} \mathscr{S}=-\mathscr{S}. \\
\end{array}%
\right.  \label{ConservationLaws2}
\end{equation}

The first equation can be deduced from
\begin{equation}\label{BJEQE}
\frac{\partial (\tau \epsilon)}{\partial \tau}=-\mathscr{P},
\end{equation}
and the second provides the cooling law obtained by Bjorken in his paper \cite{Bjorken1983}.

If the system does not have an ideal EoS but instead has an EoS with a constant speed of sound, then it follows that $(\mathscr{P}=\zeta \epsilon)$, where we have fixed the constant by demanding that the pressure goes to zero when the energy density goes to zero. The general EoS $(\mathscr{P}=\zeta \epsilon)$, which is of simple kind most commonly known in theoretical calculations, can be used to derive the most important hydrodynamic relations.
If the EoS has varying speed of sound, then one can express the pressure $\mathscr{P}$ in terms of an integral of $\zeta$.
Indeed, when considering $\zeta$ as a constant and integrating the two partial differential equations (\ref{ConservationLaws2}) with appropriate initial conditions $\tau=\tau_i$ we get,

\begin{equation}\label{BM11}
\epsilon(\tau)=\epsilon(\tau_i)\left(\frac{\tau_i}{\tau}\right)^{1+\zeta},
\end{equation}

\begin{equation}  \label{HydEntropy}
\mathscr{S} = \mathscr{S}(\tau_i) \left(\frac{\tau_i}{\tau}\right)
\end{equation}
and,
\begin{equation}\label{BM13}
\mathscr{P}(\tau)=\mathscr{P}(\tau_i)\left(\frac{\tau_i}{\tau}\right)^{1+\zeta}.
\end{equation}

In fact, the power-law scaling in different tRF is due to the Bjorken symmetry and related to the underlying EoS, except the solution given by the relation(\ref{HydEntropy}) which does not depend on the EoS.
However, using the thermodynamic relation $T\mathscr{S}=\epsilon + \mathscr{P}$, we can easily find the temperature as a function of $\tau$,
\begin{equation}\label{BM12}
T(\tau)=T(\tau_i)\left(\frac{\tau_i}{\tau}\right)^\zeta,
\end{equation}
indicating that there is a one-to-one correspondence between temperature $T$ and $\tau$.

An important feature to be noticed from these solutions that the energy density $\epsilon(\tau)$ and pressure $\mathscr{P}(\tau)$ decrease faster than the entropy $\mathscr{S}(\tau)$ under the scaling expansion of the fluid. The different solutions obtained and given by the relations(\ref{BM11}-\ref{BM13}) indicate that the system cools non linearly in time, and also, explain that the time evolution is the same in both two phases of the system
\begin{figure}
\resizebox{1.0\hsize}{!}{
\includegraphics*{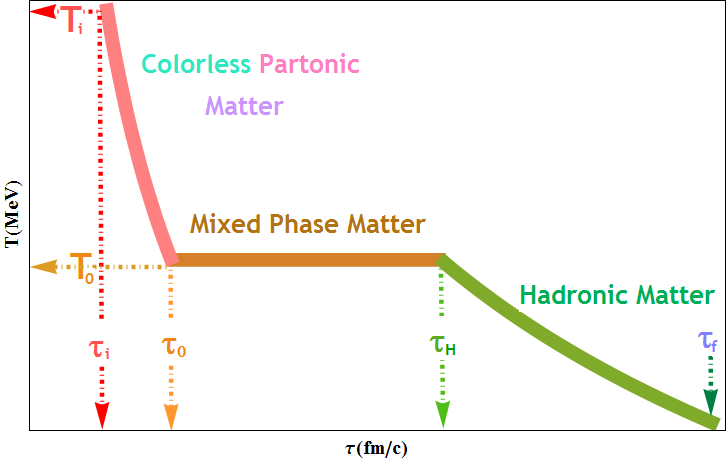}}
\caption{Diagram of temperature $T(\tau)$ as a function of the proper time $\tau$ during the evolution from PM phase to a HM phase.}\label{Fig:01}
\end{figure}

We see, thus, that with these considerations that the time evolution of any TRF is a function of only proper time $\tau$ and depends on the initial conditions for the hydrodynamic flow.
Thus the expanding matter system would appear to be similar in all frames related by different homogeneous Lorentz transformation. This feature is referred to exactly what we recall the frame-independence symmetry or boost invariance.

Now, during the mixed phase system the temperature is keep invariant and the evolution is isothermal. The relation (\ref{BM11}) remains valid and using the same thermodynamic relation as above, we can easily derive the new relation giving the entropy density as a function of $\tau$,
\begin{equation}  \label{HydEntropy2}
\mathscr{S} = \mathscr{S}(\tau_i)\left(\frac{\tau_i}{\tau}\right)^{1+\zeta}.
\end{equation}
This relation is very different from (\ref{HydEntropy}) showing the entropy density decreases more rapidly during the confining phase transition than in the PM and HM phases.

\subsection{Time Evolution, Important Times and Lifetimes }\label{Times}

For simplicity, we consider an URHIC in which there is an important Lorentz contraction in the longitudinal direction allowing to represent the two heavy ions as two thin disk. When the two nuclei cross each other and due to the color transparency, recede from each other after the collision. A large amount of energy is deposited in a small space-time volume. The hot matter created in the collision region has a very high energy density, but a small net baryon content. This is similar as in the case of the very small net baryon content of the early universe, for this reason the type hot matter which may be produced is of special astrophysical interest. Since the ground state of matter with such an energy density is in the PM phase and not in the HM phase, the quanta which carry the energy deposited in the collision region around $z=0$ can be in the form of different partons. The excited partons can rescatter and emit new particles and thereby evolve through a pre-equilibrium stage towards a thermalized CPP state, from which the system evolves further according to  the laws of RH. It is expected that most of the entropy and transverse energy are produced in this pre-equilibrium stage. Moreover, the parton dynamics during this stage determines the initial and boundary conditions for a hydrodynamic
expansion of the hot PM, so all the physical space time evolution, which comes afterwards takes its origin from these pre-equilibrium and thermalization stages. This the main idea in Bjorken's model of the space-time scenario for a URHIC, indicating that after the collision of the two nuclei at $(z, t) = (0,0)$, the energy density may sufficiently high to make it likely that a system of PM may be formed in
the central rapidity region. The PM initially may not be in thermal equilibrium, but subsequent equilibration may bring it to local equilibrium at the proper time $\tau_i$, and then the PM may then evolve according to the laws of hydrodynamics thereafter. One can imagine the process of an URHIC whose time evolution naturally splits into the following stages(early, transient and late time stages):
\renewcommand{\labelitemi}{$\bullet$}
\begin{itemize}
  \item begining of the URHIC ($\tau=0$) and a Glasma state formation if one adopt the Color Glass Condensate(CGC) model\cite{Gelis2015},
  \item equilibration of the PM at $\tau_i$ and high temperature $T_i\gg T_0$,
  \item beginning of the hydrodynamic evolution of the PM at $\tau_i$,
  \item beginning of confinment phase transition and the hadronization at $\tau_0$,
  \item end of confinement phase transition and the hadronization at $\tau_H$ and formation of the HM,
  \item beginning of the hydrodynamic evolution of the HM at $\tau_H$,
  \item beginning of the hadronic free streaming and hadronic freeze-out at $\tau_f$.
\end{itemize}

In this section we try to explain how to study the time evolution of our system when the CPP start to cool, and the HM start to form. As we know, based on the space-time diagram of Bjoken Model\cite{Bjorken1983}, the hadronization of CPP can be explained as follow: after collision and creating the CPP phase in initial proper time $\tau_i$ and high temperature $T_i\gg T_0$, the CPP starts to cool-down in time.

Under the assumption that thermal equilibrium is attained within the formation $(\tau_i)$, the time dependence of the temperature can be estimated in a hydrodynamical model. A graphical representation of this evolution is shown in (Fig.\ref{Fig:01}). This figure depicts the longitudinal time evolution of hot matter with a first-order phase transition created at the central rapidity region of an URHIC.  After a certain initial time $(\tau_i)$ the produced CPP is considered in thermal equilibrium at an initial temperature $T_i$. If we assume that no dissipation occur during both expansion and confining phase transition, meaning that these two process are assumed to be adiabatic therefore isentropic expansion of the fluid is presumed up to the transition temperature $T_0(V)$.
During the transition from the CPP to HM, the temperature in the mixed phase is maintained constant meaning that the expansion is isothermic. The latent heat of the first order confining phase transition
is absorbed in the conversion of the partonic degrees of freedom into hadronic degrees of freedom. At $(\tau_H)$ the hadronization is completed and the HM starts to cool down under isentropic expansion up to $(\tau_f)$
where the density of the system is low enough for the hadrons to escape. As the hadronization process was completed and the hadronic fireball formed in the URHIC expands, its density decreases and the hadronic mean free path
increases. Eventually, this process leads to the decoupling of hadrons, becoming non interacting and moving freely.

During the time evolution of the system undergoing the colorless confining phase transition, from the PM to the hadronic gas passing by the mixed phase and the HM, three broad and distinct periods emerge in the whole time evolution separated by four times or instants:

 \begin{itemize}
  \item the initial time $\tau_i$ at which the temperature $T_i$ is large enough to produce the CPP, $T_i>T_0$,
  \item the transition time $\tau_0$ at which the temperature drops to the transition temperature $T_0$,
  \item the hadronic time $\tau_H$ at which the hadronization process is completed. At this time a HM is formed.
  The last two times, namely $\tau_0$ and $\tau_H$ are calculated using the following relations:
\begin{equation}
\left\{
\begin{array}{c}
 \frac{\partial^2T}{\partial\tau^2}(\tau=\tau_0, V) \equiv  \mbox{Maximum or    } \notag \\
\mathcal{H}(\tau_0) = 1/2 \notag \\
 \frac{\partial^2T}{\partial\tau^2}(\tau=\tau_H,V)\equiv  \mbox{minimum or    } \\
  \tau_H=\tau_0  \Big( \frac{\mathbf{a}_{QG}}{\mathbf{a}_{HG}}\Big )^{\frac{1}{(1+\zeta)}} \tau_0 \notag \\
\end{array}%
\right.  \label{NaturalTP}
\end{equation}
The relationship relating $\tau_H$ with $\tau_0$ can be easily derived from the entropy density of the system (\ref{HydEntropy2}) as given in Ref.(\citen{Wong1994}).
  \item The freeze-out time $\tau_f$ at which the hadrons become free without interactions. At this time the hadron breakup occurs.
\end{itemize}

From these four times we can thus define three lifetimes:
\begin{itemize}
\item the lifetime of the hot matter during the pre-equilibrium stage: $\Delta\tau_{G}=\tau_i$, called Glasma in the framework of CGC model\cite{Gelis2015},
\item the lifetime of the CPP phase given by:  $\Delta\tau_{CPP}=\tau_0-\tau_i$,
\item  the lifetime of the confining phase transition given by:  $\Delta\tau_{PT}=\tau_H-\tau_0$,
\item  after the phase transition is completed at $\tau_H$, the interacting HM undergoes a hydrodynamic expansion. The lifetime of this Hadronic phase until the freeze out is given by:  $\Delta\tau_{HG}=\tau_f-\tau_H$.
\end{itemize}
The times and lifetimes, thus defined, are important for a good description of the time evolution of our system. It should be noted that our model allows us to study easily the FSE on these times and lifetimes, thus the FSS properties are investigated.

\section{Bjorken Expansion in Colorless-QCD Confining Phase Transition}\label{sec:04}
\subsection{Solution of Finite Volume Bjorken Equation: $T(\tau,V)$}\label{sec:04.01}

The time evolution of the expanding CPP, created in URHIC and undergoing the confining phase transition towards a HG, will be discussed in the context of a unified model combining our Colorless QCD-MIT Bag Model with the boost invariant Bjorken expansion. Such a model is a more physical generalization of the previous simplified models. The better way is to solve the finite volume Bjorken equation in order to deduce the variation of the temperature $T(\tau,V)$ as a function of the proper time $\tau$ and of the volume $V$\cite{Florkowski2010}. Initially, for $(\tau\ge\tau_i)$, the CPP system expands and cools down with decreasing temperature under the  relation (\ref{BM12}) with initial conditions $(T_i\sim 2T_0)$ and $\tau_i\sim 0.1-1fm/c$ until reaching $T_0(V)$ the transition temperature at time $\tau_0$. At $\tau_0$, the phase transition starts, a mixed phase system starts to develop until converting totaly PM phase into HM phase at $\tau_H$, the temperature of the total system is maintained constant ($T_0\sim T_H$). After $\tau_H$ the system is completely in hadronic phase and continues to cool down until the freeze-out time $\tau_f$.

The solution of the finite volume Bjorken equation using our EoS (\ref{pressureandDE}) is given by:
\begin{equation}\label{tauTV}
  \tau(T,V)=\tau_{i}\exp\left[\int_{T(\tau)}^{T(\tau_i)}\frac{3c_{T}(T,V)}{4(\epsilon(T,V)+\mathfrak{B}\mathscr{Q}(T,V))}dT\right],
\end{equation}

where $(\tau_i)$ and $(T(\tau_i))$ are the initial time and the initial temperature when the system is in the PM phase.
With a specific numerical method, the results of the integration of the Bjorken equation are presented in Figs.(\ref{Fig:02},\ref{Fig:03}). We show the variation of temperature as a function of the proper time $\tau$: $T(\tau,V)$ at different volumes. Also the effect of the CC is displayed.
We have considered a range of variation of the volume as $100-1000fm^3$ and analyzed the role played by the CC. We have extracted the different times and calculated the different lifetimes as defined in the section (\ref{Times}).
\begin{figure}
  \resizebox{1.0\hsize}{!}{
  \includegraphics*{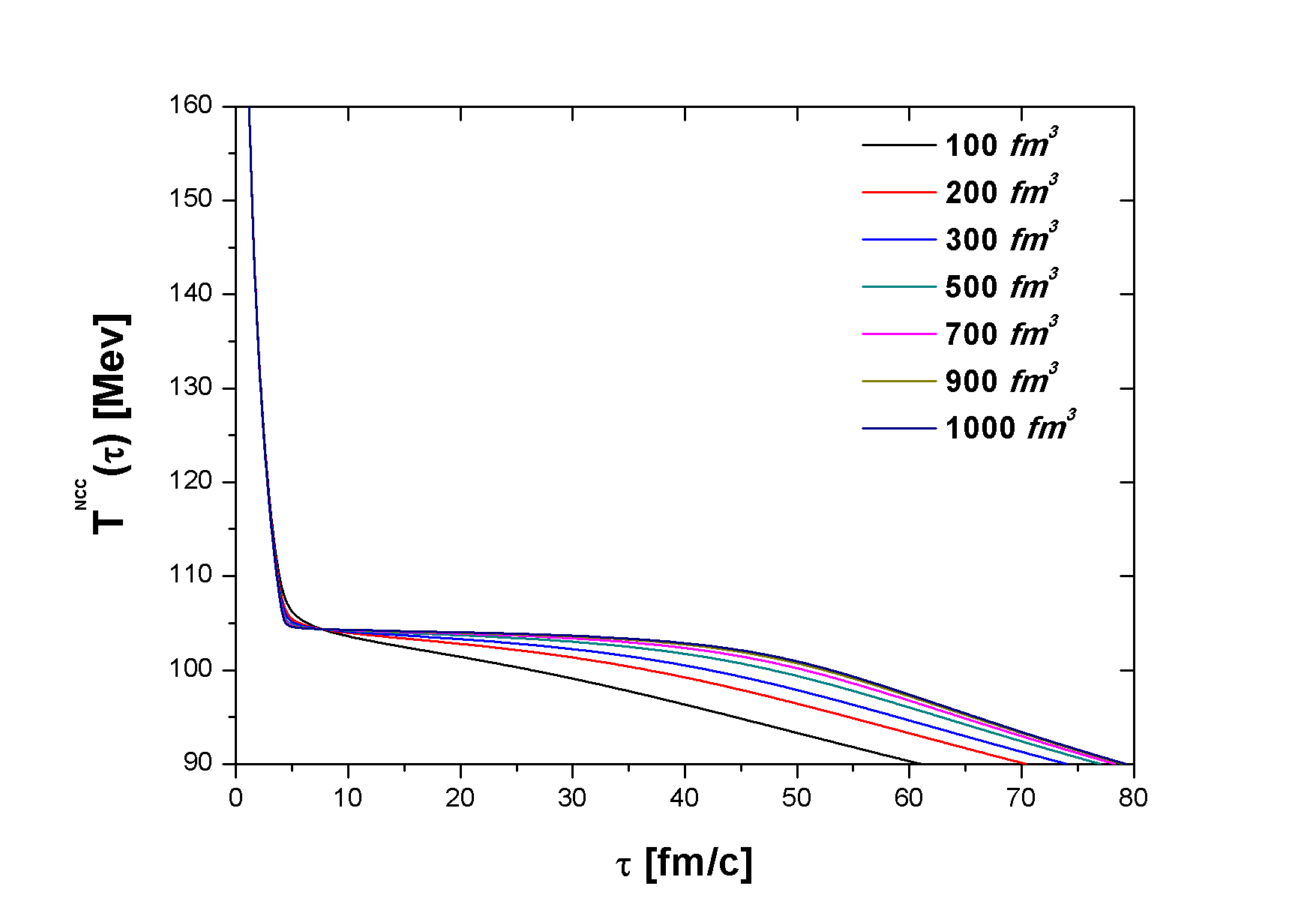}}
  \caption{Temperature evolution vs the proper time $\tau$ with initial conditions $\tau_i=0.6 fm/c$ and $T_i=200MeV$ without colorlessness condition and for different volumes $V$.}\label{Fig:02}
\end{figure}
\begin{figure}
  \resizebox{1.0\hsize}{!}{
  \includegraphics*{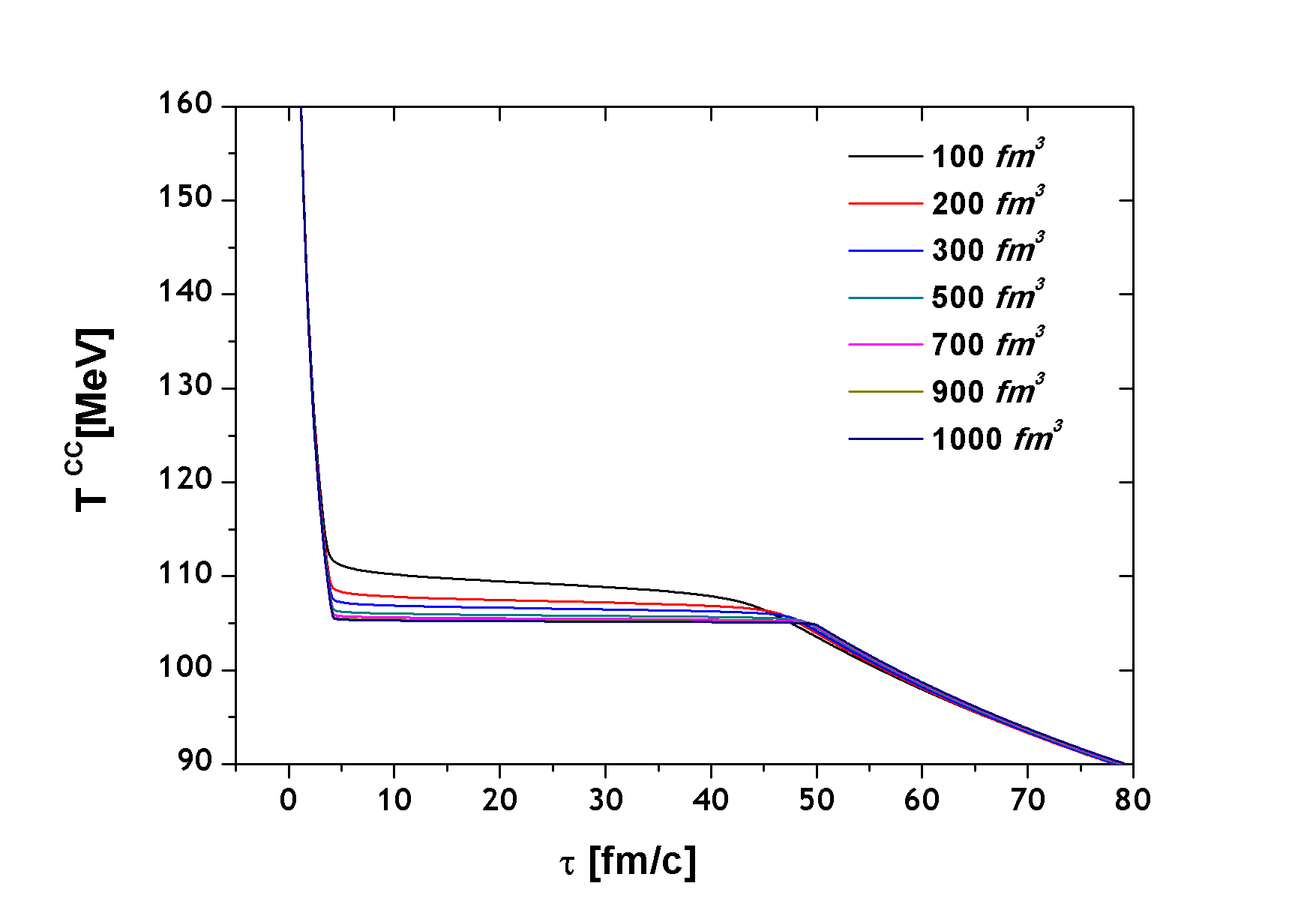}}
  \caption{Temperature evolution vs the proper time $\tau$ with initial conditions $\tau_i=0.6 fm/c$ and $T_i=200MeV$ with colorlessness condition and for different volumes $V$.}\label{Fig:03}
\end{figure}

\subsection{Results and Discussion}\label{sec:04.02}

In this paragraph we intend to summarize the results obtained from the solution of the finite volume Bjorken equation. These results concern the different times and lifetimes characterizing our system evolving from the CPP state until HG state. In the figures (\ref{Fig:02},\ref{Fig:03}), we show clearly the same behavior as in Fig.(\ref{Fig:01}) and a constancy of the temperature during the phase transition mainly in the thermodynamic limit.
Initially, the temperature of the system drops rapidly due to rapid longitudinal expansion. During the next stages of the time evolution, the rate of longitudinal expansion decreases progressively and mainly due to the confining phase transition. The time of the complete hadronization of the system is well localized and defined in the colorless case, allowing a more precise calculation of the lifetime of the CPP and getting a better knowledge of its physical meaning. The shifting effect of the transition point is clear on Fig.(\ref{Fig:03}) as resulting from the CC. In the case of the partonic plasma without the CC, the temperature during the phase transition is not really constant when the volume is finite. A slightly increasing dependence on $V$ is noticed in almost results apart from the time $\tau_{H}^{NCC}$ and the lifetime $\Delta \tau_{PT}^{NCC}$ in the case without the CC. This is related simply on the difficulty of extracting the times $\tau_{H}^{NCC}$ because the second particular point does not appear in a clear way even when the volume is large. However, we would point out that the CC increases the times and lifetimes. The PM lifetime lies in the range $2.86177fm/c \leq \Delta \tau_{PP}^{NCC} \leq 3.58577fm/c $ and the CPP lifetime lies in the range $3.03277fm/c \leq \Delta \tau_{CPP}^{CC} \leq 3.60614fm/c $. An order of increasing no bigger than  $6\%$. When we observe the results, we also note that the lifetime of the colorless phase transition lies in the range $40.79905fm/c \leq \Delta \tau_{PT}^{CC} \leq 45.76331fm/c $ and non-colorless phase transition lies $24.14182fm/c \leq \Delta \tau_{PT}^{NCC} \leq 44.44157fm/c$.\\
 Our results agree qualitatively and quantitatively  with the results obtained using different approaches \cite{Florkowski2010, QGPbblb2008, Csernai1992, Csernai1993, Mustafa1998, Rischke1996, Vogt2007, Geiger1995}. We find that in both PM and HM phases the cooling is a little slower than predicted by Bjorken's scaling solution($\theta \lesssim 1/3$). However, in terms of cooling speed the PM's cooling is considerably faster than of the HM. Another comparison makes us very confident with our timescale, from the beginning of the hadronisation process until the freezeout, which is in complete agreement with the pionic freezeout time as evaluated in the context of Hadronic Rescattering Model(HRM) for URHIC in RHIC and LHC\cite{Humanic2006}. When we compare ours results with those obtained from the hydrodynamic simulations\cite{Hirano2002} and from the fluid dynamical simulations of URHIC in full 3-Dim space\cite{Niemi2014}, a good qualitative similarity should be noted. Due to the mere presence of longitudinal expansion in our model, on can understand why the times and lifetimes are relatively long and the cooling process is slow. If one wants to develop a complete model one has to include the transverse expansion during the Bjorken type of longitudinal expansion is happening. Certainly, a 3-Dim expansion will allow the system to cool down more rapidly and to reduce the different times and lifetimes. A prolongation of times and lifetimes is noticed in case a rapid change in the degrees of freedom occurs. Thus, we conclude that the effect of the CC and the confining phase transition is to make the times and the lifetimes longer than in other cases. The effect of the CC in the CPP is the same as if the CPP system possessed a certain viscosity. The different parton-parton interactions making the PM to be colorless generates a kind of viscosity in the system and the CPP becomes non-ideal\cite{Youichi1989,Ladrem2019}.

\section{Finite Size Scaling Study of different times and lifetimes in Colorless Case} \label{sec:05}
\subsection{Theoretical Derivations} \label{ssec:5.2}
Phase transitions occur in nature in a great variety of systems and under a very wide range of conditions. Phase transitions are abrupt changes in the global behavior and in the qualitative properties of a system when certain parameters pass through particular values. At the transition point, the system exhibits, by definition, a singular behavior. As one passes through the transition region, the system moves between analytically distinct parts of the phase diagram. The singularity in a first order phase transition is entirely due to the phase coexistence phenomenon, for against the divergence in a second-order phase transition is intimately caused by the divergence of the correlation length. Now, if the volume is finite at least in one dimension with a characteristic size $\ L=V^{1/d}$, the singularity is smeared out into a peak with finite mathematical properties and Four Finite Size Effects(4FSE) can be observed \cite{Ladrem2005}:(1) the rounding effect of the discontinuities,(2) the smearing effect of the singularities,(3) the shifting effect of the finite volume transition point $T_0(V)$,(4) and the widening effect of the transition region around the transition point. These 4FSE have an important consequence putting the first and the second order phase transitions on an equal footing and can be described easily by a simple power law when neglecting the leading terms. In the case of a first order phase transition the FSS power law of the temperature-shifting effect is given by\cite{Ladrem2005,Ladrem2015},
\begin{equation}
T_{0}(V)=T_{0}(\infty)+ a V^{-1} +\mathcal{O}(V^{-2}),
\label{FSSTcfit}
\end{equation}
where $T_{0}(\infty)$ is the temperature of the bulk transition point.
Using the relation between the temperature and the proper time $T(\tau)$(\ref{BM12}) during the hydrodynamic evolution, we can easily deduce the following FSS law of the time-shifting effect,
\begin{equation}\label{FSStime1}
\tau_{0}(V)= \tau_{0}(\infty) - \frac{\tau_{0}(\infty) a}{T_{0}(\infty) \zeta}V^{-1}+\mathcal{O}(V^{-2})
\end{equation}
\begin{equation}\label{FSStime2}
\Delta \tau_{CPP}(V)=\Delta \tau_{CPP}(\infty) - \frac{\tau_{0}(\infty) a }{T_{0}(\infty) \zeta}V^{-1} +\mathcal{O}(V^{-2})
\end{equation}
with
\begin{equation}
\left\{
\begin{array}{c}
\tau_{0}(\infty)= \tau_i \left(\frac{T_i}{T_{0}(\infty)}\right)^{\frac{1}{\zeta}}  \\
\Delta \tau_{CPP}(\infty)=\tau_{0}(\infty)-  \tau_i  \\
\end{array}%
\right.  \label{FSStime3}
\end{equation}

In order to extrapolate our FSS calculation for the hadronic time $\tau_H$ we need to know how the FSE affect this time. Previously, we have obtained and discussed extensively the physical meaning of the particular points
appearing in different TRF and the emergent correlation between them.\cite{Ladrem2005,Ladrem2015,Hind2019,Hadeel2019} The transition temperatures $T_0(V)$ as a function of the hadronic temperature $T_H(V)$ is shown in figure(Fig.\ref{Fig:36}).
While the results display a linear correlation of $T_0(V)$ with $T_H(V)$  which can be described by a relationship : $T_H(V)=\lambda T_0(V) +\rho$ with $\lambda=0.24983,\rho=78.46915$. Thus, using the equation (\ref{FSSTcfit})
we can get,

\begin{equation}
T_{H}(V)=T_{H}(\infty)+b V^{-1} +\mathcal{O}(V^{-2})
\label{FSSTHfit}
\end{equation}
with
\begin{equation}
\left\{
\begin{array}{c}
T_{H}(\infty)= \lambda T_{0}(\infty)+\rho  \\
b=\lambda a  \\
\end{array}%
\right.  \label{FSSTH}
\end{equation}
In fact, by inverting the problem, starting from the same FSS's law given by the two relations (\ref{FSSTcfit},\ref{FSSTHfit}), we can derive the following relation easily,
\begin{equation}\label{THvsT0}
T_H(V)=(\frac{b}{a})T_0(V) + T_0(\infty)(1-\frac{b}{a}),
\end{equation}
a result in complete agreement with the correlation displayed in figure(Fig.\ref{Fig:36}). And with the same analytical derivation we obtain the FSS law of $\tau_{H}(V)$,
\begin{equation}\label{FSStime4}
\tau_{H}(V)=\tau_{H}(\infty) - \frac{\tau_{H}(\infty) b }{T_{H}(\infty) \zeta}V^{-1}+\mathcal{O}(V^{-2}),
\end{equation}
and,
\begin{equation}\label{FSStime5}
\Delta \tau_{PT}(V)=\Delta \tau_{PT}(\infty) - \big[\frac{b \tau_{H}(\infty)}{\zeta  T_{H}(\infty)}-\frac{a \tau_{0}(\infty)}{\zeta  T_{0}(\infty)}\big] V^{-1}+\mathcal{O}(V^{-2}),
\end{equation}

with
\begin{equation}
\left\{
\begin{array}{c}
\tau_{H}(\infty)= \tau_i \left(\frac{T_i}{T_{H}(\infty)}\right)^{\frac{1}{\zeta }}  \\
\Delta \tau_{PT}(\infty)=\tau_{H}(\infty)-  \tau_{0}(\infty)  \\
\end{array}%
\right.  \label{FSStime6}
\end{equation}

\begin{figure}
  \resizebox{0.9\hsize}{!}{
  \includegraphics*{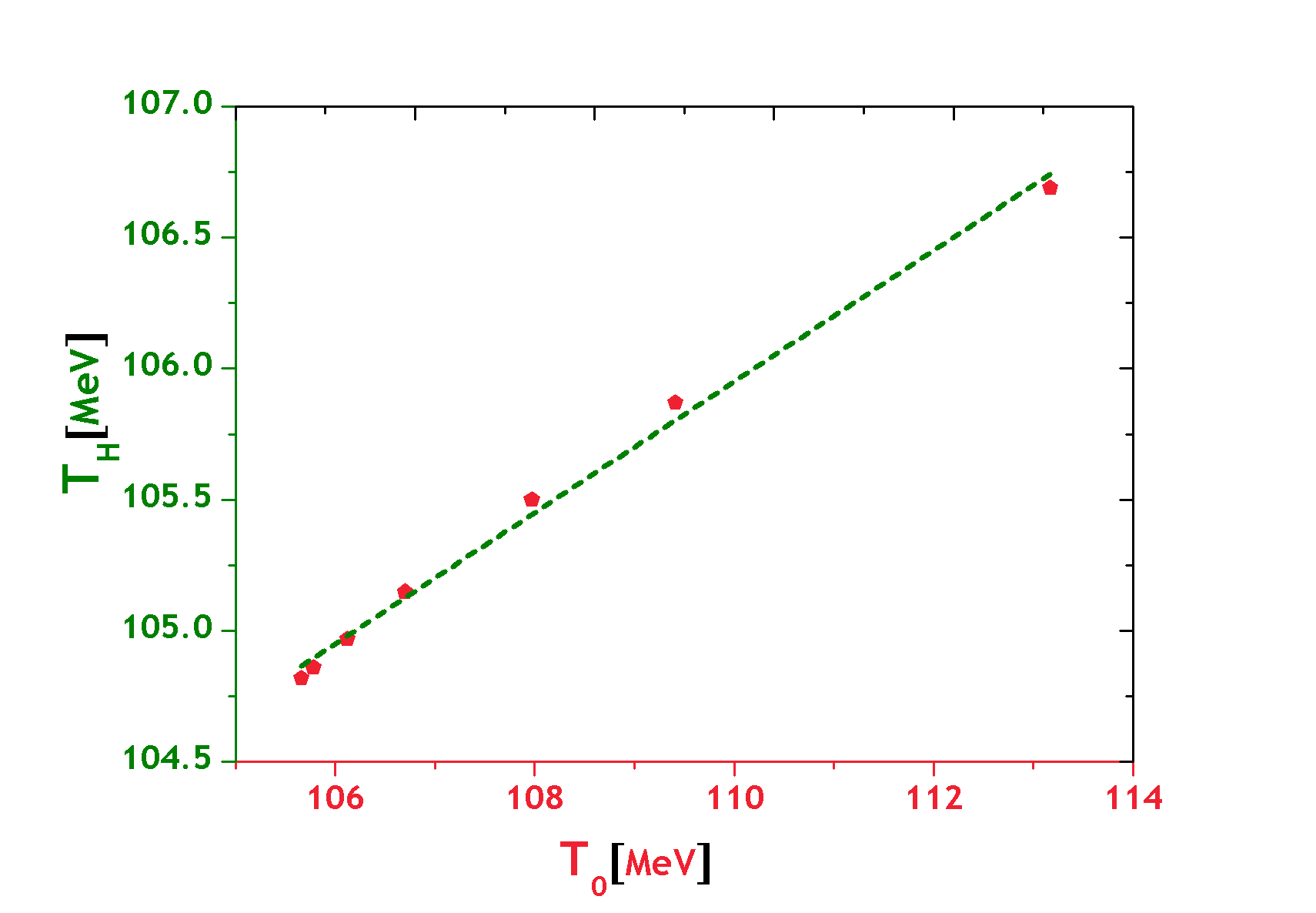}}
 \caption{Linear correlation between $T_H(V)$ and $T_0(V)$.} \label{Fig:36}
\end{figure}

We are interested in the viability of the derived time FSS's laws that could explain the FSS behaviors of times and lifetimes, then it is natural to check if our theoretical calculations fit well our results.

\subsection{Results and Discussion}
The FSE are more than clear on the whole TRF investigated in this work specially in the colorless case. For this reason, we have performed a detailed FSS study of different times and lifetimes. The results obtained in solving the finite volume Bjorken equation with the initial conditions $\tau_i=0.6 fm/c$ and $T_i=200MeV$ when the volume varies in the range of $100-1000 fm^3$ are plotted as functions of the volume on Figs.(\ref{Fig:5}-\ref{Fig:11}). The simple way to do this, is to study these quantities as functions of the volume and try to fit the FSS behavior using the relationships derived before. From the relations derived previously concerning the FSS laws of times and lifetimes (\ref{FSStime1},\ref{FSStime2},\ref{FSStime4},\ref{FSStime5}) during the hydrodynamic evolution, we can write them in more general forms,
\begin{equation}
\left\{
\begin{array}{c}
  \tau(V)=\alpha+\beta V^{-1}+\mathcal{O}(V^{-2})  \\
  \Delta \tau(V)=\gamma+\delta V^{-1}+\mathcal{O}(V^{-2})  \\
\end{array}%
\right.  \label{FSStfits}
\end{equation}

We focus on the scaling times and lifetimes with system size. In finite-size systems, times and lifetimes may converge to a finite value, which goes to bulk value when the system volume goes to infinity.
However, using an approach based on the standard FSS theory we show that the FSS of times and lifetimes should scale with inverse polynomial of the volume.

The inverse law between the temperature and the proper time transforms the shifting towards the low temperatures into shifting towards the high times. In the end, it all comes down to the simple and unique distinctive difference regarding the sign of the different fit parameters: $\alpha>0,\gamma>0$ and $\beta<0, \delta<0$. The decreasing behavior of the different temperatures becomes an increasing in different times until reaching the thermodynamic limit. In the thermodynamic limit, we notice the physical meaning of the two fit parameters $\alpha$ and $\gamma$ as being the bulk values of times and lifetimes.
\begin{equation}
\left\{
\begin{array}{c}
  \lim \limits_{V \to \infty} \tau(V) =\tau(\infty)= \alpha   \\
   \lim \limits_{V \to \infty} \Delta \tau(V)= \Delta \tau(\infty) = \gamma  \\
\end{array}%
\right.  \label{FSStParam}
\end{equation}

\begin{figure}
  \resizebox{1.0\hsize}{!}{
  \includegraphics*{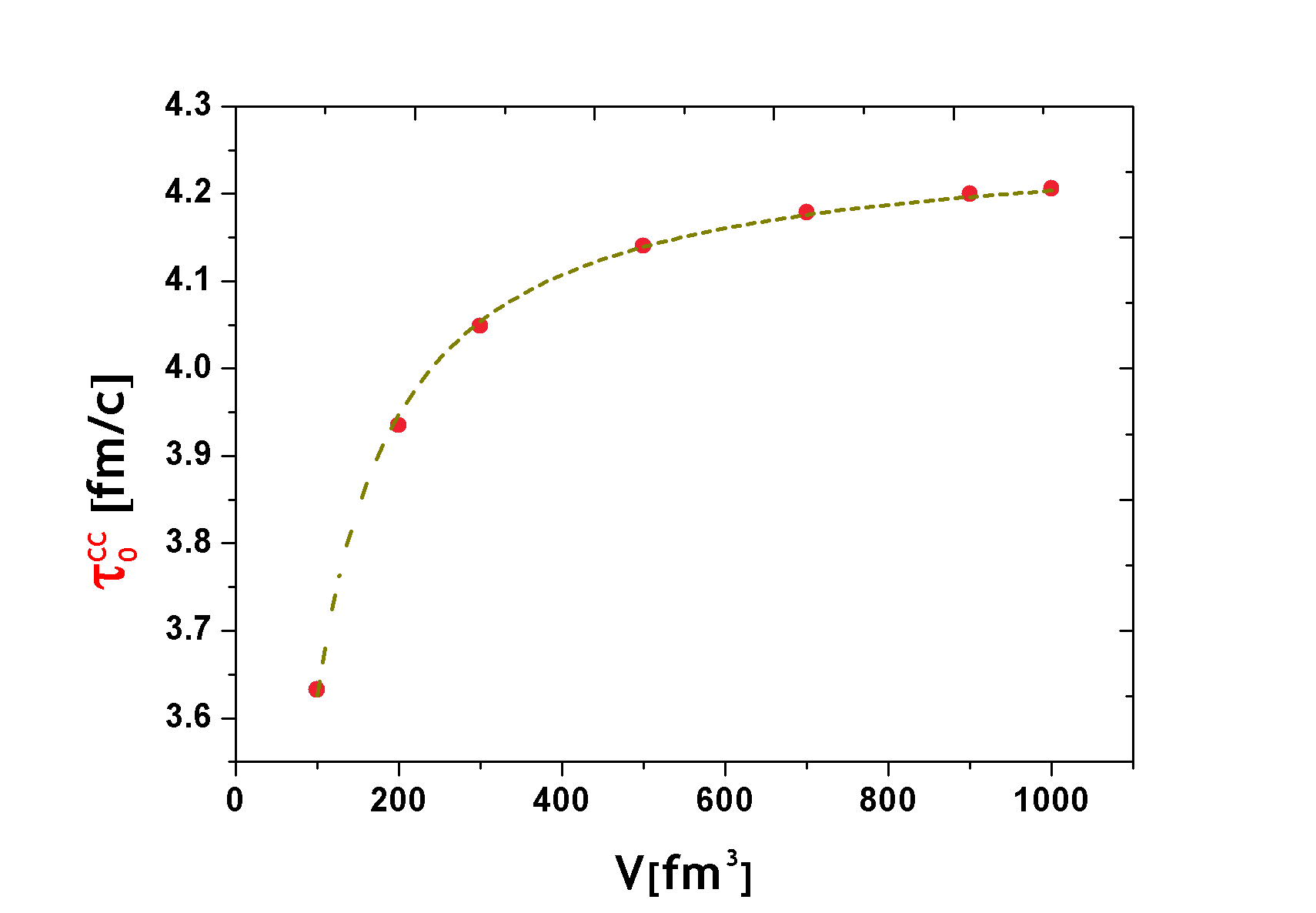}}
  \caption{The second-order FSS fitted curve (Rel. \ref{FSStfits}) of $\tau^{CC}_0(V)$.}\label{Fig:5}
   \resizebox{1.0\hsize}{!}{
  \includegraphics*{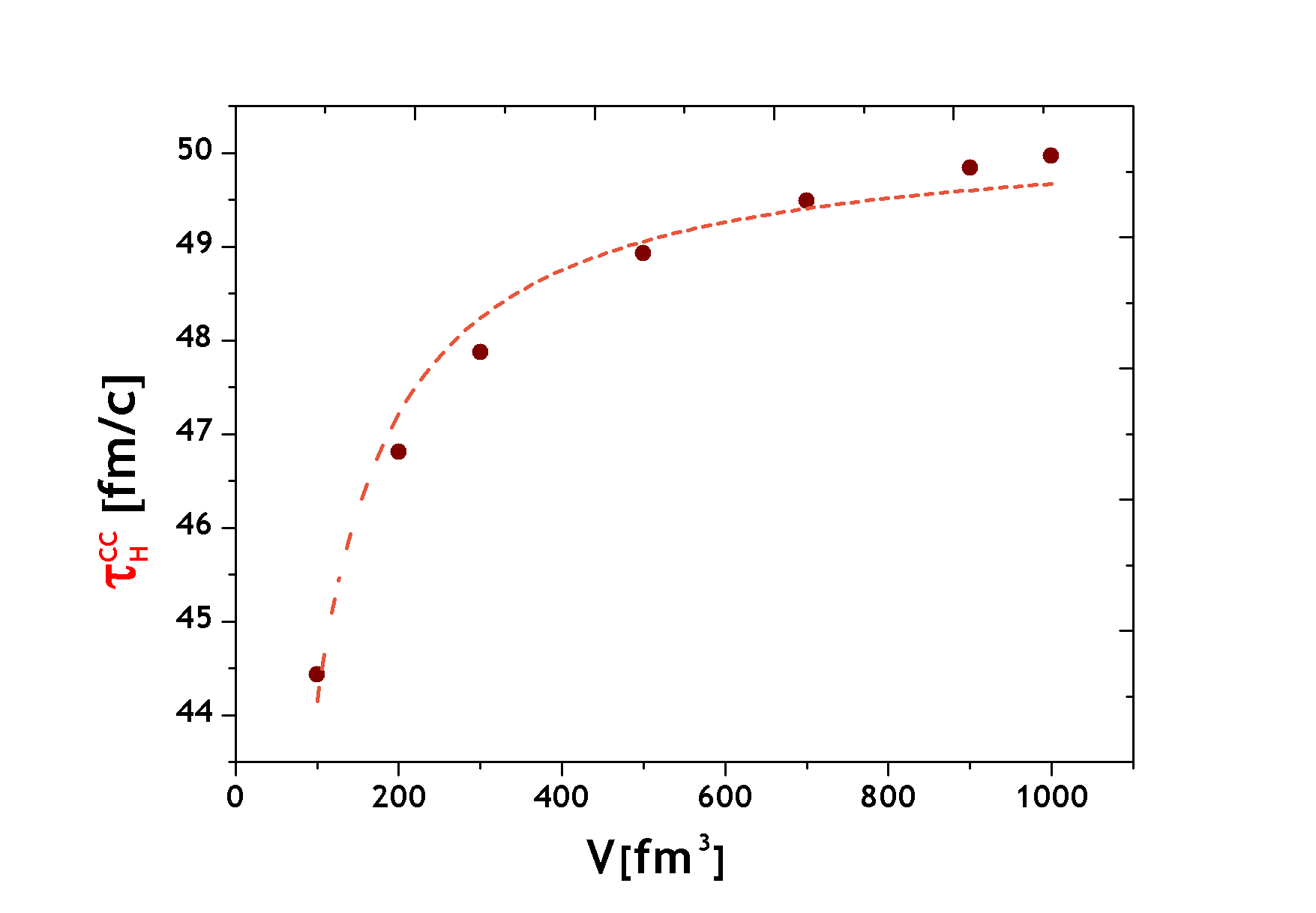}}
  \caption{The second-order FSS fitted curve (Rel. \ref{FSStfits}) of $\tau^{CC}_H(V)$. }\label{Fig:7}
\end{figure}
\begin{figure}
  \resizebox{1.0\hsize}{!}{
   \includegraphics*{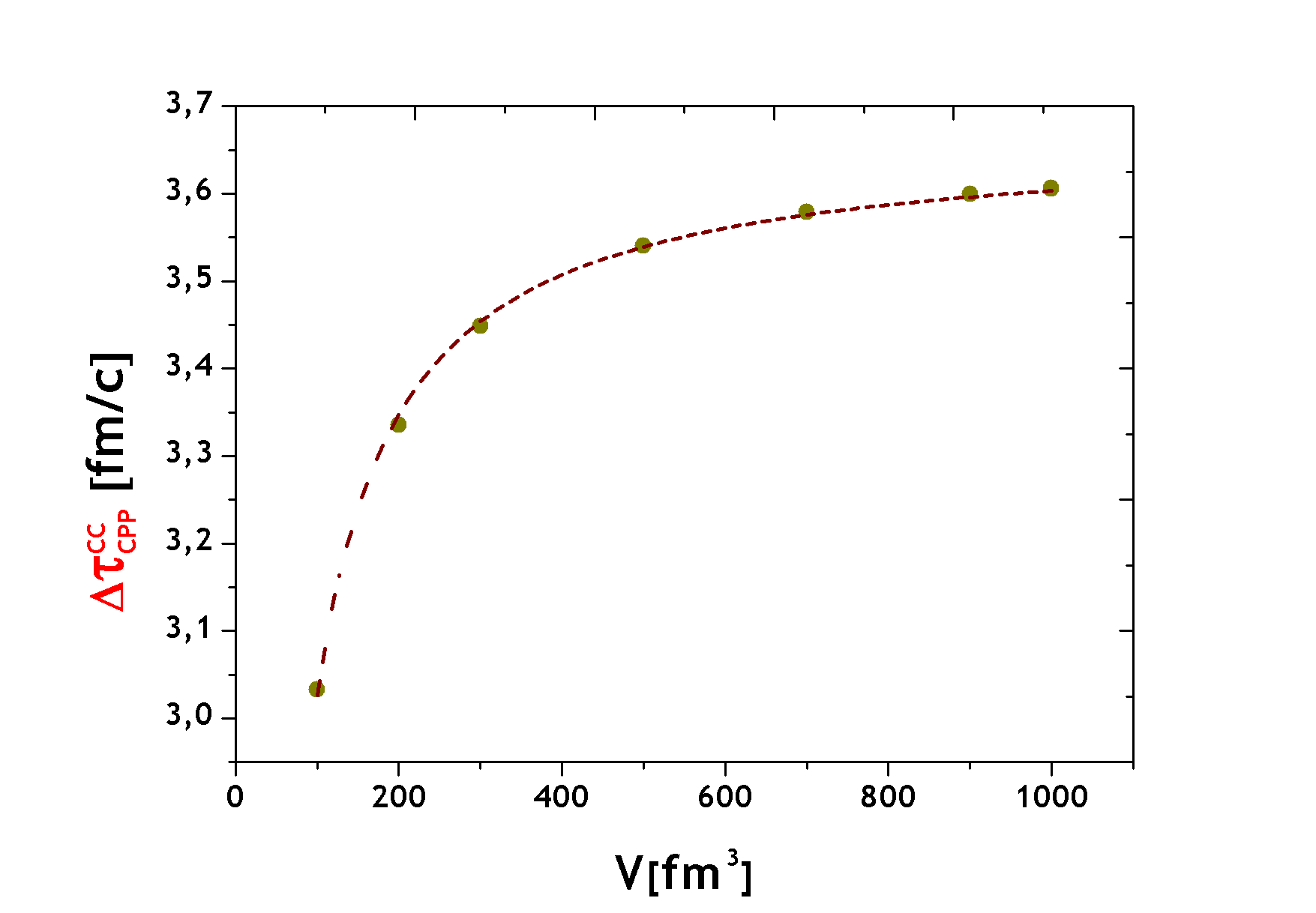}}
  \caption{The second-order FSS fitted curve (Rel. \ref{FSStfits}) of $\Delta\tau^{CC}_{CPP}(V)$ .}\label{Fig:9}
   \resizebox{1.0\hsize}{!}{
    \includegraphics*{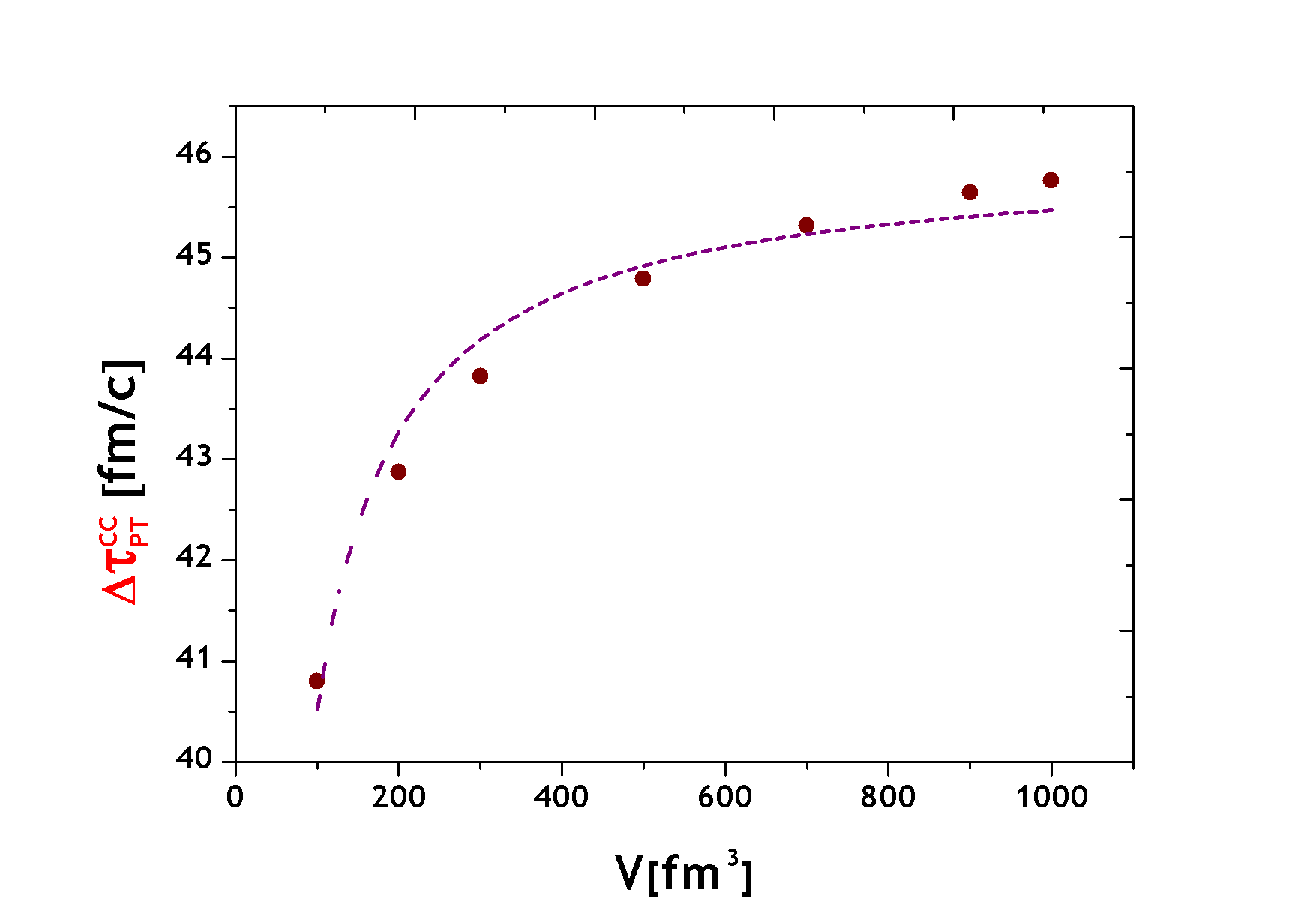}}
  \caption{The second-order FSS fitted curve (Rel. \ref{FSStfits}) of $\Delta\tau^{CC}_{PT}(V)$ .}\label{Fig:11}
\end{figure}
\begin{table}
\vspace*{1cm}  % with the correct table height
\centering
\caption{Fit parameters of the four times $\tau^{CC}_{0}(V)$ , $\tau^{CC}_{H}(V)$ , $\Delta \tau^{CC}_{CPP}(V)$ and  $\Delta \tau^{CC}_{PT}  (V)$.}\label{tab:5}
\begin{tabular}{llllll}
\hline\noalign{\smallskip}
fit s & $\tau^{CC}_{0}$ & $\tau^{CC}_H$ & $\Delta\tau^{CC}_{CPP}$ & $\Delta\tau^{CC}_{PT}$ \\
parameter & & & & \\
\noalign{\smallskip}\hline\noalign{\smallskip}
$\alpha (\boldsymbol{\gamma}) \left[ fm/c\right]$ &$4.267$ & $50.280$  &$\boldsymbol{3.667}$ & $\boldsymbol{46.013}$ \\
$-\beta (\boldsymbol{-\delta}) \left[ fm^4/c\right]$ &$63.8$ & $613.7$   &$\boldsymbol{63.85}$ & $\boldsymbol{549.945}$\\
$\chi^2$ &$0.999$ & $0.972$  &$0.999$ & $0.9668$\\
\noalign{\smallskip}\hline
\end{tabular}
\vspace*{1cm}  % with the correct table height
\end{table}

We have to notice that the FSE are more evident and clear in the CC than in the NCC and the quality of the fits in the colorless case is better than the case without the CC. Probably this is due to the fact that the CC make a larger contribution in the FSE. As we see in Figures (\ref{Fig:5}-\ref{Fig:11}), the derived time-FSS's laws (Rel. \ref{FSStfits}) fit very well our results. It is important to note that the results of $\tau_{0}^{CC}(V)$ and $\Delta \tau_{CPP}^{CC}(V)$ are well described than those of $\tau_{H}^{CC}(V)$ and $\Delta \tau_{CPP}^{PT}(V)$. It appears that this can be essentially explained by the effect of CC which is predominately present in the PM phase. The different values of the $\chi^2$ are all larger than 0.94. The table (\ref{tab:5}) contains the numerical values of the fit parameters. The lifetime of our CPP is about $3.667fm/c$ and the duration of the colorless confining phase transition is about $46.013fm/c$. We find that the system spends much of its time in such a mixed phase during the confining phase transition. This means that the expansion rate in both hadronic and partonic matter is fast compared to the hadronization rate\cite{VonGersdorff1986}.
After the CPP reaches the transition temperature $T_0(V)$, it must convert the entropy stored in the partonic matter  into the entropy of a hadron matter. Since the degrees of freedom of the CPP are an order of magnitude larger than those of a HM, this conversion takes a long time. If the system is only longitudinally expanding during this time, the ratio of the time at which the plasma began the phase transition, $\tau_0(V)$, to that at which it completes the phase transition, $\tau_H(V)$ is given by the ratios of these degrees of freedom $\tau_H/\tau_0\approx a_{QG}/a_{HG}\approx 12-13$. The system spends a very long time in mixed phase at a temperature close to the transition temperature $T_0(V)$. We have obtained somewhere long lifetime, indicating that the longitudinal expansion alone is not sufficient. Thus the inclusion of a transverse expansion would certainly cool the system more rapidly and reduce its lifetime. It is remarkably interesting to note that our lifetimes are in good agreement with those, averaging the values of the initial conditions, obtained by other models\cite{Florkowski2010, QGPbblb2008, Csernai1992, Mustafa1998,Vogt2007}.

\section{The Effects of Initial Conditions $(T_i,\tau_i)$} \label{sec:06}
\subsection{Early Stages and Importance of the Initial Conditions} \label{ssec:6.0}
As shown previously, our hydrodynamic model contains a set of differential equations for each fluid element. The integration of these differential equations necessitates initial conditions for the system consisting of the distributions of energy, charge densities, and the velocity fields at an initial time $(\tau_i)$  appropriately chosen. It is important to remember that once the EoS, and initial conditions for the differential equations are fixed, the hydrodynamic equations govern the space-time evolution and described by their solutions. In the ideal fluid approximation, only the EoS of the fluid system contains the complete information about the nature of the constituents of the fluid and their mutual microscopic interactions.

We know that many works have been published trying to draw a bridge from the original quantum state of the two colliding ultra-relativistic heavy ions towards the macroscopic matter that defines the initial conditions, in order to initiate the hydrodynamic expansion. It is not the topic of the present work related to the energy density of the system $(\epsilon_i)$ without detailed features. Only, we focused on trying to learn more about the hydrodynamic expansion, given by the simplest version of the 1-Dim space Bjorken flow, using our colorless EoS without any approximation. Only, in order to keep a clear picture and to study the effect of the CC on the hydrodynamic expansion, that our first choice was a simple 1-Dim expansion meaning that we do not consider transverse expansion.

 Thus the initial conditions needed to solve the hydrodynamic equations were simple, initial values of time and temperature: $(T_i,\tau_i)$. The 3-Dim expansion, including all necessary details, is planned for future work.

 Many of early works on hydrodynamic properties from the produced PM used this kind of initial condition. Some (3 + 1)Dim codes for such purposes also have been developed (see Ref. \citen{Derradi2016}).
Later, phenomenological or theoretical approaches on the physics of the URHIC advanced, and several realistic models have been developed. However, with the advent of the experimental program at RHIC and LHC, early studies of the experimental results, which started in 2001, have been conducted using hydrodynamic calculations based on different approaches like,
\begin{itemize}
\item The simulation approach using the geometrical point of view of the URHIC, called Monte-Carlo Glauber model (MC-G)\cite{Miller2007}. The centrality is an important property of an URHIC, and given by the impact parameter of the collision. It is relatively well-estimated on an event-by-event basis. The impact parameter is a crucial factor in determining the overlap geometry of the nuclear collision, but unfortunately alone is not enough to determine more detailed features of initial conditions.
\item More elaborated microscopic transport models were built, based on nucleon-nucleon event generators such as HIJING, PYTHIA, NEXUS, UrQMD, etc\cite{Derradi2016}. Full 3-Dim space hydrodynamic simulations were performed (See Table 3 of reference \citen{Hirano2013}).

\item The theoretical ideas evolve, the vision of phenomena change and new models are emerging. In this context, a new approach appeared, may be the most interesting because it enables us to relate the initial conditions of the system easily to the state of its fundamental structure. This approach is very different and is based on the fundamental property of gluon saturation in QCD at high energies, assimilating an URHI as a Lorentz-contracted of highly coherent and dense gluon matter, called Color Glass Condensate (CGC)\cite{Gelis2015}, a universal form of hadronic matter at extremely high energies. Then, an URHIC can be viewed as a collision of two CGC sheets creating between them a coherent classical gluon field, called Glasma, which eventually decays into different partons and thermalizes, forming the PM. In the context of this model, the initial energy-momentum tensor is estimated from the glasma state.
\end{itemize}
The conceptual difference between the different approaches mentioned above leads to very different pictures for the initial state in order to start the hydrodynamic expansion. Therefore, different models can exhibit quantitative differences and depending on the initial condition model, the initial profile of the energy density distribution changes appreciably, and the final observables, like particle multiplicities and their fluctuations, can be affected also.

\subsection{Initial Conditions in Our Model} \label{ssec:6.0}
In studying the space-time evolution of a hot QCD matter, once it has been formed in an URHIC, it is usually assumed that RH is applicable. Furthermore, even if a hydro-dynamical description is appropriate, one needs to fix
the boundary and initial conditions. As it is well known that our time-Bjorken equation describing the hydrodynamic expansion of our system its integration should be supplied with the appropriate initial conditions. These initial conditions are very important to start the time evolution of the system under consideration. The initial conditions are the necessary ingredient for solving the hydrodynamic equations and the resulting solutions depend intimately on their choice. Therefore, the effect of initial conditions of proper time and temperature $(T_i,\tau_i)$ on hydrodynamical evolution of the system undergoing the colorless confining phase transition is important to investigate. The lack of information concerns the initial conditions, since these depend on the pre-equilibrium stage of the URHIC, a crucial little known stage, that is, the space-time evolution of the system from the moment of nuclear overlap between the two heavy ions to the establishment of an equilibrated PM. The initial conditions for the hydrodynamic description $(T_i,\tau_i)$, at a specified time to when the fluid dynamical expansion of the plasma starts, depend sensitively on the preceding space-time evolution during the pre-equilibrium stage. They are usually chosen on the basis of rough but acceptable estimates based on different approaches. The pre-equilibrium phase is of crucial importance because it has proved to have an influence on the hydrodynamic expansion of the PM and on the yield of certain physical observables. In this section we discuss in more detail the effect of the initial conditions, in proper time and temperature, used by us to analyze the behavior of the time evolution of the system.

\subsection{Results and Discusion} \label{ssec:6.1}
The finite volume Bjorken equation is solved with some particular initial conditions. As well known that, these initial conditions are sufficiently chosen to ensure that the PM is formed in an equilibrium state.
Like the proper time $\tau_i$ which expected to be around $0.1-1fm/c$\cite{Liu2014}, and the initial energy density $\epsilon_i$  which, according to Bjorken work \cite{Bjorken1983}, is estimated to $\epsilon_i\sim 1-10 GeV/fm^3$ for the RHIC collider and $\epsilon_i\sim 15 GeV/fm^3$ for Pb-Pb collision in LHC collider\cite{Krajczar2011}. In this work, we choose the range of initial temperature $T_i$ between $180-300 MeV$ which equivalent to initial energy density in $\epsilon_i\sim 1.7-12.7 GeV/fm^3$. Because we have two initial conditions $(\tau_i,T_i)$,  in order to investigate the effect of them one has to fix one and let the second variable. We perform the work with a fixed volume $V=1000fm^3$ in the two cases : without and with the CC.
The plots displayed in figures(Fig.\ref{Fig:12}-\ref{Fig:15}) represent the results obtained concerning the relation $T(\tau)$ for different initial conditions:$(T_i,\tau_i)$. We see that the behavior of the whole plots is the same as depicted by the first ones(Fig.\ref{Fig:02}-\ref{Fig:03}), only there is some dilation in times with increasing the initial conditions, which consequently is reflected by the increase in different lifetimes.
The plots displayed in figures(Fig.\ref{Fig:30}-\ref{Fig:33}) represent the different times and lifetimes as a function on the initial conditions $(T_i,\tau_i)$. However and in order to avoid overlapping, similar curves have been shifted vertically for clarity. We notice an increasing of the different times and lifetimes with increasing the initial conditions. Even in the case of 3-Dim fluid dynamical simulations of URHIC\cite{Niemi2014}, similar behavior is observed. The dependence of times and lifetimes on the initial conditions is not affected by the CC.

\begin{figure}
  \resizebox{1.0\hsize}{!}{
  \includegraphics*{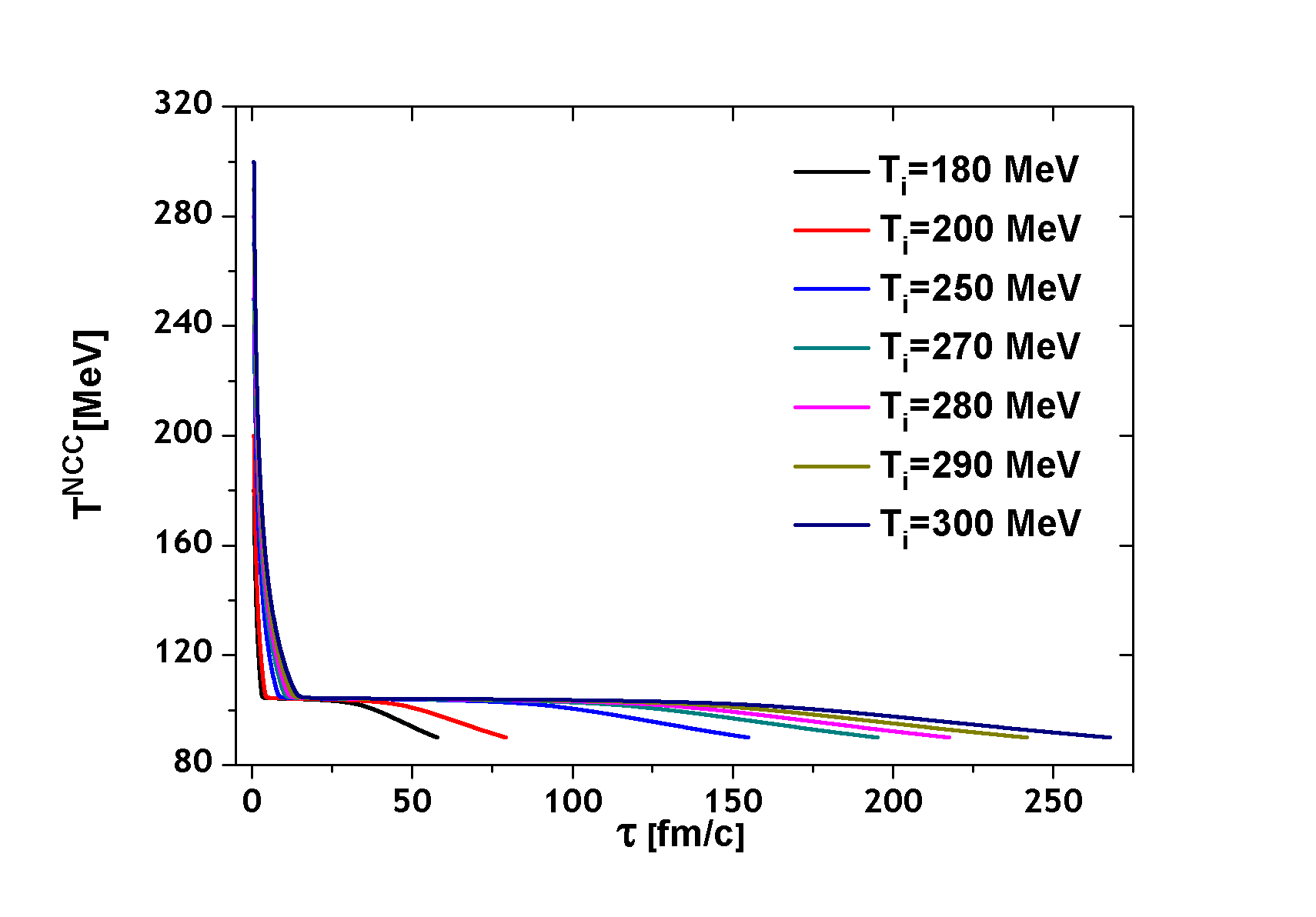}}
  \caption{Time evolution of $T^{NCC}(\tau)$ vs the proper time $\tau$ for different values of $T_i$ ($V=1000fm^{3}$ and $\tau_i=0.6fm/c$).}\label{Fig:12}
\end{figure}
\begin{figure}
  \resizebox{1.0\hsize}{!}{
  \includegraphics*{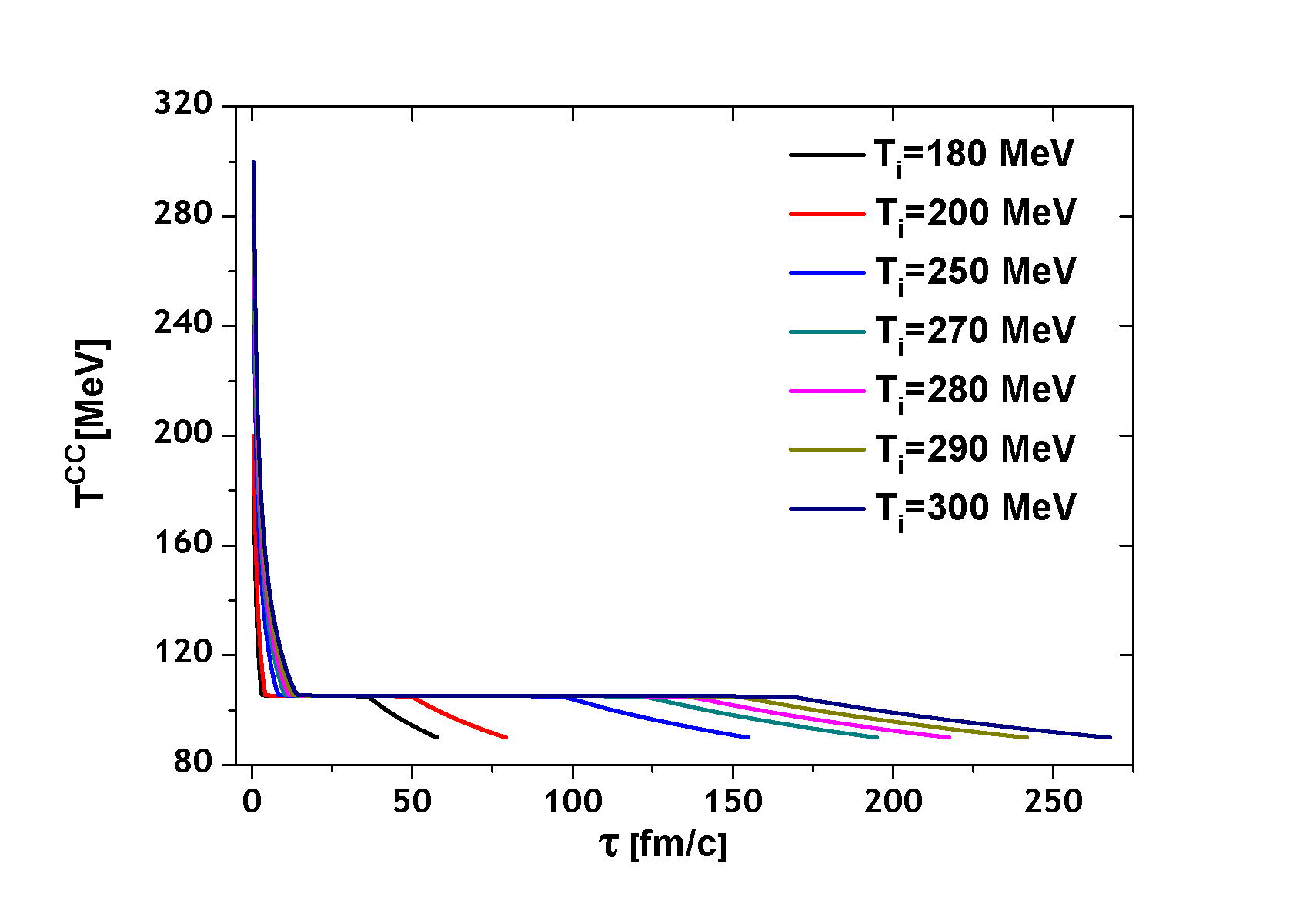}}
  \caption{Time evolution of $T^{CC}(\tau)$ vs the proper time $\tau$ for different values of $T_i$ ($V=1000fm^{3}$ and $\tau_i=0.6fm/c$).}\label{Fig:13}
\end{figure}
\begin{figure}
  \resizebox{1.0\hsize}{!}{
  \includegraphics*{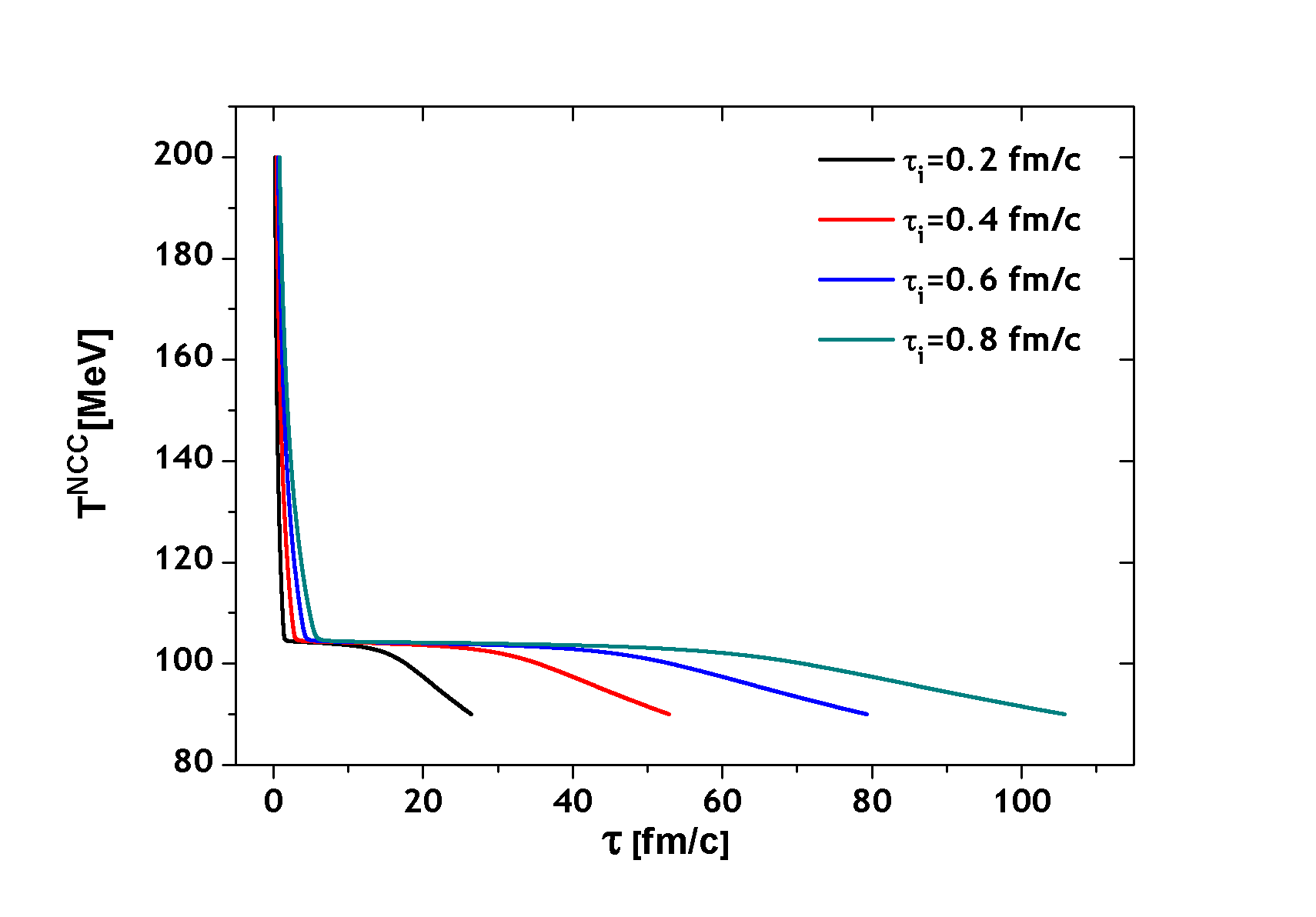}}
  \caption{Time evolution of $T^{NCC}(\tau)$ vs the proper time $\tau$ for different values of $\tau_i$ ($V=1000fm^{3}$ and $T_i=200MeV$).}\label{Fig:14}
  \end{figure}
\begin{figure}
   \resizebox{1.0\hsize}{!}{
  \includegraphics*{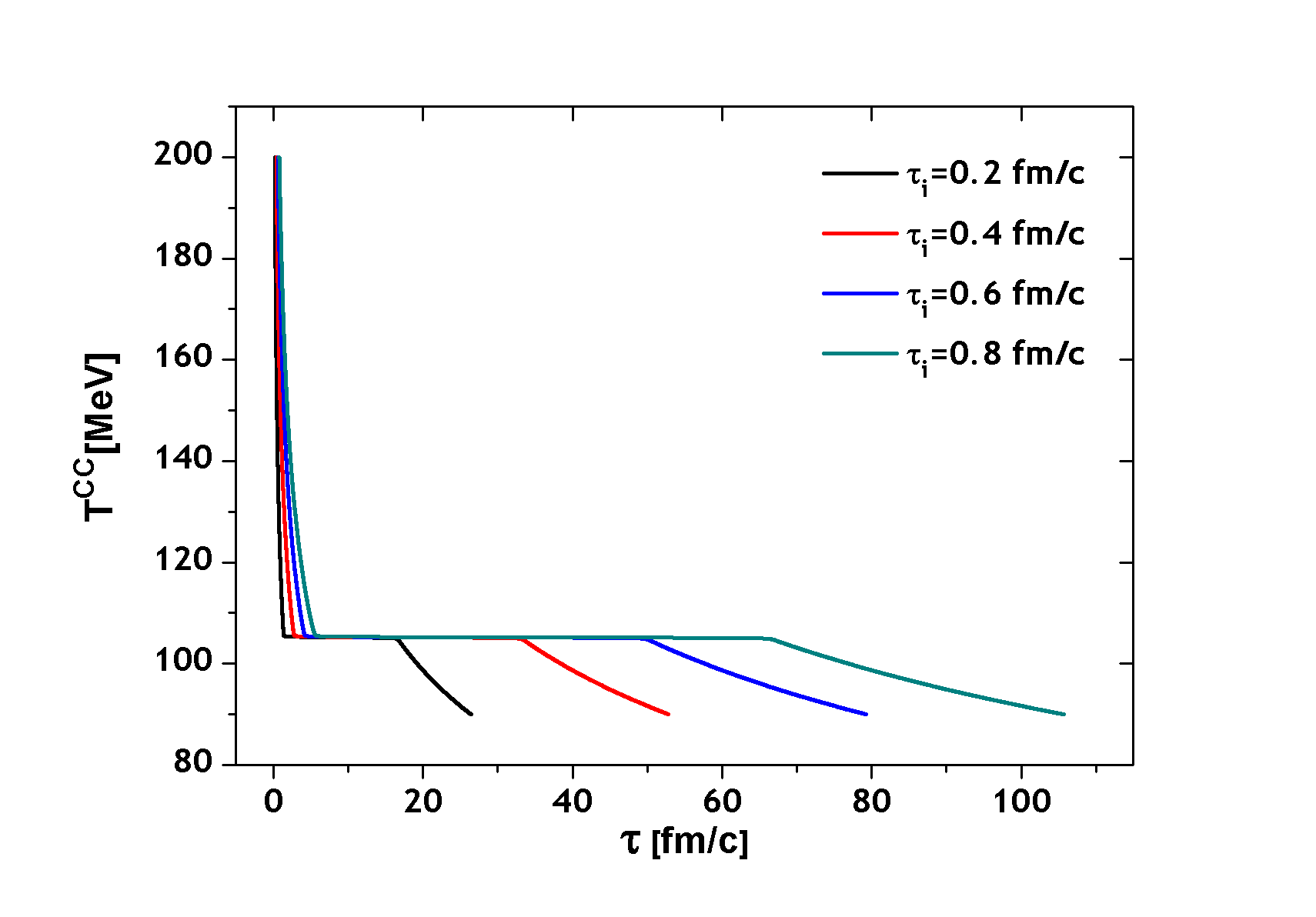}}
  \caption{Time evolution of $T^{CC}(\tau)$ vs the proper time $\tau$ for different values of $\tau_i$ ($V=1000fm^{3}$ and $T_i=200MeV$).}\label{Fig:15}
\end{figure}

When trying to fit these different plots using functions given by the following power laws,
\begin{equation}
\left\{
\begin{array}{c}
  \tau(T_i)\propto (T_i)^{\nu_T}\\
  \Delta  \tau(T_i)\propto (T_i)^{\mu_T}  \\
\end{array}%
\right.  \label{tauTTi}
\end{equation}
\begin{equation}
\left\{
\begin{array}{c}
  \tau(\tau_i)\propto (\tau_i)^{\nu_{\tau}}\\
  \Delta  \tau(\tau_i)\propto (\tau_i)^{\mu_{\tau}},  \\
\end{array}%
\right.  \label{tauTtaui}
\end{equation}

we obtain excellent results which are summarized in the table (\ref{tab:7}).
\begin{table}
\vspace*{1cm}  % with the correct table height
\centering
\caption{Fit parameters of the four lifetimes $\Delta\tau^{NCC}_{PP}$, $\Delta\tau^{NCC}_{PT}$, $\Delta\tau^{CC}_{CPP}$ and $\Delta\tau^{CC}_{PT}$  and the four times $\tau^{NCC}_{0}$, $\tau^{CC}_{0}$, $\tau^{NCC}_{H}$ and $\tau^{CC}_{H}$. }\label{tab:7}
\begin{tabular}{llllll}
\hline\noalign{\smallskip}
fit parameters & $\Delta\tau^{NCC}_{PP}$ & $\Delta\tau^{CC}_{CPP}$ & $\Delta\tau^{NCC}_{PT}$ & $\Delta\tau^{CC}_{PT}$ \\
\noalign{\smallskip}\hline\noalign{\smallskip}
$\mu_T$ &$3.02625$ &$3.00008$ &$2.99526$ & $3.00068$ \\
$\mu_{\tau}$ &$0.98727$ &$1.0000$ &$0.98727$ & $1.000$\\
\hline\noalign{\smallskip}
      & $\tau^{NCC}_{0}$ & $\tau^{CC}_{0}$ & $\tau^{NCC}_{H}$ & $\tau^{CC}_{H}$ \\
\noalign{\smallskip}\hline\noalign{\smallskip}
$\nu_T$ &$3.0279$ &$3.00067$ &$2.99809$ & $3.00067$ \\
$\nu_{\tau}$ &$0.99369$ &$1.0000$ &$0.99320$ & $1.000$\\ \hline
\end{tabular}
\vspace*{1cm}  % with the correct table height
\end{table}

These behaviors can be explained from the general forms (Rel.\ref{FSStime3},\ref{FSStime6}), predominantly due to the $\tau_{0}(\infty)$ and $\tau_{H}(\infty)$ as functions of $\tau_{i}$ and $T_{i}$ and therefore a good agreement is noticed. A similar behavior between the lifetime and the initial energy density was observed in the context of different approach \cite{Rischke1996,Rischke1995} and indeed confirms our results. As the initial conditions become large, it takes longer to cool the system and the different lifetimes become longer. Finally, we can say that the numerical values of the different exponents are basically,
\begin{equation}
\left\{
\begin{array}{c}
  \mu_T=\nu_T\simeq 3\\
  \mu_{\tau}=\nu_{\tau}\simeq 1.  \\
\end{array}%
\right.  \label{munuNV}
\end{equation}

\begin{figure}
  \resizebox{1.0\hsize}{!}{
  \includegraphics*{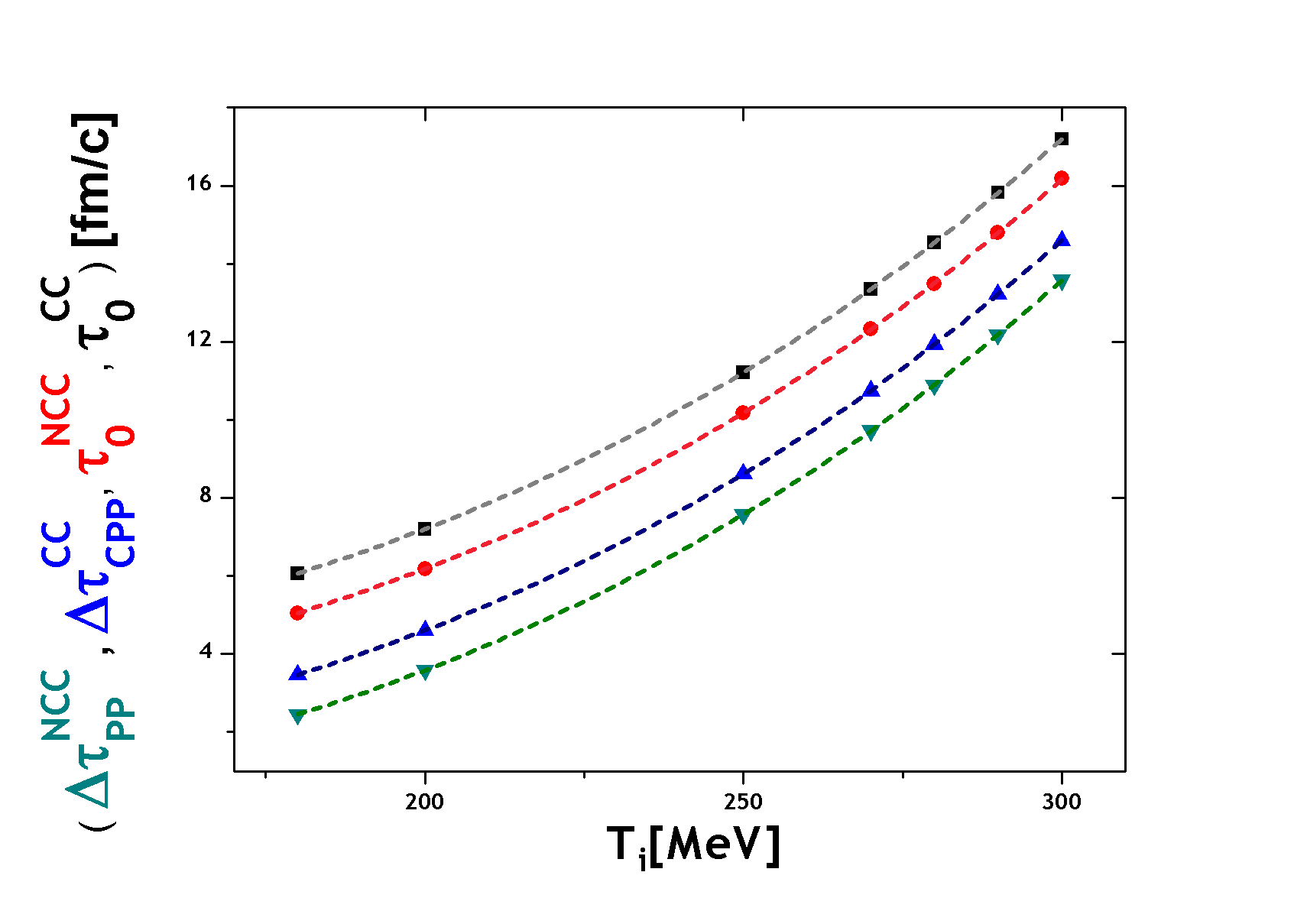}}
 \caption{Variation of $\Delta \tau^{NCC}_{PP}$, $\Delta \tau^{CC}_{CPP}+1fm/c$ , $\tau^{NCC}_{0}+2fm/c$ and $\tau^{CC}_{0}+3fm/c$ vs the initial temperature $T_i$.}\label{Fig:30}
  \end{figure}
\begin{figure}
  \resizebox{1.0\hsize}{!}{
  \includegraphics*{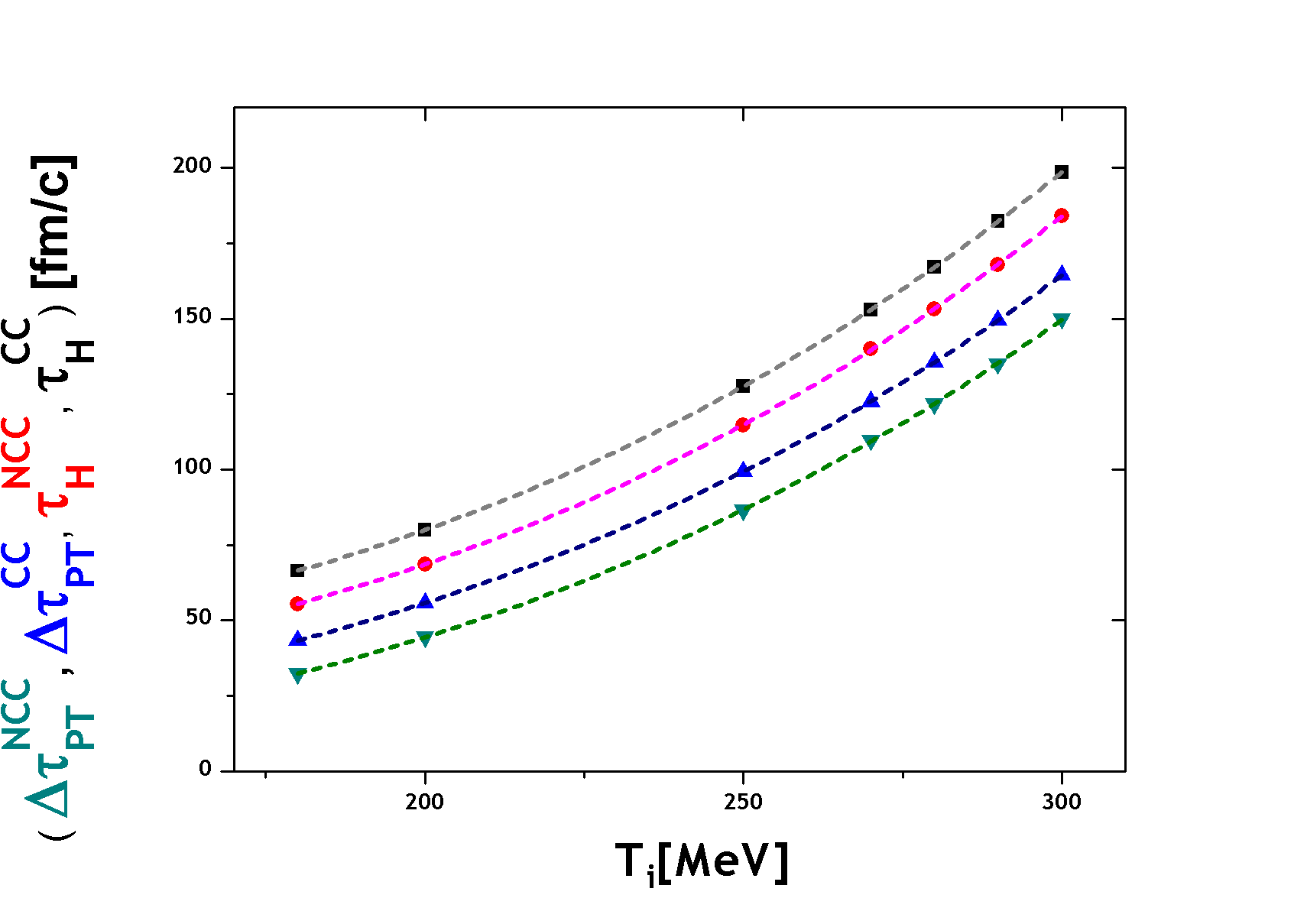}}
 \caption{Variation of $\Delta \tau^{NCC}_{PT}$, $\Delta \tau^{CC}_{PT}+10fm/c$ , $\tau^{NCC}_{H}+20fm/c$ and $\tau^{CC}_{H}+30fm/c$ vs the initial temperature $T_i$.}\label{Fig:31}
  \end{figure}
\begin{figure}
  \resizebox{1.0\hsize}{!}{
  \includegraphics*{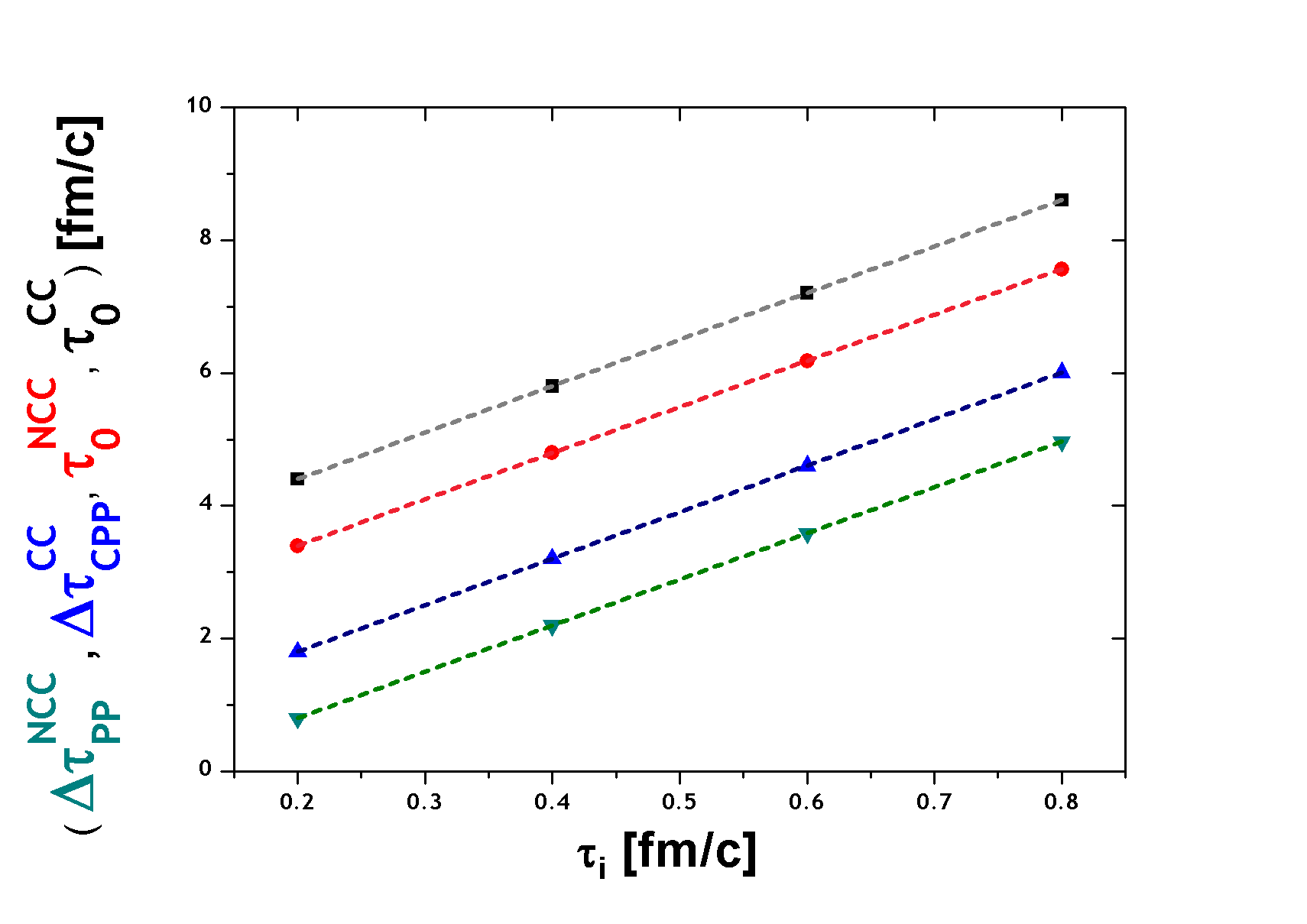}}
 \caption{Variation of $\Delta \tau^{NCC}_{PP}$, $\Delta \tau^{CC}_{CPP}+1fm/c$ , $\tau^{NCC}_{0}+2fm/c$ and $\tau^{CC}_{0}+3fm/c$ vs the initial time $\tau_i$.}\label{Fig:32}
  \end{figure}
\begin{figure}
  \resizebox{1.0\hsize}{!}{
  \includegraphics*{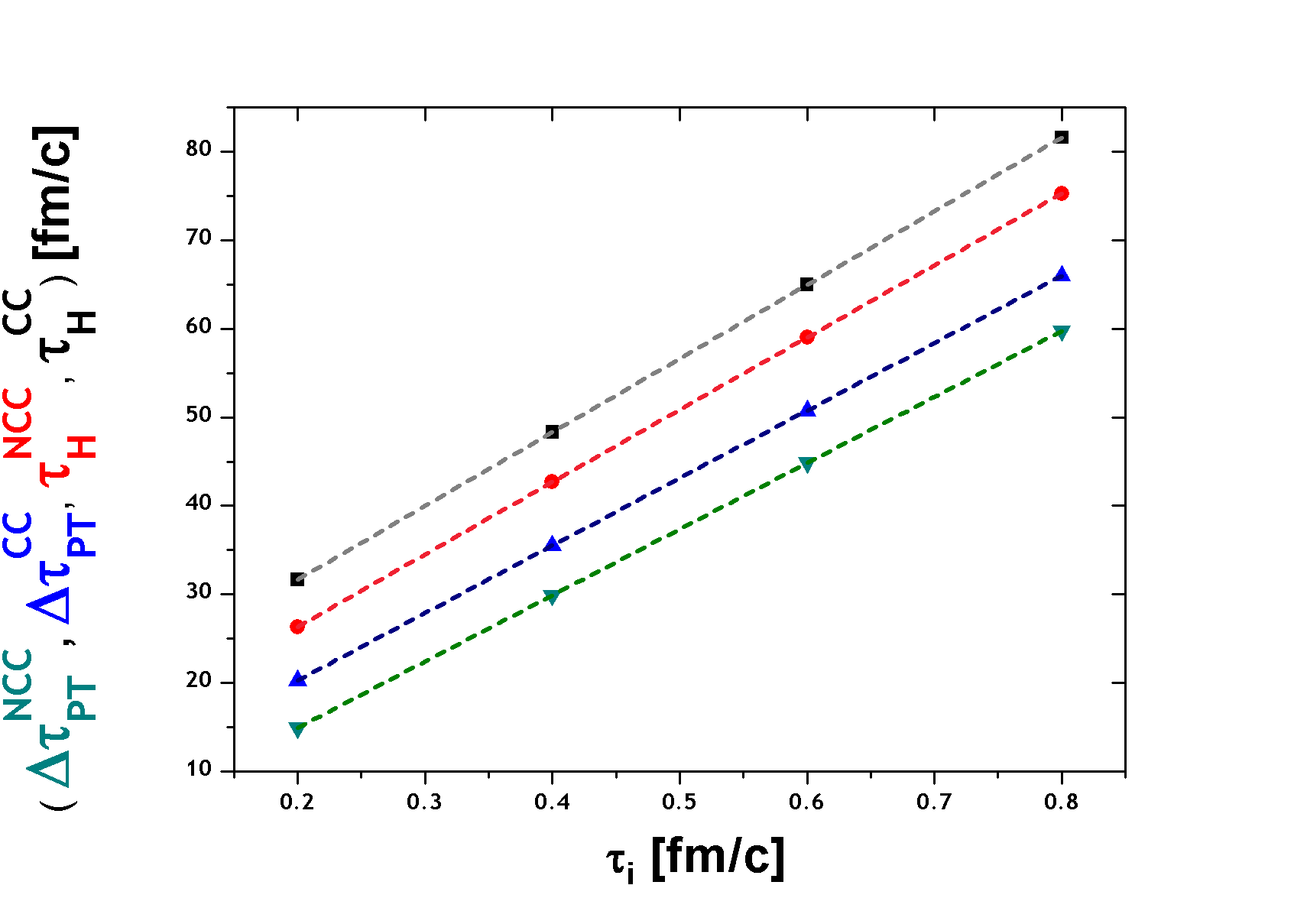}}
 \caption{Variation of $\Delta \tau^{NCC}_{PT}$, $\Delta \tau^{CC}_{PT}+5fm/c$ , $\tau^{NCC}_{H}+10fm/c$ and $\tau^{CC}_{H}+15fm/c$ vs the initial time $\tau_i$.}\label{Fig:33}
  \end{figure}

\section{Time Evolution of Thermal Response Functions : time Response Functions}\label{sec:07}
\subsection{Mathematical Property of Functions Composition} \label{ssec:7.1}
In a system within our Colorless QCD MIT-Bag Model, some TRF are discontinuous functions (like $\epsilon(T,V),\\ \mathcal{H}(T,V), ...$) and other TRF are continuous (like $\mathscr{P}(T,V),\\\mathscr{C}_{s}^{2}(T,V), ...$ ).
In any discontinuous TRF due to the  first order confining phase transition, which is related to the latent heat, a finite discontinuity which is delimited by two particular points. The solution of the Bjorken equation that we have obtained has revealed that the discontinuity in $T(\tau)$ is delimited by the same particular points. Mathematically speaking in order to get the time dependence of any TRF $\mathscr{F}(\tau)$  we shall compose the first TRF $\mathscr{F}(T)$ with $T(\tau)$. In any case, it is worth pointing out two important properties of the
composition of discontinuous functions :
\begin{itemize}
\item  $1^{st}$ property
\begin{equation}
\left\{
\begin{array}{c}
  \mbox{If     } \mathscr{F}(T) \mbox{      is discontinuous}  \\
  \mbox{and if } T(\tau) \mbox{        is discontinuous}  \\
  \Rightarrow   \mathscr{F}\circ T=\mathscr{F}(\tau) \mbox{         is continuous      } \\
\end{array}%
\right.  \label{FSSfits}
\end{equation}
The composition of two functions with the same finite discontinuity generates a continuous function.
\item  $2^{nd}$ property
\begin{equation}
\left\{
\begin{array}{c}
  \mbox{If     } \mathscr{F}(T) \mbox{      is continuous}  \\
  \mbox{and if } T(\tau) \mbox{        is discontinuous}  \\
  \Rightarrow   \mathscr{F}\circ T=\mathscr{F}(\tau) \mbox{         is discontinuous      } \\
\end{array}%
\right.  \label{FSSfits1}
\end{equation}
\end{itemize}
The composition of a continuous function with a discontinuous function with a finite discontinuity generates a discontinuous function with the same discontinuity.
One intriguing mathematical property being explored in any functional analysis textbook is that of the continuity property in the composition of functions. The continuity property between composing functions and the resulting
function may help us to explain the behavior of the time evolution of different tRF investigated in this part of the present work.

\subsection{Results of $\mathcal{H}(\tau,V)$, $\epsilon(\tau,V)$, $\mathscr{P}(\tau,V)$ and $\mathscr{C}_{s}^{2}(\tau,V)$: Analysis and Discusion} \label{ssec:7.2}
We have used the duality relation $T(\tau,V)$ in order to transform the order parameter $\mathcal{H}(T,V)$, the energy density $\epsilon(T,V)$, the pressure $\mathscr{P}(T,V)$ and the sound velocity $\mathscr{C}_{s}^{2}(T,V)$ into the corresponding tRF $\mathcal{H}(\tau,V)$, $\epsilon(\tau,V)$, $\mathscr{P}(\tau,V)$ and $\mathscr{C}_{s}^{2}(\tau,V)$.
The results we obtained were presented graphically and displayed on different plots given by the figures (Figs.\ref{Fig:16}-\ref{FIG23}). We start our discussion with the plot of the order parameter $\mathcal{H}(\tau,V)$ which increases monotonically in both cases (colorless case and the non-colorless case) without discontinuities, Figs.(\ref{Fig:16},\ref{Fig:17}).
During the longitudinal expansion and as the confining phase transition progresses, the available space will be progressively invaded by HM, thus the order parameter increases as a function of proper time. The delay in completion of QCD confining phase transition due to CC is seen clearly. Our order parameter has a behavior in complete agreement with those obtained and quoted in Refs.(\citen{Csernai1992, Mustafa1998}).
The same behavior is noticed in Figs.(\ref{Fig:18},\ref{Fig:19}) showing the plots of the energy density $\epsilon (\tau,V)$ with a decreasing feature. It is logical that the energy density decreases when the volume increases, following the hydrodynamical expansion of the system. With a system that looks like ours, namely a system undergoing a confinement phase transition, using a 3-Dim hydrodynamic simulation model, the time evolution of the energy density obtained is very similar to ours\cite{Hirano2002}. A time evolution that can be easily described by a power law of type: $\epsilon(\tau,V)\propto \tau^{-\theta}$, with a slight dependence of the exponent $\theta$ with the volume.

However, on the plots of the pressure $\mathscr{P}(\tau,V)$ the shape of the curves is very different, a knee structure appears when we take into account the CC, which manifests itself as a second particular point in
Figs.(\ref{Fig:20},\ref{Fig:21}). We notice that the FSE are more clear in the colorless case than in the non-colorless case. The two particular points emerging in the behavior of $\mathscr{P}(\tau,V)$ are nothing the points
marking the beginning of the confining phase transition occuring at $\tau_0$ and its end of the mixed phase period $\tau_H$. The change in the cooling law for the pressure when entering the mixed phase period is clearly visible. A similar behavior and an agreement is noticed with the result obtained in Ref.(\citen{Rischke1996}) using simpler EoS than ours.

In the plots of pressure $\mathscr{P}(\tau,V)$ the FSE are important around the first particular point in the colorless case than in the non-colorless one. However, this behavior is reversed around the second
particular point. The appearance of the second particular point is more obvious in the colorless case rendering the emergence of the softest region more clear in the colorless case and the time duration of this region is more
definite which is nothing that the $\Delta \tau_{PT}^{CC}$.

Most of these physical properties of the system control well the origin of the disturbances that are produced in the system and the way of their propagation due to the high energy collision. This fact has led to extensive studies of the hydrodynamics associated with the propagation of the sound. The time evolution in the first-order confining phase transitions involves nontrivial hydrodynamics. For that reason, the study of the propagation of sound in that system is very interesting. The speed of sound has long been considered as a sensitive hydrodynamic function to the behavior of the strongly interacting matter undergoing a variation in its EoS (phase transition). As a consequence, from a hydrodynamic point of view, the sound velocity varies in space and time during the confining phase transition.

From the definition of $\mathscr{C}_{s}^{2}$ (Rel. \ref{FirstSoundV}) we can write it as a function of $(\frac{\mathscr{P}}{\epsilon})$,
\begin{equation}
\mathscr{C}_{s}^{2}=\left(\dfrac{\partial \mathscr{P}}{\partial \epsilon }\right)=\left(\frac{\mathscr{P}}{\epsilon} \right)+\epsilon \dfrac{\partial }{\partial \epsilon }\left(\frac{\mathscr{P}}{\epsilon} \right)
\end{equation}%
Let us now return to the curves displayed in Figs.(\ref{FIG22},\ref{FIG23}) illustrating the variation of the speed of sound squared as a function of time $\tau$ for different volumes. When approaching the thermodynamic limit the speed of sound takes the value for an Ultra-Relativistic(UR) ideal gas $\mathscr{C}_{s}^{2}=1/3$ at $\tau \leq \tau_{0}$\ and $\tau \geq \tau _{H}$.
On a range of evolution time between $\tau _{0}$\ and $\tau _{H},\ $ i.e., commonly known as the softest region, during the mixed phase the speed of sound is nearly vanishing,\ reflecting the first order character of the transition. In our case and from figure \ref{FIG23}, we can extract the numerical values of  $\tau_{0}\sim 4.267\frac{fm}{c}$ and $\tau _{H}\sim 50.280\frac{fm}{c}$. It has been interpreted that the state of the mixed phases does not expand due to its internal pressure, even if there are strong gradients in the energy density, which has the consequence of not performing mechanical work and therefore cools less quickly. Thus, the expansion of the system is delayed and its lifetime is considerably prolonged. For small systems, the sound velocity is damped during the phase transition and does not vanish, since pressure gradients are finite, but they are still smaller than for an ideal gas EoS, and therefore the tendency of the system to expand is also reduced.

During the lifetime $\Delta \tau_{PT}$ of the confinement phase transition, the energy density  $\epsilon (\tau)$ changes significantly, while the pressure $\mathscr{P} (\tau)$, which is discontinuous, varies slowly. The sound velocity $\mathscr{C}_{s}^{2}(\tau)$ must therefore become very small in this range of time. The EoS of the system, relating energy density $\epsilon (\tau)$ with pressure $\mathscr{P} (\tau)$, is called stiff when the sound velocity is high and soft when it is low. Thus, our PM EoS is therefore stiff well before and well after the confinement phase transition, but becomes soft near the finite volume transition point $T_0(V)$.

The exact value $\mathscr{C}_{s}^{2}=1/3$ is reached following different ways in the two regions(PM and HM phases). In the HM the tendency to reach the UR limit is slower in the non colorless case than in the
colorless case, however the FSE are more pronounced in the colorless case as always.

Also, in these plots we notice that the FSE are more important in the colorless case than in the non colorless case, leading the sound velocity at the softest point to be distinctly higher in the colorless case. For example $ \mathscr{C}_{s}^{2}(\tau)^{CC}=0.050$ and $ \mathscr{C}_{s}^{2}(\tau)^{NCC}=0.020$ at $V=100fm^3$ \cite{Ladrem2019}.

We also find, when approaching the thermodynamic limit, that the sound velocity, in the PM phase, $\mathscr{C}_{s}^{2}$ reaches the UR value $\mathscr{C}_{s}^{2}=1/3$ for the time values larger than in the finite volume case. This what it is displayed in Fig.(\ref{FIG23}). The minimum in the sound velocity squared is found to be $(\mathscr{C}_{s}^{2})_{min}\simeq0.00744$ at $V=1000fm^{3}$ and $(\mathscr{C}_{s}^{2})_{min}\simeq 0.04252$ at $V=100fm^{3}$. The sound velocity value drops by a factor of 7 at $V=1000fm^{3}$ and by a factor of 3 at $V=100fm^{3}$ when the time is between the two limits.

\begin{figure}
  \resizebox{1.0\hsize}{!}{
  \includegraphics*{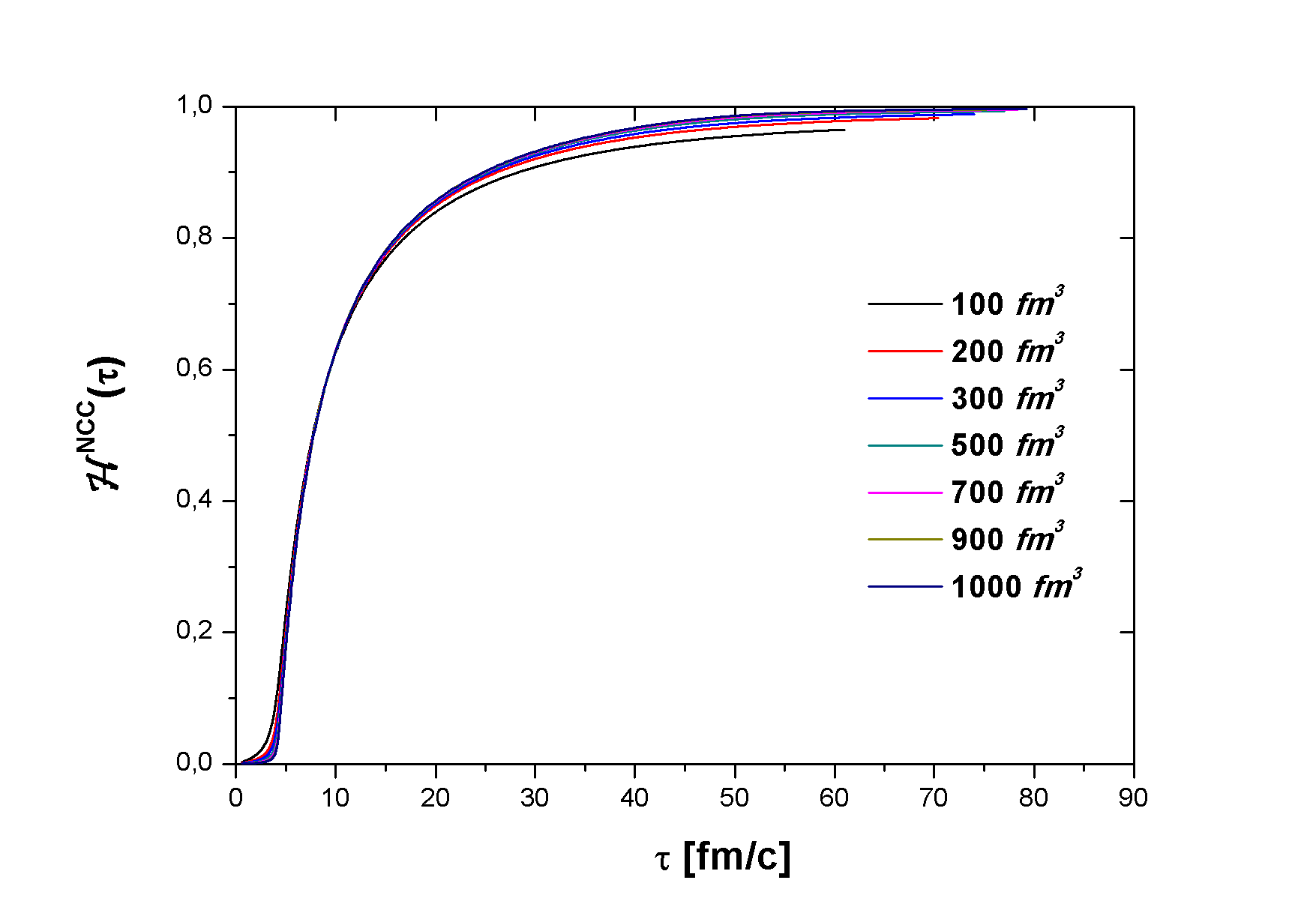}}
  \caption{The order parameter $\mathcal{H}^{NCC}(\tau,V)$ vs the proper time $\tau$, without CC, for different volumes $V$.}\label{Fig:16}
\end{figure}
\begin{figure}
  \resizebox{1.0\hsize}{!}{
  \includegraphics*{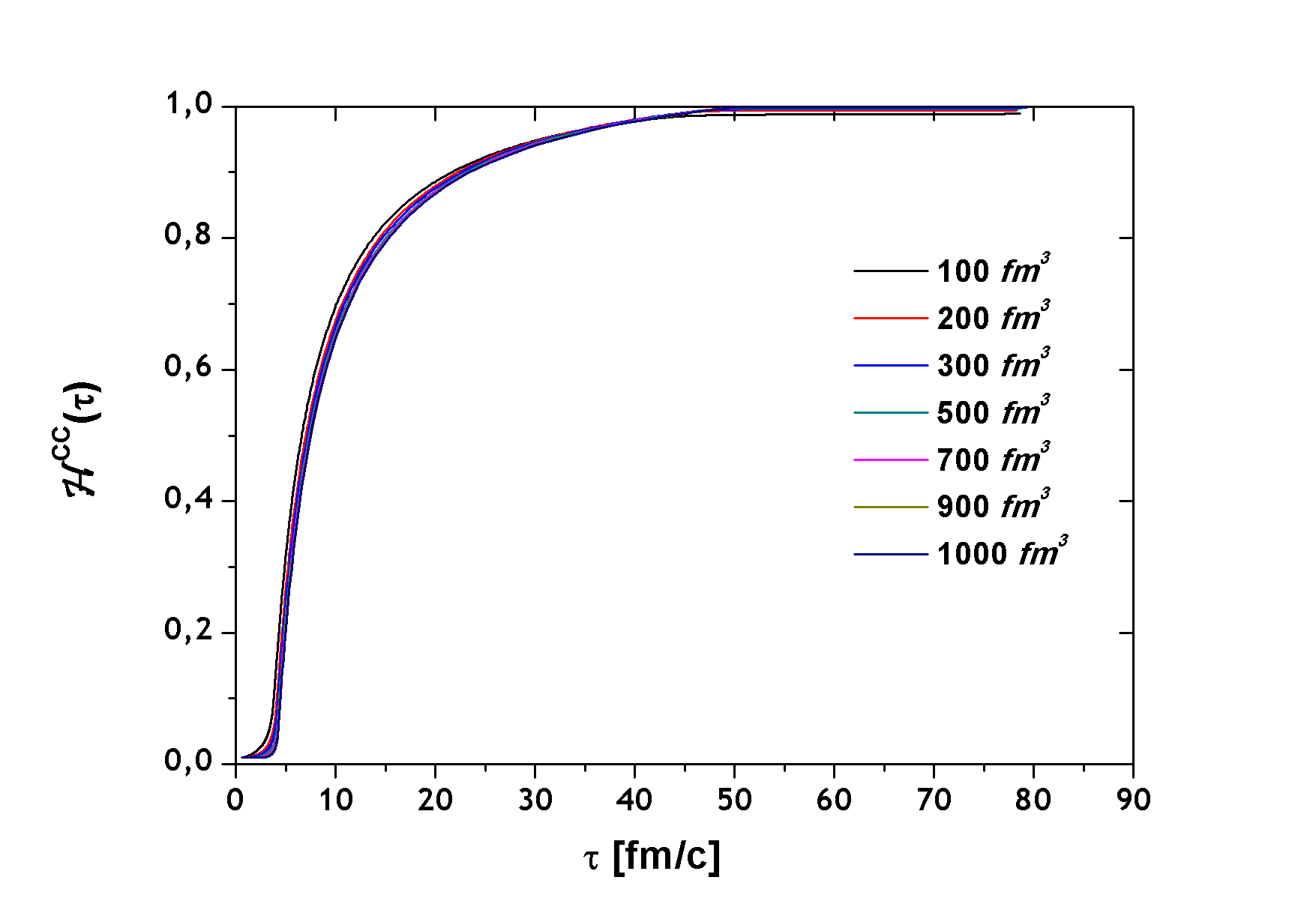}}
  \caption{The colorless order parameter $\mathcal{H}^{CC}(\tau,V)$ vs the proper time $(\tau)$ for different volumes $V$.}\label{Fig:17}
\end{figure}

\begin{figure}
  \resizebox{1.0\hsize}{!}{
  \includegraphics*{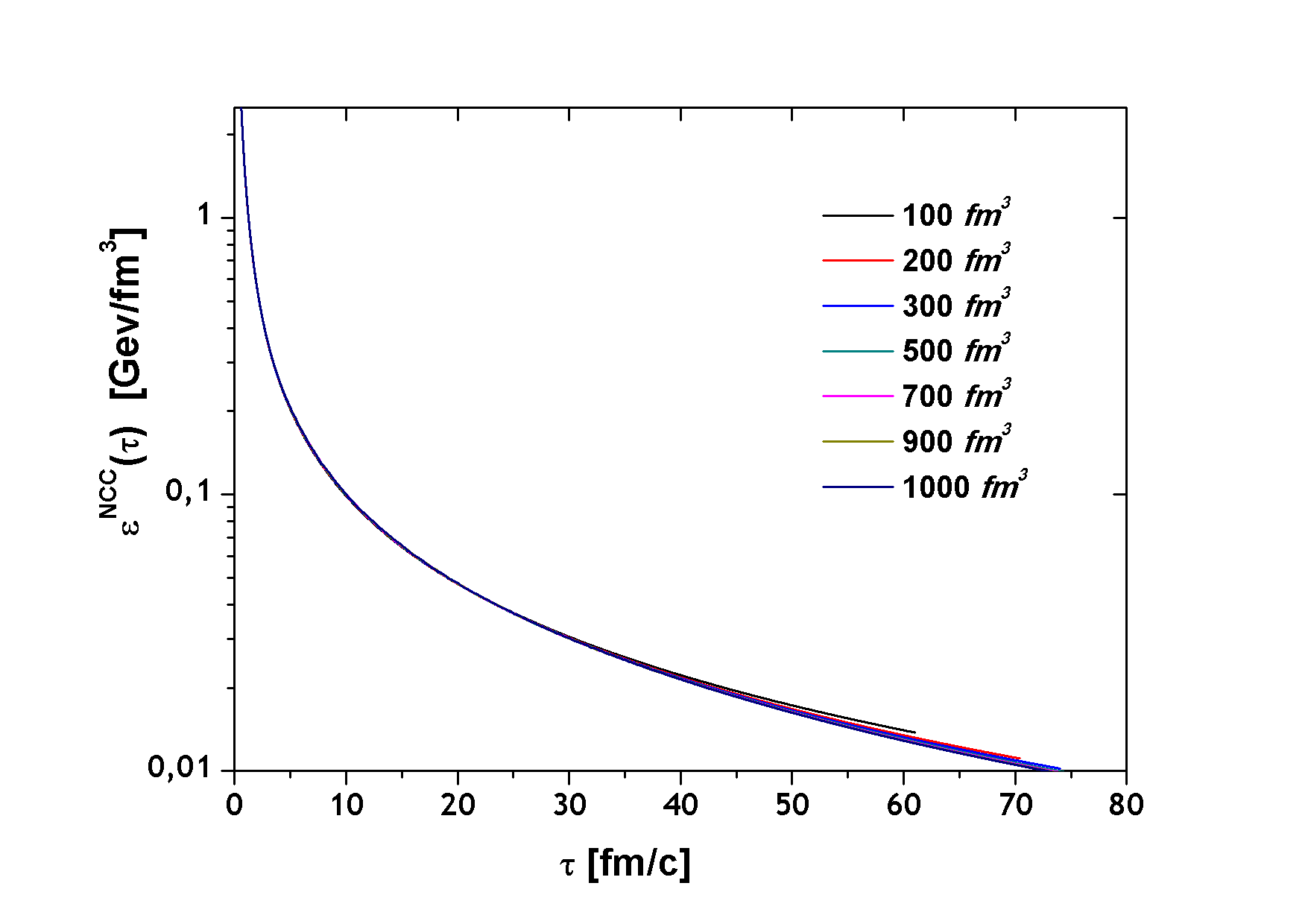}}
 \caption{The energy density $\epsilon^{NCC}(\tau,V)$ vs the proper time $\tau$, without CC, for different volumes $V$.}\label{Fig:18}
\end{figure}
\begin{figure}
  \resizebox{1.0\hsize}{!}{
  \includegraphics*{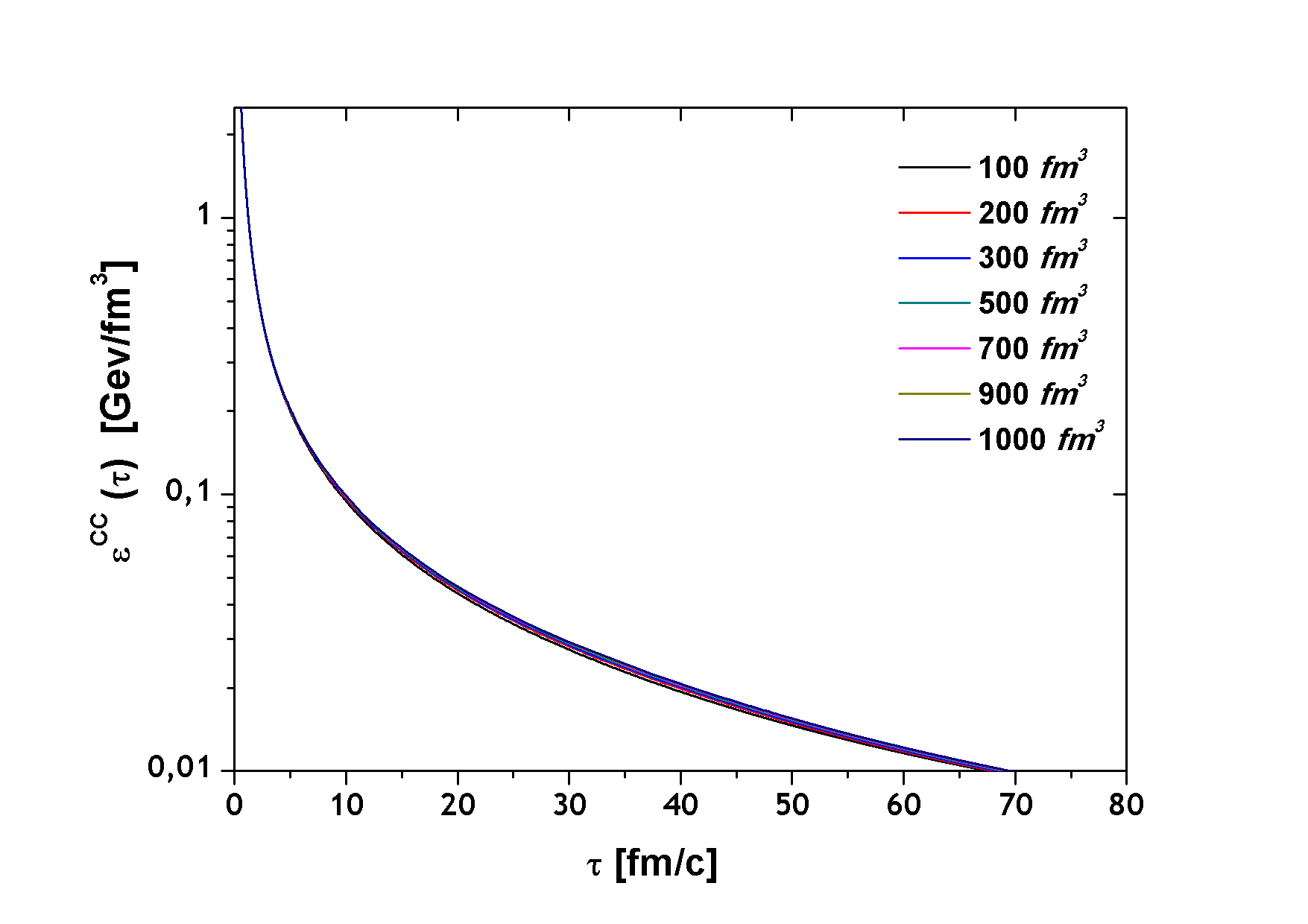}}
  \caption{The colorless energy density $\epsilon^{CC}(\tau,V)$ vs the proper time $\tau$ for different volumes $V$.}\label{Fig:19}
\end{figure}
\begin{figure}
  \resizebox{1.0\hsize}{!}{
  \includegraphics*{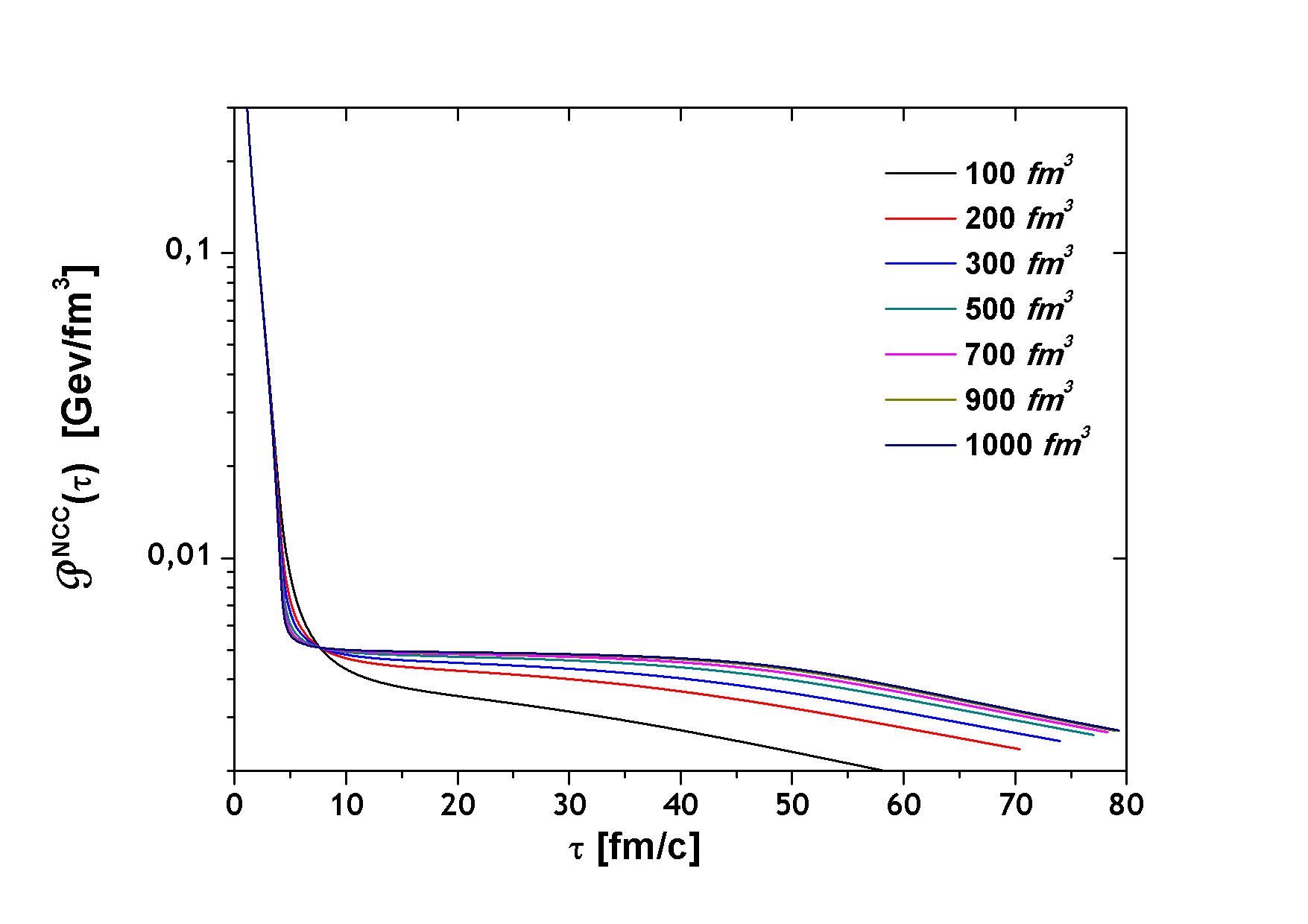}}
  \caption{The pressure $\mathscr{P}^{NNC}(\tau,V)$ vs the proper time $\tau$, without CC, for different volumes $V$.}\label{Fig:20}
\end{figure}
\begin{figure}
  \resizebox{1.0\hsize}{!}{
  \includegraphics*{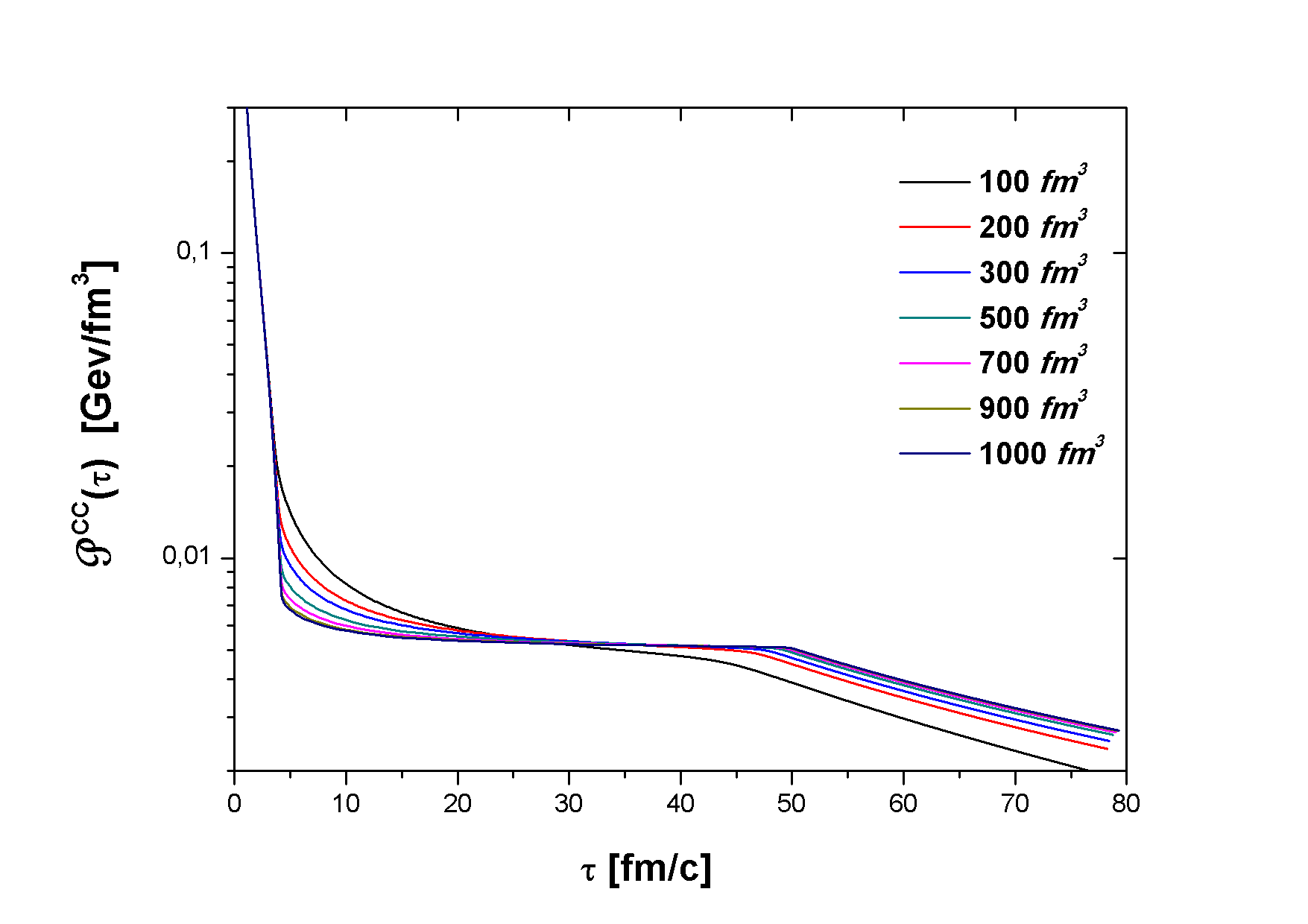}}
  \caption{The colorless pressure $\mathscr{P}^{CC}(\tau,V)$ vs the proper time $\tau$ for different volumes $V$.}\label{Fig:21}
\end{figure}

\begin{figure}
  \resizebox{1.0\hsize}{!}{
  \includegraphics*{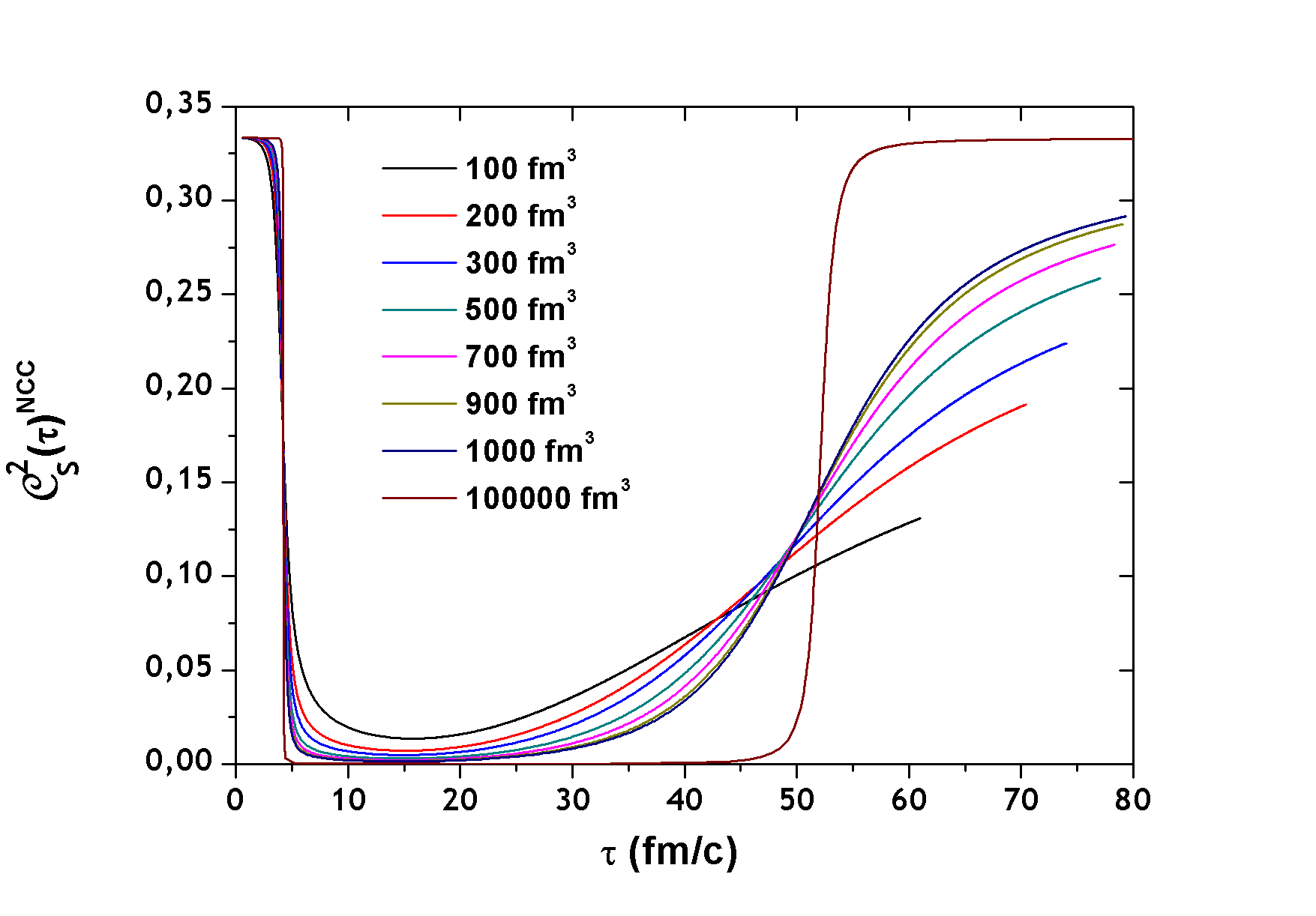}}
  \caption{The sound velocity $(\mathscr{C}_{s}^{2})^{NCC}(\tau,V)$ vs the proper time $\tau$, without CC, for different volumes $V$.} \label{FIG22}
\end{figure}
\begin{figure}
  \resizebox{1.0\hsize}{!}{
  \includegraphics*{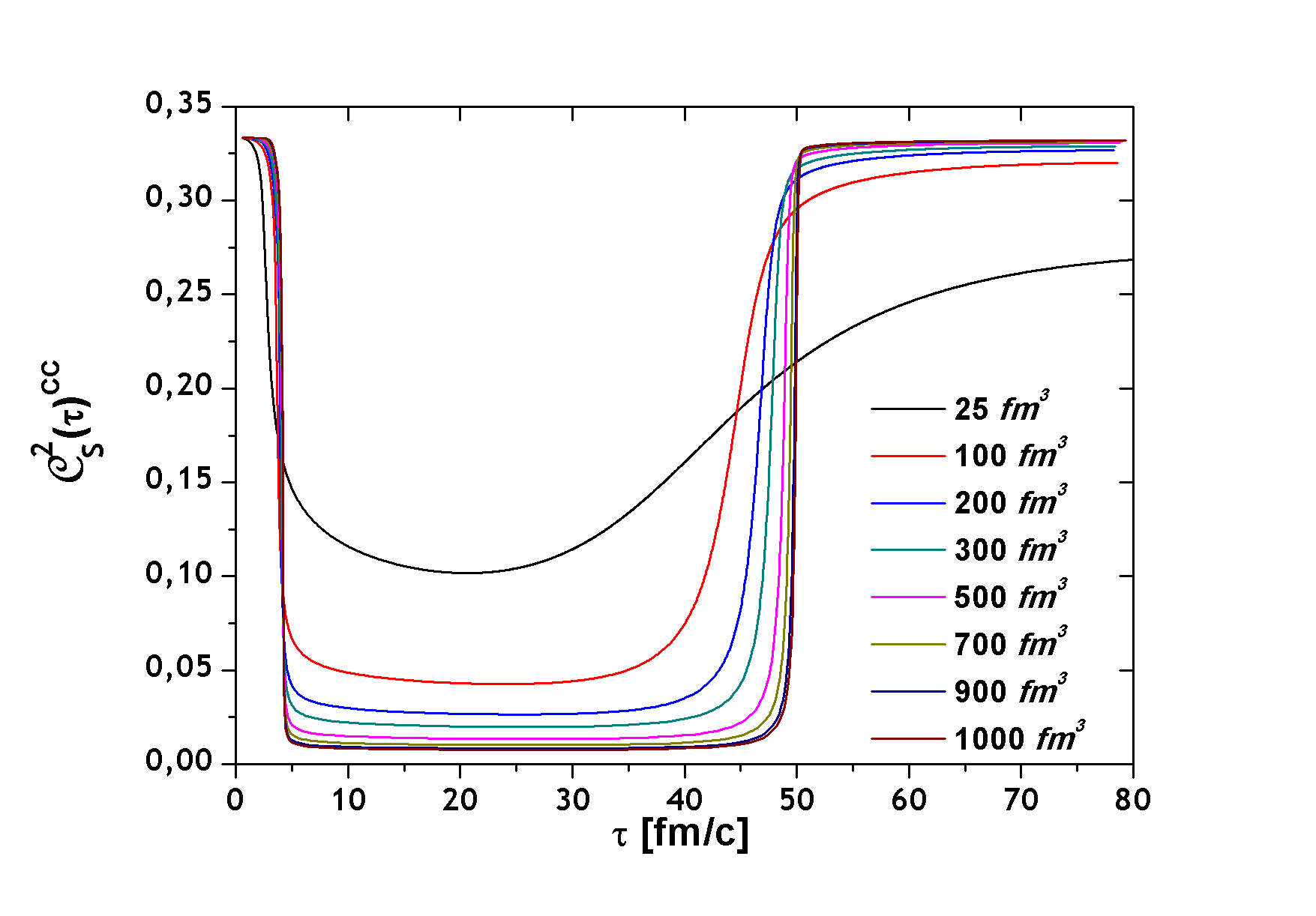}}
  \caption{The colorless sound velocity $(\mathscr{C}_{s}^{2})^{CC}(\tau,V)$ vs the proper time $\tau$ for different volumes $V$.} \label{FIG23}
\end{figure}

The softest region in EoS of our system is expected to have a significant influence on the collective dynamics of the hot and dense matter formed in URHIC. In particular, a small sound velocity delays the expansion of
the compressed matter and also leads to a reduced transverse collective flow. The plot of $ \mathscr{C}_{s}^{2}(\tau)$ as a function of $\tau$ is very similar to the plot of $ \mathscr{C}_{s}^{2}(\epsilon)$ as a function of $\epsilon$ (see the Ref. \citen{Ladrem2019}). This similarity is essentially due to the continuous decreasing behavior of energy density $\epsilon(\tau)$ as a function of proper time $\tau$ only the order between HG and CPP phases is inverted.

\subsection{Time Evolution of Energy Density $\epsilon(\tau,V)$ : Detailed Analysis and Discussion} \label{ssec:7.3}
In the context of our model the time evolution of the system, undergoing the confining phase transition from a CPP through the mixed phase to HG, experience in three stages. Obviously and because of the hydrodynamic expansion, the energy density decreases with time. The plots displayed in figures (Figs.\ref{Fig:35},\ref{Fig:366} and \ref{Fig:37}) depict this decreasing of the energy density with time during each stage of the hydrodynamical expansion, starting from the CPP phase until the final HM phase and going through the mixed phase at $V=1000fm^3$. These figures contain different sketches representing the different phases taken from the reference\cite{QGPHP}. Within each stage the system follows its time evolution. We have analyzed each stage individually. During the first stage, starting from $\tau_i$ until $\tau_0$, the colorless PM evolves hydrodynamically following the power law : $\epsilon \propto \tau^{-\theta_{CPP}}$. The second stage starts when the hadronic conversion is triggered and the mixed phase system evolves with a power less than the first one: $\epsilon \propto \tau^{-\theta_{Mixed}}$. This stage corresponds to the confining phase transition. In the final stage the system consist in pure HM, follows an hydrodynamical evolution with a freeze out towards a hadronic gaz. The density energy in this stage is similar to the first one $\epsilon \propto \tau^{-\theta_{HG}}$ with a power greater than one.
From the fitting work, the numerical values of the different powers obtained are: $\theta_{CPP}=1.26087,\theta_{Mixed}=1.00865$ and $\theta_{HP}=1.33135$. We see from the first value $\theta_{CPP}=1.26087 < 4/3$ meaning that the CPP is not a perfect gaz, may be in a correlated state like a liquid state and confirming our previous result\cite{Ladrem2019}. What is interesting to mention in this case concerns this temporal decrease in the energy density of the PM, deduced from two completely different approaches, namely the Partonic Cascade Model(PCM)\cite{Geiger1995} and its modified version($\theta_{RHIC} \thicksim 1.31093, \theta_{LHC} \thicksim 1.27265$)\cite{Biro1993}, which is in agreement with our result. Also, the same energy density decreasing with time is obtained using the Hot Glue Model(HGM)\cite{Xiong1994}.
Concerning the mixed phase, during the confining phase transition, the time evolution is slowed down. The system evolves with a power close to unity which is in good agreement with the fact that during the phase transition
the sound velocity is fundamentally zero. This explains why the phase transition make the lifetime of the system longer and why the time evolution is linear. However, when the system becomes a pure HM the trend of time
evolution becomes again similar to the first one. The evolution becomes faster and the numerical value of the power is greater than one but remains below the ideal value: $4/3$.
In the case of a 3-Dim space expansion of the system one can expect the numerical values of the $\theta$ exponent greater than $4/3$\cite{Cooper1975,Baym1983}. In order to compare our results with a hydrodynamic expansion in 3-Dim space, we have extracted the numerical of $\theta$ in each of the three stages of the time evolution of the energy density result: $\epsilon \propto \tau^{-\theta}$, obtained in the context of the hydrodynamic simulations\cite{Hirano2002}. After performing the fitting work, we obtained the following numerical values of the different powers: $\theta_{QGP}=1.488,\theta_{Mixed}=1.973$ and $\theta_{HP}=3.160$. These exponents reflect a a 3-Dim expanding system with a more rapid cooling process than in 1-Dim space expansion.

In each stage of the hydrodynamical evolution, the energy density decreasing process is due to the longitudinal scaling expansion of the system and to a mechanical work if it is possible. In the mixed phase stage only the geometrical dilution plays its role, leading to the cooling law described by a power $\theta_{Mixed} \thicksim 1 $. However, during the first and the third stages, additional work is performed and the system is in possession of a trend to cool more faster leading to cooling law with different powers $\theta_{CPP} \thicksim \theta_{HG} > 1$.

Now, when think over about the figure (Fig.\ref{FIG23}) we see that $\mathscr{C}_{s}^{2}$ becomes gradually zero when approaching the thermodynamic limit. This tendency affects the variation of the exponent
$\theta_{Mixed}$ as a function of the volume. After extracting the numerical values of $\theta_{Mixed}$ at different volumes and plot them as a function of $V$, we obtain the graph displayed by the figure (Fig.\ref{Fig:38}). We notice a logical decreasing in agreement with the known result according to which the sound velocity should converge towards zero value during the mixed phase in the thermodynamic limit.
The fitting function is nothing that the FSS power law with an exponent close to one:
\begin{equation}
\theta_{Mixed}(V)- 1 \propto  V^{-1} + \mathcal{O}(V^{-2})
\label{FSSTHETA}
\end{equation}

\begin{figure}
  \resizebox{0.9\hsize}{!}{
  \includegraphics*{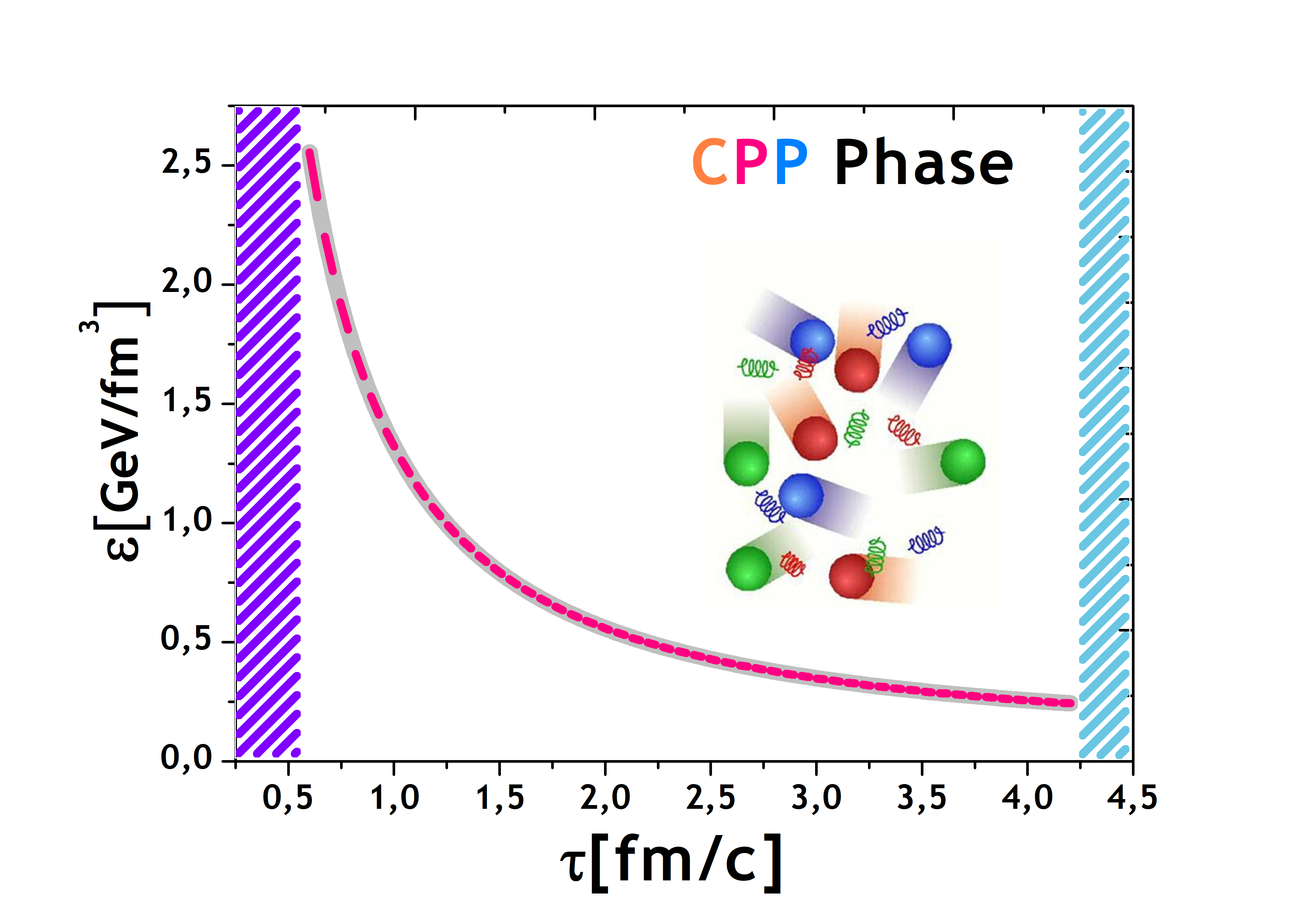}}
 \caption{Time evolution of the CPP phase for a volume $V=1000fm^3$.}\label{Fig:35}
\end{figure}
\begin{figure}
  \resizebox{0.9\hsize}{!}{
  \includegraphics*{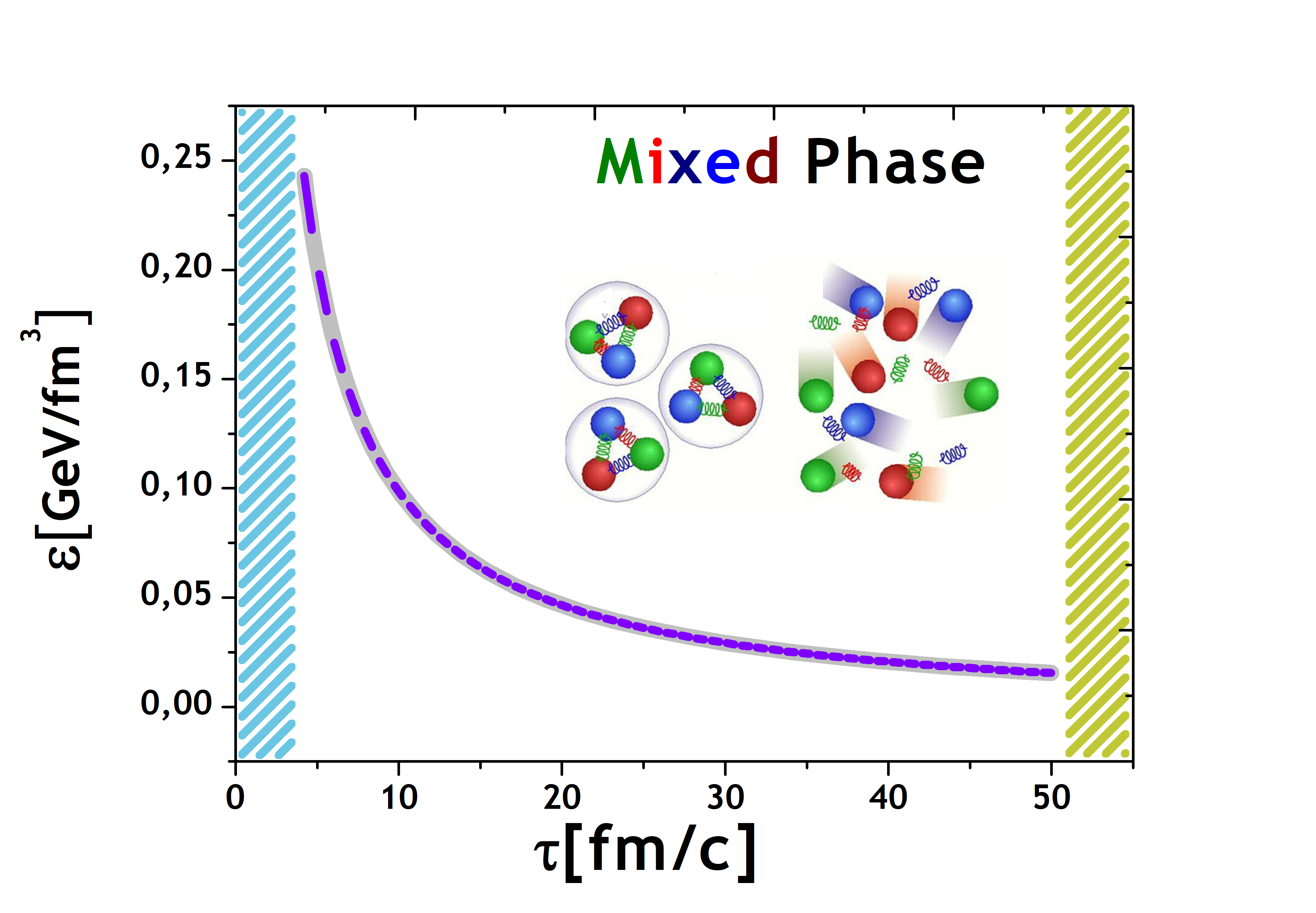}}
 \caption{Time evolution of the mixed phase for a volume $V=1000fm^3$.}\label{Fig:366}
\end{figure}
\begin{figure}
  \resizebox{0.9\hsize}{!}{
  \includegraphics*{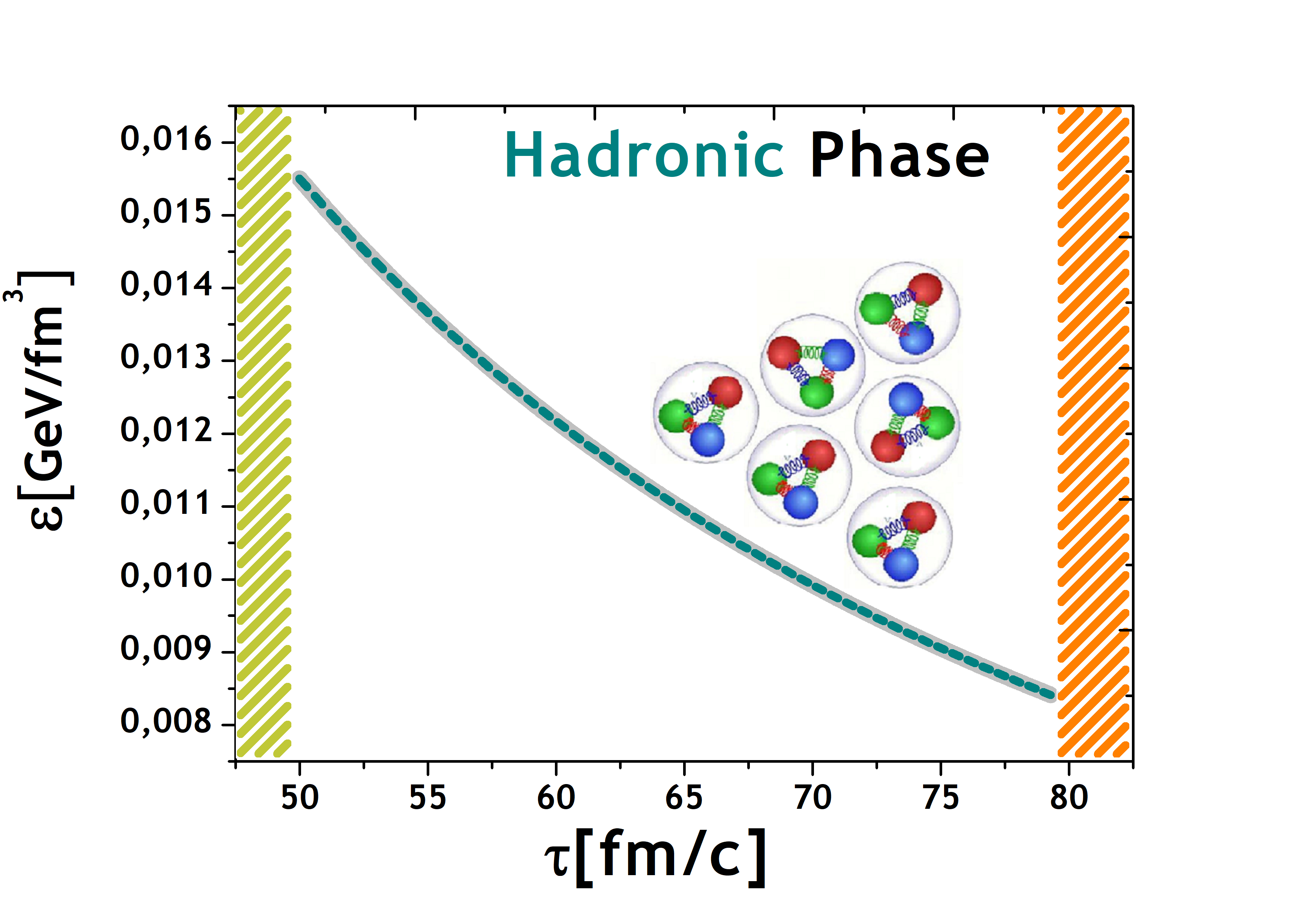}}
 \caption{Time evolution of the HM phase for a volume $V=1000fm^3$.}\label{Fig:37}
\end{figure}
\begin{figure}
  \resizebox{0.9\hsize}{!}{
  \includegraphics*{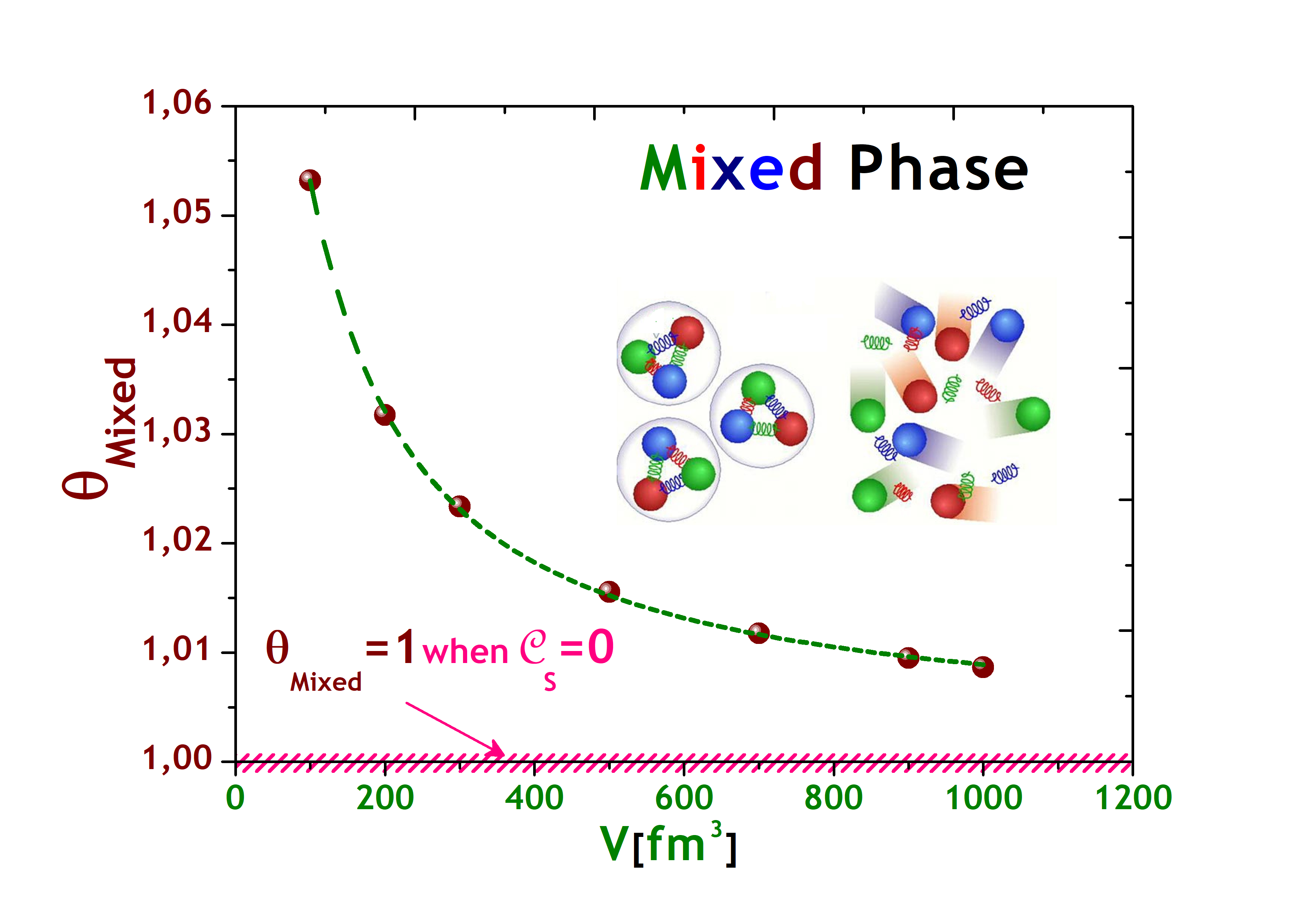}}
 \caption{FSS behavior of the exponent in the hydrodynamical evolution during the mixed phase: $\epsilon \propto \tau^{-\theta_{Mixed}}$.}\label{Fig:38}
\end{figure}

We show that the CPP is maintained for a remarkably long lifetime in colorless case than that in non-colorless case. Consequently, we say that the effect of the confining phase transition and the effect of the CC, makes the lifetime of the partonic matter remarkably longer than other cases. The same result was already obtained in a different context\cite{Youichi1989}. The CC has the same effect as the viscosity \cite{Mustafa1998}. Such a colorless state would exhibit some viscosity effects which can substantially affect the EoS by making it not ideal\cite{Ladrem2019} and slow down the cooling rate of the colorless PP mainly during the two first stages of the hydrodynamic expansion\cite{Biro1993}. The emergence of the color charge confinement from parton-parton interactions renders the CPP as a system having some viscosity and as a strongly correlated system. Also, under other models\cite{Vogt2007,QGPbblb2008}, the evaluation of the lifetime of the confining phase transition has provided results in agreement with ours.

\subsection{Periodic Cycle in Time Evolution, Pressure Anisotropy, and Viscosity.} \label{Cycle}
If we analyse the time evolution of the energy density $\epsilon(\tau)$ during the whole expansion, starting from zero the instant of the collision until the total freeze-out, one can notice a periodic cycle between different time scaling laws as clearly seen on the table(\ref{TableCycle}).
We can summarize that the time behavior of the energy density $\epsilon(\tau)$ as a function of proper time $\tau$ reveals a different scaling law in each stage of the time evolution. During the first stage which corresponds to the pre-equilibrium stage very well described by the Glasma model, the system manifests a free streaming behavior signaled by $\propto \tau^{-1}$\cite{Kovchegov,Krasnitz}. It can be understood as being due to non-interacting particles away from the collision point: the system expands freely without losing any energy. This scaling law means that the energy density multiplied by the time, $\tau \epsilon (\tau)$, saturates at a certain time $\tau_G$, sometimes called the formation time, i.e. the time at, which partons are liberated from the nucleon's wave-function, where the free-streaming expansion starts\cite{Fukushima2012}. This is natural on physical grounds, because the classical fields of the Glasma keep interacting strongly until $\tau_G$ , and become quasi-free at $\tau>>\tau_G$. The same time behavior is obtained for the energy density $\epsilon(\tau) $ during both the confinement phase transition stage and the hadronic free streaming stage. However, in the second and the fourth stages the energy density is described by another time behavior: $\epsilon(\tau) \propto \tau^{-4/3}$. Therefore, one can notice the periodicity of the behavior of the time evolution of the energy density $\epsilon (\tau)$ with  a definite time cycle. In our Colorless QCD-MIT Bag model, the transition from the hydrodynamic behavior $\epsilon(\tau) \propto \tau^{-4/3}$ during the colorless partonic phase into a free streaming behavior during the confinement phase transition $\epsilon(\tau) \propto \tau^{-1}$ normally occurs since the total partition function of the system contains the different parts of the system. Even so, the transition from the pre-equilibrium stage towards the hydrodynamic stage remains somewhere mysterious.

The traceless and conserved energy-momentum tensor (Rel.\ref{ConservationLaws}) of our system with no transverse coordinate dependence is uniquely determined in terms of the energy density $\epsilon(\tau)$ takes the form ,
\begin{equation} \label{TensorAnisotropy}
 \mathscr{T}^{\nu}_{\mu}= diag \big[\epsilon(\tau),\mathscr{P}_T(\tau),\mathscr{P}_T(\tau),\mathscr{P}_L(\tau)\big].
\end{equation}

The longitudinal pressure $\mathscr{P}_L(\tau)$ and the transverse pressure $\mathscr{P}_T(\tau)$ are consequently given by\cite{Baym1984,Blaizot2017} :
\begin{equation}
\left\{
\begin{array}{c}
  \mathscr{P}_L(\tau)=-\epsilon(\tau)-\tau \frac{d\epsilon(\tau)}{d\tau}  \\
 \mathscr{P}_T(\tau)=+\epsilon(\tau)+\frac{1}{2}\tau \frac{d\epsilon(\tau)}{d\tau}  \\
\end{array}%
\right.  \label{PLPT}
\end{equation}

The precise form of the energy density as a function of time $\epsilon(\tau)$ depends on the initial state and is governed by our complicated EoS is given by the plots (\ref{Fig:18},\ref{Fig:19}).

In addition to this, the first equation in (Rel. \ref{ConservationLaws2}) can be rewritten in the form,
\begin{equation}
\tau \partial_{\tau} \epsilon=- (\epsilon+\mathscr{P}_L) .
\label{BjorkenPL}
\end{equation}

It is in the case where $\mathscr{P}_L$ is exactly zero that the system reaches the limit of the free-streaming expansion leading to $\epsilon (\tau) \propto \tau^{-1}$. On the other hand, if the system is completely thermalized, meaning that the pressure is completely isotropic:$\mathscr{P}_L=\mathscr{P}_T$ leading to an expansion with positive $\mathscr{P}_L$ and producing work against the expansion of the matter, the energy density decreases faster than in the free-streaming case: $\epsilon (\tau) \propto \tau^{-4/3}$ using the relativistic EoS $\epsilon=3\mathscr{P}$. This time behavior is also obtained in the case of a purely longitudinal (i.e., Bjorken expansion). Therefore, the isotropization process cannot be attained using a Glasma alone as an initial state; thus, one should look for another mechanism occurring in the higher-order corrections, leading to the growth of the longitudinal pressure $\mathscr{P}_L$. An example of this mechanism is the instability due to the $\eta$-dependent fluctuations in the Glasma state\cite{Fukushima2012}. There is an alternative mechanism by which the time scaling law of the energy density is given by: $\epsilon (\tau) \propto \tau^{-1}$. This mechanism corresponds to a phase transition phenomenon. One can imagine that a kind of a phase transition occurs just before thermalization, during which the sound velocity can be reduced to zero. In this case, we can get $\tau \epsilon(\tau)\simeq constant$. We know that during the pre-equilibrium stage, the internal structure evolves enormously and quickly; it consists of pure gluonic state. As time proceeds, the phenomenon of creation of pairs comes into play, the gluonic pure state turns into another state, namely partonic state, which thermalizes afterward. Consequently, we can say that somewhere a phase transition is responsible for the scaling law $\epsilon (\tau) \propto \tau^{-1}$ during the pre-equilibrium stage.
Also in the case of expansion with dissipative corrections to first order, where $\sigma(T,V)$ represents the shear viscosity, the equation of motion may be rewritten in the form,
\begin{equation}
\tau \partial_{\tau} \epsilon=- (\epsilon+\mathscr{P}-\frac{4}{3}\frac{\sigma}{\tau}) .
\label{BjorkenViscosity}
\end{equation}
Assuming the constancy of $\sigma(T,V)$, the solution for the energy density is given by $\epsilon (\tau) \propto \tau^{-1}$ \cite{Danielewicz1985,Muronga2002,Baier2006}. Meaning that the partonic plasma cools more slowly than in the absence of dissipative effects $\sigma(T,V)=0$, and the corresponding expansion is just a constant-energy one $\tau \epsilon(\tau)\simeq constant$ rather than isentropic one. It also coincides with the maximum-entropy expansion considered in Ref \citen{Gyulassy1984}.

It appears clearly that the variation cycle of the time behavior in energy density is related to the anisotropy of the pressure. Generally, the study of the time growth of the longitudinal pressure $\mathscr{P}_L$ is done by the following ansatz: $\mathscr{P}_L(\tau)=\kappa(\tau)\mathscr{P}_T(\tau)$, where $\kappa(\tau)$ represents the anisotropy function, measuring the pressure anisotropy of the fluid. Also, one can quantifies this pressure anisotropy by introducing another dimensionless parameter: $\Delta(\tau)=\kappa^{-1}(\tau)-1$ \cite{Martinez2010}. The pressure anisotropy emerges at various stages of the hydrodynamic evolution of the hot matter created in the URHIC inducing even a variation in the geometric form of the system until its final isotropisation\cite{Strickland2014}. The isotropic case corresponds to the value $\kappa(\tau)=1$ and the free streaming behavior is obtained with $\kappa(\tau)=0$. Finally, one can say that the pressure anisotropy and the dissipative effects in the strongly interacting matter evolving from the first instants of the URHIC are crucial and important ingredients for the time evolution of the system.

\begin{table}
\centering
\caption{Different Time Scales}\label{TableCycle}
\begin{tabular}{|p{9mm}|p{9mm}|p{11mm}|p{10mm}|p{12mm}|p{12mm}|}
\hline
Time  & &  & & &  \\
Interval & $0-\tau_{i}$ & $\tau_{i}-\tau_{0}$ & $\tau_{0}-\tau_{H}$ & $\tau_{H}-\tau_{F}$ & $\tau_{F}-\infty$\\
\hline
Time  & &  & & &  \\
Scaling  & &  & & &  \\
Law  & $ \tau^{-1}$ & $\tau^{-4/3}$  &$\tau^{-1}$ & $\tau^{-4/3}$ & $\tau^{-1}$  \\
of $\epsilon(\tau)\propto$ & &  & & &  \\
\hline
State  & Glasma & Partonic & Mixed  & Hadronic & Hadronic  \\
       & & Matter & Phase  & Matter & Gaz  \\
\hline
$\kappa(\tau)$  & $   0$ & $  1$  &$  0$& $  1$ & $  0$  \\
\hline
\end{tabular}
\end{table}

\section{Conclusion}\label{sec:08}

The hydrodynamic evolution of a system undergoing the QCD deconfinement phase transition exhibits in different time intervals, quite different physics. This is a part of the richness of the QCD theory at positive temperature.
The time evolution of the confining phase transition from a CPP created in URHIC towards a Hadronic Gas, dominated by the hadronization phenomena, is discussed in the context of the hydrodynamical Bjorken expansion with our
Colorless QCD-MIT Bag Model. The finite volume Bjorken equation in the case of a longitudinal expansion scenario of an ideal relativistic medium in finite volume is solved using certain initial conditions and their effect is studied in detail showing a good agreement with the theoretical predictions. The evolution of the temperature as a function of the proper time $T(\tau,V)$ is then obtained at different volumes. Different times characterising different scales of the whole time evolution, like the time of transition point $\tau_0$, the hadronic time $\tau_H$, the lifetime of the CPP $\Delta \tau_{CPP}$ and the lifetime of the confining phase transition
$\Delta \tau_{PT}$ are calculated and their finite size scaling properties are studied in detail. In particular, we showed that our model with and without CC makes different predictions about times and lifetimes.

In the colorless case, these times and lifetimes scale with Taylor expansion depending on the inverse system size: $V^{-1}$ as a variable. The latter results have been rigorously confirmed from theoretical calculation based on the standard FSS theory of the thermal phase transition in which we insert the duality relation $T(\tau,V)$ and from which a new finite size scaling law is derived.

Also, the time evolution of some tRF as the order parameter $\mathcal{H}(\tau,V)$, energy density $\epsilon(\tau,V)$, pressure $\mathscr{P}(\tau,V)$ and the sound velocity
$\mathscr{C}_{s}(\tau,V)$ are investigated. The continuity properties of these tRF as a function of $\tau$ is discussed. Indeed, these properties result from the composition of the TRF with the duality relation $T(\tau,V)$.
We have also investigated the implications of having a variable speed of sound in the time evolution.

The confinement phenomenon and the underlying CC in any many-parton system are considered as resulting from the color interaction between partons, rendering the system colorless. This manifests itself as a
non-ideal character in EoS. The results of this work show that the CC has a reasonable mechanism for inducing liquid behavior in the PM system. The numerical value of $\theta_{CPP} < 4/3$ characterizing the negative power of the time decay of the energy density $\epsilon (\tau) \propto \tau^{-\theta_{CPP}}$ is consistent with our estimate of the plasma parameter $\Gamma_{CPP}$ \cite{Ladrem2019}.

On the other hand, the restriction to colorless states leads to an increased lifetimes and times, which may play a significant role in the confining phase transition, as well as for the process of hadronization.
The colorless QCD confining phase transition has three important consequences (i)to extend the different times and lifetimes (ii)to significantly slow down the cooling of the system  and (iii) to produce a (longitudinally) large volume of a hot hadronic gas at $T_0(V)$. It is natural that longitudinal expansion dominates early stages of an URHIC because simply in the Bjorken picture the system is characterized by an anisotropic initial condition. If one wants to develop a complete model one has to include the transverse expansion during the Bjorken type of longitudinal expansion is happening. During the confining phase transition and since the bulk sound velocity is zero: $\mathscr{C}_{s}(\tau,\infty)=0$, the expansion of the mixed phase does not contribute to the transverse expansion\cite{Milyutin1998}.

Finally, let us give some comments concerning the comparison between our results and those obtained from the predictions of other models(like Nucleation Model\cite{Csernai1992,Csernai1993,Mustafa1998}, Partonic Cascade Model\cite{Geiger1995}, Phenomenological Model based on an Operator-Field Langevin Equation\cite{Mizutani1988,Youichi1989} and Hot Glue Model\cite{Shuryak1992,Xiong1994}) used in this work. The mathematical approach in our model and its physical content are clearly different. Our study is focused on the CC, on the FSE and on the initial conditions which mark the end of the very important pre-equilibrium stage, in which pressure anisotropy is considered as a crucial and an important ingredient for the subsequent time evolution of the system. In spite of that, numerically and qualitatively our results are consistent with results obtained by these different models.

\section{Acknowledgement}\label{sec:09}
This research work was supported in part by the Deanship of Scientific
Research at Taibah University (Al-Madinah Al-Munawwarah, KSA).
M.L.H.L. would like to dedicate this work in living memory of
his daughter Ouzna Ladrem (Violette) died suddenly in March 24,2010(spring season). May
Allah has mercy on her and greets her in his vast paradise.


\begin{thebibliography}{9999}

\bibitem{QGP2018} H. Satz, {\it Extreme States of Matter in Strong Interaction Physics}. (Springer Verlag, 2018).
\bibitem{QCD2013} J. Collins, {\it Foundation of Perturbative QCD}, 1st edn. (CUP, 2013).
\bibitem{QCD1998} T. Muta, {\it Foundation of Quantum Chromodynamics: an Introduction to Perturbative Methods in Gauge Theories}, $2^nd$ edition. (World Scientific Pub Co Inc, 1998).
\bibitem{HAG2016} J. Rafelski (Ed.), {\it Melting Hadrons, Boiling Quarks - From Hagedorn Temperature to Ultra-Relativistic Heavy-Ion Collisions at CERN}. (Springer Verlag, 2016).
\bibitem{JetQuenchingExp} K. Adcox et al., Phys. Rev. Lett. {\bf 88, 2}, 022301(2002); G. Aad et al. {\it Phys. Rev. Lett.} {\bf 105, 25}, 252303-1(2010).
\bibitem{EllipticFlowExp} K.H. Ackermann, et al., Phys. Rev. Lett. {\bf 86, 3}, 402(2001); K. Aamodt, et al., {\it Phys. Rev. Lett.} {\bf 105, 25}, 252302(2010).
\bibitem{ParticleProduction} ALICE Collaboration, Phys. Lett. B {\bf 790, 1}, 35(2019)
\bibitem{Ladrem2011} M. Chekerker, M. Ladrem, F.C. Khanna, A.E. Santana, Int. J. Mod. Phys. A {\bf 26, 17}, 2881(2011).
\bibitem{Ladrem2013} M. Ladrem, M. Chekerker, F.C. Khanna, A.E. Santana, Int. J. Mod. Phys. A {\bf 28, 10}, 1350032(2013).
\bibitem{DeTar1985}  C. De Tar, Phys. Rev. D \textbf{32, 1}, 276(1985).
\bibitem{DeTar1987}   C. De Tar and J. Kogut, Phys. Rev. Lett. \textbf{59, 4}, 399(1987).
\bibitem{Milton1983}  K. A. Milton et al., Phys. Rev. D \textbf{27, 4}, 958(1983); Phys. Rev. D \textbf{27, 6}, 1348(1983).
\bibitem{Ladrem2019}  M. L. H. Ladrem et al., Int. J. Mod. Phys. A \textbf{34, 9}, 1950051(2019).
\bibitem{Ladrem2005} M. Ladrem et al., Eur. Phys. J. C \textbf{5, 2}, 711(2005).
\bibitem{Ladrem2015} M. Ladrem, M. A. A. Ahmed, Z. Z. Alfull, S. Cherif, Eur. Phys. J. C \textbf{75, 9}, 431(2015).
\bibitem{Redlich1980} K. Redlich and L. Turko, {\it Z. Phys. C} {\bf 5,3}, (1980) 201.
\bibitem{Turko1981} L. Turko,{\it Phys. Lett. B} {\bf 104,2},(1981)153.
\bibitem{ElzeCC} Hans-Thomas Elze et al., {\it Phys. Lett. B} {\bf 124,6}, 515 (1983); {\it Z. Phys. C} {\bf 24,3}, 361 (1984);{\it Phys. Lett. B} {\bf 179,4}, 385 (1986); {\it Phys. Rev. A} {\bf 33,3},(1986) 1879; {\it Phys. Rev. D} {\bf 35,2}, 748 (1987); Doktorarbeit, Frankfurt am Main (1985), unpublished.
\bibitem{Islam2014} C. A. Islam et al., J. Phys. G:Nuclear and Particle Physics   \textbf{41, 02}, 025001(2014).
\bibitem{Zakout2} I. Zakout et al., {\it Phys. Rev. C} {\bf 78,03}, 034916 (2008); Nucl. Phys. A \textbf{781,1-2}, 150(2007).
\bibitem{Abir2009} R. Abir and M. G. Mustafa, {\it Phys. Rev. C} {\bf 80,05}, 051903(R)(2009)
\bibitem{CGS1986} J. Cleymans, R. V. Gavai and E. Suhonen, Phys. Reports {\bf 130, 4}, 217(1986).
\bibitem{Mhamed2014}   M. A. A. Ahmed, \textit{Master thesis in theoretical physics}, Taibah University, Al-Madinah Al-Mounawwarah, KSA (2014).
\bibitem{Ejiri2006} S. Ejiri et al., {\it Phys. Rev. D} {\bf 73,5}, 054506 (2006).
\bibitem{Cheng2008} M. Cheng et al., {\it Phys. Rev. D} {\bf 77,1}, 014511 (2008).
\bibitem{Feynman1969}  R. Feynman, Phys. Rev. Lett. \textbf{23, 24}, 1415(1969).
\bibitem{Bjorken1983}  J. D. Bjorken, Phys. Rev. D \textbf{27, 1}, 140(1983).
\bibitem{Gelis2015}  F. Gelis, Int. J. Mod. Phys. E \textbf{24, 10}, 1530008(2015).
\bibitem{RFD2019} P. Romatschke and U. Romatschke, {\it Relativistic Fluid Dynamics In and Out of Equilibrium And Applications to Relativistic Nuclear Collisions}, 1st edn. (CUP, 2019).
\bibitem{Wong1994} Cheuk-Yin Wong , {\it Introduction to High-Energy Heavy-Ion Collisions} (World Scientific Publishing, 1994).
\bibitem{Florkowski2010} W. Florkowski, {\it Phenomenology of Ultra-relativistic Heavy-ion Collisions} (World Scientific Publishing, 2010).
\bibitem{QGPbblb2008} K. Yagi, T. Hatsuda and Y. Miake, {\it Quark-Gluon Plasma: From Big Bang to Little Bang }, $1^{st}$edition. (CUP, 2008).
\bibitem{Csernai1992}  L. P. Csernai and J.I. Kapusta, Phys. Rev. Letters \textbf{69, 5}, 737(1992).
\bibitem{Csernai1993}  L. P. Csernai et al., Z. Phys. C \textbf{58, 3}, 453 (1993).
\bibitem{Mustafa1998}  M. G. Mustafa, D. K. Srivastava, B. Sinha, Eur. Phys. J. C \textbf{5}, 711(1998).
\bibitem{Rischke1995} D. H. Rischke, S. Bernard and J. A. Maruhn, Nucl. Phys. A \textbf{595, 3}, 346(1995).
\bibitem{Rischke1996} D. H. Rischke and M. Gyulassy, Nucl. Phys. A \textbf{597, 4}, 701(1996); Nucl. Phys. A \textbf{608, 4}, 479(1996).
\bibitem{Geiger1995} K. Geiger, Phys. Rep. C \textbf{258, 4-5}, 237(1995).
\bibitem{Humanic2006}  T. J. Humanic, Int. J. Mod. Phys. E \textbf{15, 1}, 197(2006).
\bibitem{Hirano2002} Tetsufumi Hirano and Keiichi Tsuda, Phys. Rev. C \textbf{66, 5}, 054905 (2002).
\bibitem{Niemi2014} H. Niemi and G. S. Denicol, arXiv:1404.7327v1 [nucl-th](2014).
\bibitem{Youichi1989} Y. Akase et al., Progress of Theoretical Physics, {\bf 82, 3}, 591(1989).
\bibitem{Mizutani1988} M. Mizutani et al., Phys. Rev. D \textbf{37, 10}, 3033(1988)
\bibitem{Hind2019}   S. A. Hind, \textit{Master thesis in theoretical physics}, Taibah University, Al-Madinah Al-Mounawwarah, KSA (2019).
\bibitem{Hadeel2019}   A. K. Hadeel, \textit{Master thesis in theoretical physics}, Taibah University, Al-Madinah Al-Mounawwarah, KSA (2019).
\bibitem{VonGersdorff1986}  H. Von Gersdorff et al., Phys. Rev. D \textbf{34, 3}, 794(1986).
\bibitem{Vogt2007} R. Vogt, {\it Ultrarelativistic Heavy-Ion Collisions} (Elsevier, 2007).
\bibitem{Hirano2013} T. Hirano et al., Progress in Particle and Nuclear Physics \textbf{70}, 108(2013).
\bibitem{Derradi2016} R. Derradi de Souza, T. Koide and T. Kodamaa, Progress in Particle and Nuclear Physics \textbf{86}, 35(2016).
\bibitem{Miller2007}  M. L. Miller et al., Annu. Rev. Nucl. Part. Sci \textbf{57}, 205(2007).
\bibitem{Liu2014} F. M. Liu and S. X. Liu, Phys. Rev. C \textbf{89, 3}, 034906(2014).
\bibitem{Krajczar2011} CMS Collaboration, K. Krajczar, J. Phys. G  \textbf{38, 12}, 124041(2011).
\bibitem{QGPHP} kakudan.rcnp.osaka-u.ac.jp/jp/overview/world/QGP.html.
\bibitem{Biro1993} T. S. Biro et al., Phys. Rev. C \textbf{48, 3}, 1275 (1993).
\bibitem{Kovchegov} Y. V. Kovchegov,  Nucl. Phys. A \textbf{762, 3-4}, 298(2005); Nucl. Phys. A \textbf{774, 1-4}, 869(2006); Eur. Phys. J. A \textbf{29, 1}, 43(2006); Nucl. Phys. A \textbf{830,1-4}, 395c(2009).
\bibitem{Krasnitz} A. Krasnitz and R. Venugopalan,  Nucl. Phys. B \textbf{557, 1-2}, 237(1999); Phys. Rev. Letters \textbf{84, 19}, 4309(2000).
\bibitem{Krasnitz2003} A. Krasnitz et al.,  Nucl. Phys. A \textbf{717, 3-4}, 268(2003).
\bibitem{Krasnitz2001} A. Krasnitz et al.,  Phys. Rev. Letters \textbf{87, 19}, 192302-1(2001).
\bibitem{Fukushima2012} K. Fukushima and F. Gelis, Nucl. Phys. A \textbf{874}, 108(2012).
\bibitem{Lappi} T. Lappi, Phys. Rev. C \textbf{67, 5}, 054903(2003); {\it Phys. Lett. B} {\bf 643,1}, 11(2006).
\bibitem{Fukushima2007} K. Fukushima, Phys. Rev. C \textbf{76, 2}, 021902(2007).
\bibitem{Danielewicz1985} P. Danielewicz and M. Gyulassy, Phys. Rev. D \textbf{31, 1}, 53(1985).
\bibitem{Muronga2002} A. Muronga,  Phys. Rev. Letters \textbf{88, 06}, 062302-1(2002)+ Erratum Phys. Rev. Letters \textbf{89, 15}, 159901(2002); Acta Physica Hungarica (A) Heavy Ion Physics \textbf{15, 03-04}, 337(2002).
\bibitem{Baier2006} R. Baier, P. Romatschke,and U. A. Wiedemann  Phys. Rev. C \textbf{73, 06}, 064903(2006).
\bibitem{Gyulassy1984} M. Gyulassy and T. Matsui, Phys. Rev. D \textbf{29, 03}, 419(1984).
\bibitem{Janik2006} R. A. Janik and R. Peschanski, Phys. Rev. D \textbf{73, 04}, 045013(2006).
\bibitem{Blaizot2017} J. P. Blaizot and Li Yan, Journal of High Energy Physics \textbf{2017, 11}, 161(2017).
\bibitem{Milyutin1998} P. Milyutin, V. E. Fortov and N. Nikolaev, JETP Letters \textbf{68, 03}, 191(1998).
\bibitem{Shuryak1992}  E. Shuryak, Phys. Rev. Letters \textbf{68, 22}, 3270(1992).
\bibitem{Xiong1994} Li Xiong and Edward V. Shuryak, Phys. Rev. C \textbf{49, 4}, 2203(1994).
\bibitem{Cooper1975}F. Cooper, G. Frye and E. Schonberg, Phys. Rev. D \textbf{11, 1}, 192(1975).
\bibitem{Baym1983} G. Baym et al., Nucl. Phys. A \textbf{407, 3}, 541(1983).
\bibitem{Baym1984} G. Baym, Phys. Letters B \textbf{138, 1-3}, 18(1984).
\bibitem{Muller2019} D. M$\ddot{u}$ller, \textit{PhD thesis in theoretical physics}, Technische Universit$\ddot{a}$t Wien, $\ddot{O}$sterreich (2019).
\bibitem{Strickland2014} M. Strickland, Acta Physica Polonica B \textbf{45, 12}, 2355(2014).
\bibitem{Policastro2001} G. Policastro, D.T. Son, A.O. Starinets, Phys. Rev. Letters \textbf{87, 8}, 081601(2001).
\bibitem{Martinez2010}  M. Martinez and M. Strickland  Phys. Rev. C \textbf{81, 2}, 024906(2010).

\end{thebibliography}
\end{document}